\documentclass[preprint,12pt]{elsarticle}
\usepackage{amssymb}
\usepackage{amsmath}
\usepackage[utf8]{inputenc} 
\usepackage[T1]{fontenc}    
\usepackage{hyperref}       
\usepackage{url}            
\usepackage{booktabs}       
\usepackage{xcolor}

\newcommand{\rv}[1]{{\color{black}{#1}}}
\newcommand{\rrv}[1]{{\color{black}{#1}}}
\newcommand{\rrrv}[1]{{\color{black}{#1}}}
\usepackage{bm}

\usepackage{subcaption}
\usepackage[export]{adjustbox}
\usepackage{wrapfig}
\usepackage{overpic}
\journal{Acta Materialia}
\usepackage{multirow}
\usepackage[normalem]{ulem} 
\usepackage{comment}

\usepackage{subfiles}

\newcommand{\braket}[1]{{\left\langle{#1}\right\rangle}}
\begin{document}

\begin{frontmatter}
\title{\rrrv{Point defect design in (Ba,Sr)TiO$_3$ -- an insight on agglomeration}}
\author[1]{Sheng-Han Teng}
\author[1,2]{Anna Gr\"{u}nebohm}
\affiliation[1]{organization={Interdisciplinary Centre for Advanced Materials Simulation (ICAMS) and \\ Center for Interface-Dominated High Performance Materials (ZGH), Ruhr-University Bochum}, 
addressline={Univerist\"atstr. 150},
city={Bochum}, 
postcode={44780},
country={Germany}}
\affiliation[2]{organization={Faculty of Physics and Astronomy, Ruhr-University Bochum},
addressline={Univerist\"atstr. 150},
city={Bochum}, 
postcode={44780},
country={Germany}}


\begin{abstract}
Functional properties of ferroelectrics and their change\rrrv{s} with time depend crucially on the defect structure.
In particular, point defects and bias fields induced by defect dipoles modify the field hysteresis and play an important role in fatigue and aging.
However, a full understanding on how order, agglomeration and strength of defect dipoles affect phase stability and functional properties is still lacking.
To close these gaps in knowledge, we screen these parameters by \textit{ab\ initio} based molecular dynamics simulations with the effective Hamiltonian method for the prototypical ferroelectric material (Ba,Sr)TiO$_3$.
Our findings suggest that the {\it{active surface area}} of the defects, rather than the defect concentration is the decisive factor. 
For a fixed defect concentration, clustering reduces the {\it{active surface area}} and thus the defect-induced changes of phase stability and field hysteresis. Particularly planar agglomerates of defects appear as promising route for the material design as their impact on the field hysteresis can be controlled by the field direction and \rrrv{as} their impact on the phase stability shows a cross-over with the strength of the defect dipoles. 
For this agglomeration, we show that \rrrv{the recoverable stored energy can outperform the response of pristine (Ba,Sr)TiO$_3$ even in its paraelectric phase due to a pinched double-loop field hysteresis.}
\end{abstract}

\begin{graphicalabstract}
\includegraphics[width=\linewidth]{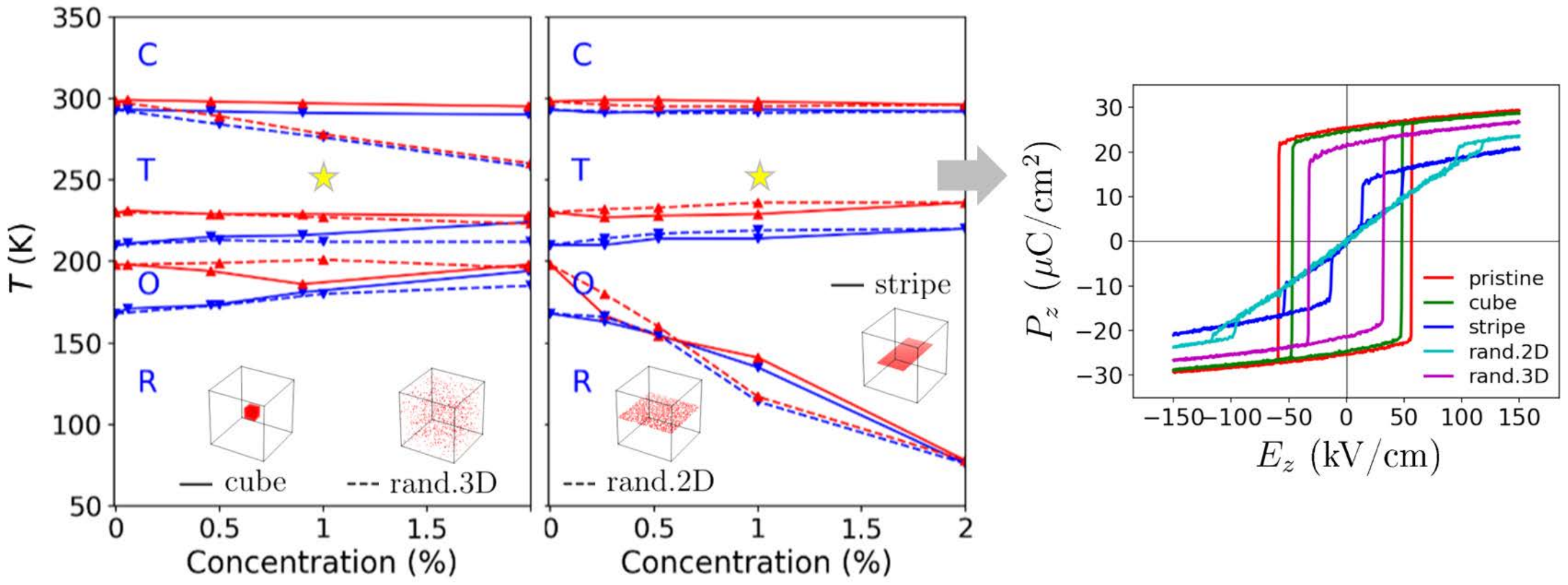}
\end{graphicalabstract}





\end{frontmatter}


\section{Introduction}
The exceptional dielectric, caloric, and piezoelectric functional responses of ferroelectric perovskites in the vicinity of structural phase transitions, as well as their adjustable field hysteresis in the ferroic phases, make these materials important for a wide range of applications \cite{whatmore100YearsFerroelectricity2021,grunebohmInterplayDomainStructure2021} including energy storage and electronic devices \cite{veerapandiyanStrategiesImproveEnergy2020,khoslaIntegrationFerroelectricMaterials2021}.
Adjusting transition temperatures and functional properties is commonly done by chemical substitution \cite{michelInterplayFerroelectricityMetallicity2021, veerapandiyanOriginRelaxorBehavior2022, acostaBaTiO3basedPiezoelectricsFundamentals2017,kleinFermiEnergyCommon2023}.
Thereby, one has to distinguish between isovalent substitution, e.g.\ substitution of Ba in BaTiO$_3$ by Sr or Ca ($\rm Sr_{Ba}^{\times}$, $\rm Ca_{Ba}^{\times}$), or Ti by Zr ($\rm Zr_{Ti}^{\times}$),  and aliovalent substitution, e.g.\ substitution of Ti by Fe or Nb ($\rm Fe_{Ti}'$, $\rm Nb_{Ti}^{\cdot}$).
In the former case, the substituents introduce local distortions and commonly reduce transition temperatures \cite{acostaBaTiO3basedPiezoelectricsFundamentals2017, dimouInitioBasedStudy2023}.
Although changes of the local polarization \cite{dimouInitioBasedStudy2023} and even non-polar cells may occur \cite{hillWhyAreThere2000, maStateTransitionElectrocaloric2017, veerapandiyanOriginRelaxorBehavior2022}, the dynamics of the local switching are not changed significantly for small defect concentrations, as all dipoles switch by ionic displacements happening on the ps-scale \cite{li90degreePolarizationSwitching2013, huangBehaviour180Polarization2014}.  
In the latter case, depending on the type and distribution of the dopant and the thermal treatment of the system, changes of transition temperatures and internal bias fields with shifting and pinching of hysteresis loops have been reported \cite{genenkoMechanismsAgingFatigue2015, krupska-klimczakOverviewNonpiezoelectricProperties2022, liEnhancingPropertiesLeadfree2022}. 
As \rrrv{these} internal restoring forces and pinched \rv{or antiferroelectric-like} hystereses are beneficial for energy storage \cite{veerapandiyanStrategiesImproveEnergy2020, daiCombinatorialOptimizationPerovskitebased2024, zhuNovelFerroelectricPolymers2012, rabeAntiferroelectricityOxidesReexamination2013, aramberriFerroelectricParaelectricSuperlattices2022} as well as \rrrv{for} large reversible piezoelectric \cite{renLargeElectricfieldinducedStrain2004} and electrocaloric responses \cite{liDopinginducedPolarDefects2021,grunebohmInfluenceDefectsFerroelectric2016}, this field arose considerable attention.
It is accepted that these observations are related to charge compensation by vacancies and the formation of defect complexes with local dipole moments, such as (Fe$_{\text{Ti}}'$-V$_{\text{O}}^{\cdot\cdot}$)$^{\cdot}$ and (Mn$_{\text{Ti}}''$-V$_{\text{O}}^{\cdot\cdot}$)$^{\times}$\cite{renLargeElectricfieldinducedStrain2004, zhangReorientationMnTiVODefect2008}.
Switching of these defect complexes requires slow ion migration which may take more than $10^7$~s at room temperature \cite{zhangReorientationMnTiVODefect2008, erhartFormationSwitchingDefect2013, zhangFlexoelectricAgingEffect2023}.
Thus these defects align with the surrounding polarization during aging and later act as restoring forces on domain and phase structure.
It is furthermore accepted that changes of the functional properties with time, e.g., functional fatigue, are related to ion migration and redistribution of defects \cite{verdierEffectThermalAnnealing2004, genenkoMechanismsAgingFatigue2015}.
In experimental samples, the distribution of defects depends on processing, thermal treatment, field-loading\rv{,} and internal and external boundaries, e.g.\ electrodes, grains, or dislocations \cite{lupascuAgglomerationMicrostructuralEffects2004, genenkoMechanismsAgingFatigue2015}. 
For example, \rv{their} uniform random distribution after ion-bombardment \cite{saremiLocalControlDefects2018}, planar arrays of oxygen vacancies in reduced atmosphere \cite{suzukiDislocationLoopFormation2001}, defect planes in a severely fatigued sample \cite{yangNanoscaleObservationTimeDependent2012}, \rv{segregated oxygen vacancies at crystallographic shear planes \cite{batukTrappingOxygenVacancies2015}}, clustering of oxygen vacancies \cite{akbarianUnderstandingInfluenceDefects2019}, and agglomeration of defects at dislocation lines \cite{adepalliTunableOxygenDiffusion2017} have been reported.

To tap the full potential of defect-engineering and to improve the understanding of functional fatigue, a microscopic understanding is mandatory. 
Since most microscopic modeling so far focuses on the alignment of single defect dipoles, even after decades of research, this understanding is still incomplete.
In the present study, we use an \textit{ab\ initio} derived effective Hamiltonian to screen the impact of defect distribution on ferroelectric phase diagrams and field hystereses in BaTiO$_3$ solid solutions.
Thereby we focus on point defects without local polarization and defect dipoles which re-orient orders of magnitude slower than the switching by Ti off-centering and can thus be approximated by fixed dipoles \cite{zhangReorientationMnTiVODefect2008, zhangFlexoelectricAgingEffect2023}, see Fig.~\ref{fig:batio3_unit_cell}.
We show that although the impact of the defects on phase transition and field hysteresis depends on defect strength and concentration, the distribution and agglomeration of defects play an even larger role. 


\section{Methods}
\label{sec:Methods}
\subsection{Molecular dynamics simulations}
We use the effective Hamiltonian from Ref.~\citenum{zhongFirstprinciplesTheoryFerroelectric1995}, which is parameterized based on first-principles calculations for the solid solution of (Ba,Sr)TiO$_3$ \cite{nishimatsuMolecularDynamicsSimulations2016}:
\begin{align}
	H^{\text{eff}}&=\frac{M_{\text{dpl}}^{*}}{2}\sum_{\bm{R},i}\dot{u}_i^2(\bm{R}) \nonumber\\ 
	&+V^{\text{self}}(\{\bm{u}\})+V^{\text{dpl}}(\{\bm{u}\})+V^{\text{short}}(\{\bm{u}\}) \nonumber\\
	&+V^{\text{elas,homo}}({\eta_\alpha})+V^{\text{elas,inho}}(\{\bm{w}\}) \nonumber\\
	&+V^{\text{coup,homo}}(\{\bm{u}\},{\eta_\alpha})+V^{\text{coup,inho}}(\{\bm{u}\},\{\bm{w}\}) \nonumber\\
	&+V^{\text{mod,inho}}(\{\bm{w}\},\{s\})-Z^{*}\sum_{\bm{R},i}\mathcal{E}\cdot u_i(\bm{R}), \label{eq:Heff}
\end{align}
where $\{\bm{u}\}$, $\{\bm{w}\}$, $\eta_\alpha (\alpha=1,...,6)$, $\{s\}$, $Z^{*}$ and $\mathcal{E}$ are local soft mode vectors in Cartesian coordinates $i(=x,y,z)$, local acoustic displacement vectors, the tensor of homogenous strain, the relative number of Ba and Sr atoms in each unit cell (u.c.), the Born effective charge associated with the soft mode, and an external electrical field, respectively.
The local mode gives the local polarization as \rv{$\bm{P}=Z^{*}\bm{u}/a^3$} where $a$ is the lattice constant. 

\rv{In (Ba,Sr)TiO$_3$, the transition temperatures and the polarization in all ferroelectric phases are reduced with the Sr concentration, see Fig.~\ref{app:pt_st_pristine}.}
Without loss of generality, we focus on Ba$_{0.7}$Sr$_{0.3}$TiO$_3$ with its Curie temperature close to room temperature using a random distribution of Sr in combination with the chemical modulation $V^{\text{mod,inho}}$ which depends on the Sr concentration in the unit cell and a pressure correction of $-1.2$~GPa, cf.~Ref.~\citenum{nishimatsuMolecularDynamicsSimulations2016}. 

\begin{figure}[tb]
  \centering
  \includegraphics[width=0.75\linewidth]{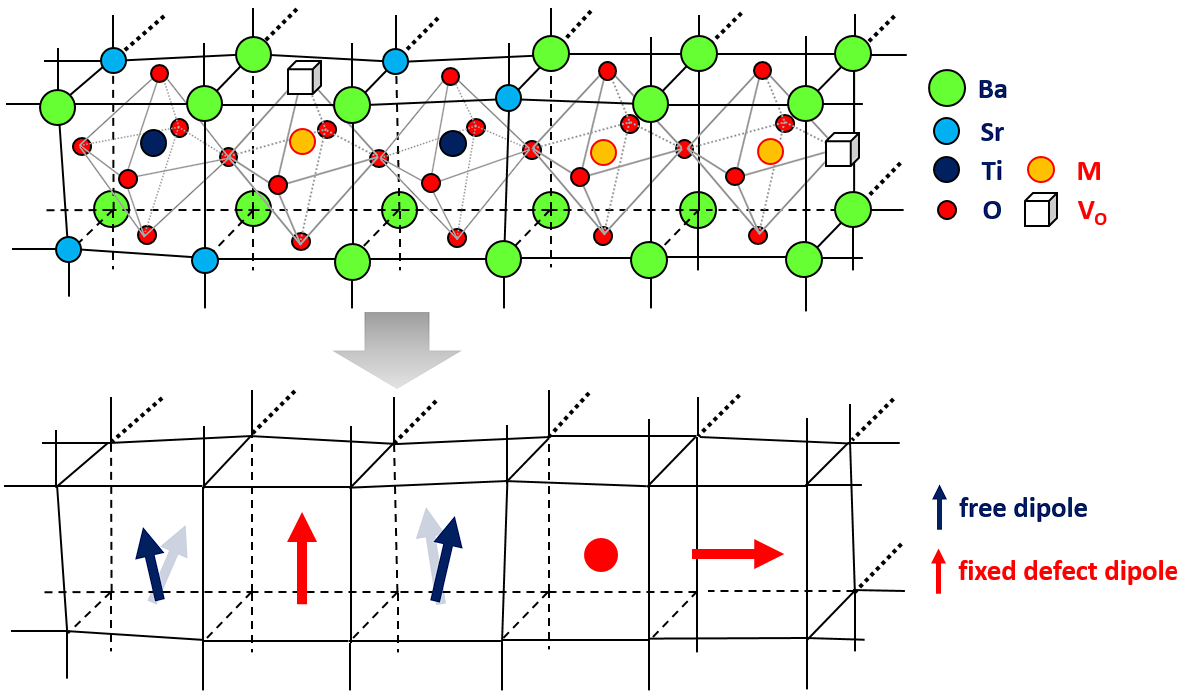}
  \caption{Coarse-graining and modeling of defects: The atomic displacements in all unit cells of (Ba,Sr)TiO$_3$ are coarse-grained to free dipoles (dark blue arrows) and local strain \rrrv{which is optimized internally}. If Ti is substituted by another atom (M), the local dipole moment can be reduced. We consider the limiting case of a non-polar cell (red dot). Defect complexes formed by a dopant M and an oxygen vacancy (\rrrv{V}$_{\text{O}}$) carry finite dipole moments which switch orders of magnitude slower than the surrounding material and are modeled by frozen-in dipoles (red arrows).}
  \label{fig:batio3_unit_cell}
\end{figure}

The other terms in Eq.~\eqref{eq:Heff} are the self-energy, $V^{\text{self}}$, the long-range dipole-dipole interaction, $V^{\text{dpl}}$, the short-range harmonic interaction between local modes, $V^{\text{short}}$, the elastic energy from homogeneous and inhomogeneous strains, $V^{\text{elas,homo}}$ and  $V^{\text{elas,inho}}$, the coupling terms between local soft mode and strain, \rrrv{$V^{\text{coup,homo}}$} and $V^{\text{coup,inho}}$.
As the strain is optimized internally, only the dynamics of the dipole moments are explicitly treated in molecular dynamics simulations, using $M^{*}_{\text{dpl}}$ the effective mass of the local soft mode. This coarse-graining is illustrated in Fig.~\ref{fig:batio3_unit_cell}.

A supercell with $L_x\times L_y\times L_z=48\times48\times48$~u.c.\ is combined with periodic boundary conditions and the dipoles are initialized with a normal distribution with $\braket{u_x}=\braket{u_y}=\braket{u_z}=0.00$~\AA{} and $\braket{u_i^2}-\braket{u_i}^2=0.03$~\rv{\AA$^2$}{} at 350~K or with $\braket{u_x}=\braket {u_y}=\braket{u_z}=0.10$~\AA{} and $\braket{u_i^2}-\braket{u_i}^2=0.01$~\rv{\AA$^2$}{} at 50~K to perform cooling and heating simulations.
For each temperature, we use a time step of $\Delta t=$2~fs and thermalize the system for 40000~steps in the canonical ensemble using the Nos\'{e}-Poincar\'{e} thermostat before we calculate the average properties in additional 20000~steps. The final configuration is the input for the next temperature step using  $\Delta T=\pm1$~K.
We monitor the absolute values of the macroscopic polarization along direction $i$ ($|\braket{P_i}|$), and the mean of the magnitude of the local polarization $\braket{|P_i|}$. Both quantities differ in the presence of domains. 
Unlike stated otherwise, the temperatures of ferroelectric to ferroelectric transitions are identified from the maximum/minimum of strain in the high/low-temperature phase. The paraelectric-ferroelectric transition temperature is determined from the maximal derivative of polarization with respect to temperature.

\begin{figure}[tb]
    \centering
    \includegraphics[width=0.6\textwidth]{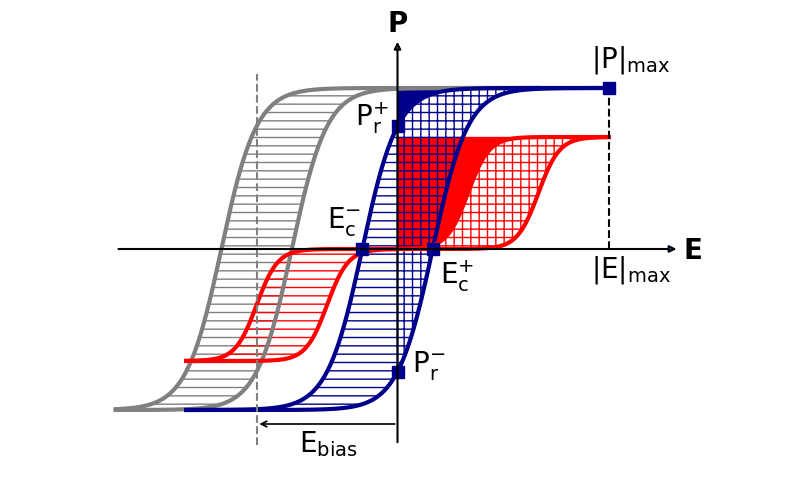}
    \caption{Characteristic field hysteresis representative for the pristine material (dark blue) compared to the pinched hysteresis in the presence of 2D defect agglomerates (red) without remanent polarization ($P_{\text{r}}^{\pm}=0$), and the shifted hysteresis with an internal bias field ($E_{\text{bias}}$) induced by defect dipoles collinear with the applied field (gray). $E_{\text{c}}^{\pm}$, and $P_{\text{r}}^{\pm}$ are coercive fields and remanent polarizations for both field directions ($E^{\pm}$) and $|P|_{\text{max}}$ is the polarization at the maximal field strength $E_{\text{max}}$. Hatched and colored regions illustrate work loss ($W_{\text{loss}}$) and recoverable energy ($W_{\text{rec}}$) in a unipolar field loop.}
    \label{fig:pehys}
\end{figure}

For chosen equilibrated configurations the field-hysteresis in $\braket{100}$-type fields with a magnitude of $150$~kV/cm is recorded starting with the negative field direction at the ramping rate of \rv{$1.5\times 10^{16}$}~V/m/s. 
Note that the virgin curve for this first poling is not discussed in the following. 
As shown in Figure~\ref{fig:pehys}, each hysteresis loop is characterized by the values of polarization without external bias ($P_{\text{r}}^{\pm}$) and in the maximal applied field ($P_{\text{max}}^{\pm}$), the coercive fields ($E_{\text{c}}^{\pm}$), which we determine by the maxima of $dP/dE$, the widths of the hysteresis loops ($\Delta E_{\text{c}}=|E_{\text{c}}^{(\rightarrow)}-E_{\text{c}}^{(\leftarrow)}|$), \rrrv{and} the bias fields ($E_{\text{bias}}=\frac{1}{N}\sum_i^N E_{\text{c}}^{(i)}$). \rv{Based on these quantities, one can determine \cite{veerapandiyanStrategiesImproveEnergy2020} the energy density which can be reversibly stored in a unipolar field-cycle 
\begin{equation}
W_{\text{rec}}^{\pm}=\int_{P_{\text{r}}^{\pm}}^{P^{\pm}_{\text{max}}} E dP\;,
\label{eq:rec}
\end{equation}
 the work losses for the switching from remanent to saturation polarization 
\begin{equation}
W_{\text{loss}}^{\pm}=\int_{P_{\text{r}}^{\mp}}^{P_{\text{max}}^{\pm}} E dP-W_{\text{rec}}^{\pm}\;,
\label{eq:loss}
\end{equation}
and the energy storage efficiency $\eta$  
\begin{equation}
\eta=\frac{W_{\text{rec}}}{(W_{\text{rec}}+W_{\text{loss}})}\;.
\label{eq:eta}
\end{equation}
}
 
Data as well as Jupyter notebooks used for the analysis are available via Zenodo \cite{tengImpactDefectsPhase2024} and Gitlab \cite{tengShengHanTengP1defect_study2024}\rrrv{, respectively}.

\subsection{Defect dipoles}
To depict the range of reasonable defect strengths and concentrations, we first categorize reports on defects from literature.
Typical oxygen vacancy and doping concentration are in the range of 0--2~\% (i.e.\ 10$^{14}$--10$^{21}$~cm$^{-3}$) \cite{genenkoMechanismsAgingFatigue2015, saremiLocalControlDefects2018, acostaBaTiO3basedPiezoelectricsFundamentals2017, krupska-klimczakOverviewNonpiezoelectricProperties2022}.
It has furthermore been reported that defect dipoles increase \cite{liuMultiscaleSimulationsDefect2017, machadoSiteOccupancyEffects2019} or decrease \cite{machadoSiteOccupancyEffects2019} the remanent polarization of the pristine material ($P_0$). 
To cover this range, we model defect strengths from $0$ to $1.8P_0$ with $P_0=28.41$~$\mu$C/cm$^2$ the polarization we find for (Ba$_{0.7},$Sr$_{0.3}$)TiO$_3$ at 210~K (tetragonal phase). If not stated otherwise, we use $u=0.1$~\AA, i.e.\ $0.9P_0$ as example. 
\begin{figure}[ht]
    \centering
    \resizebox{.8\linewidth}{!}{
    \begin{Overpic}[abs]{\begin{tabular}{p{0.3\textwidth}}\vspace{1.5cm}\\\end{tabular}}
        \put(15,5){\includegraphics[width=1.5cm]{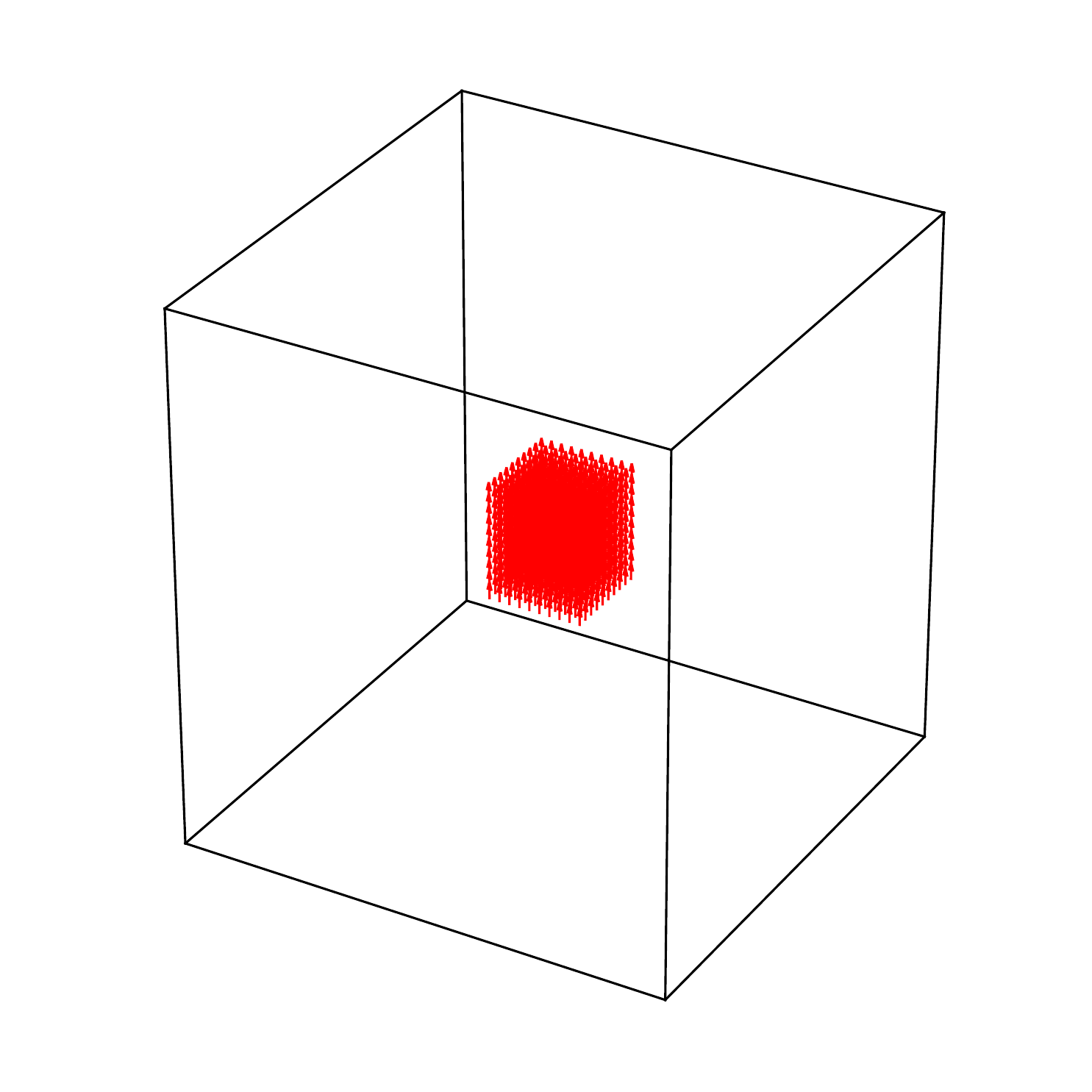}}
        \put(15,45){\normalsize \text{(a)}}
	\put(5,5){\includegraphics[width=.4cm]{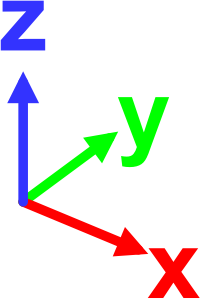}}
	  \put(65,25){\Large{$\Leftrightarrow$}}
        \put(85,5){\includegraphics[width=1.5cm]{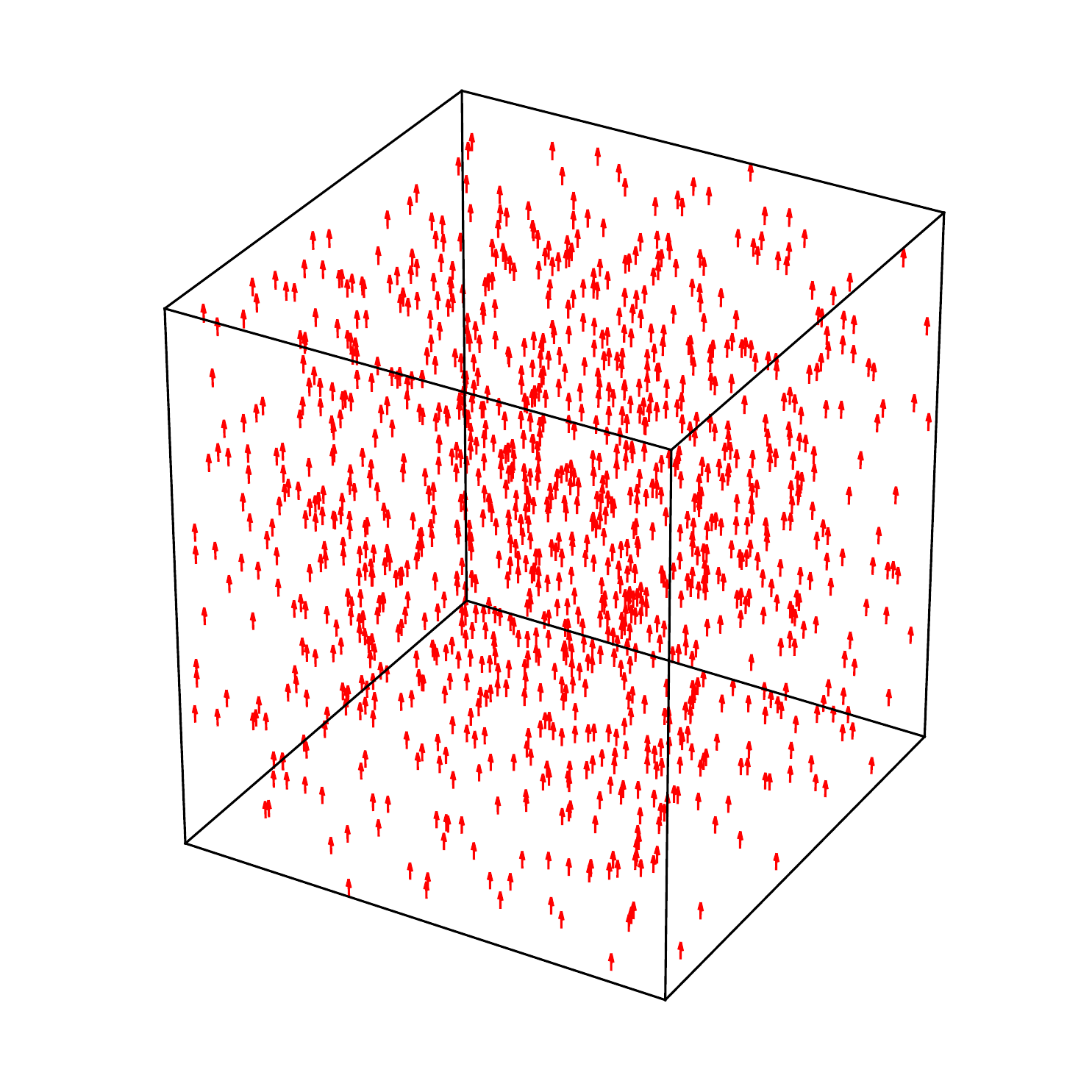}}
        \put(85,45){\normalsize \text{(b)}}
    \end{Overpic}
    \begin{Overpic}[abs]{\begin{tabular}{p{0.45\textwidth}}\vspace{1.5cm}\\\end{tabular}}
        \put(15,5){\includegraphics[width=1.5cm]{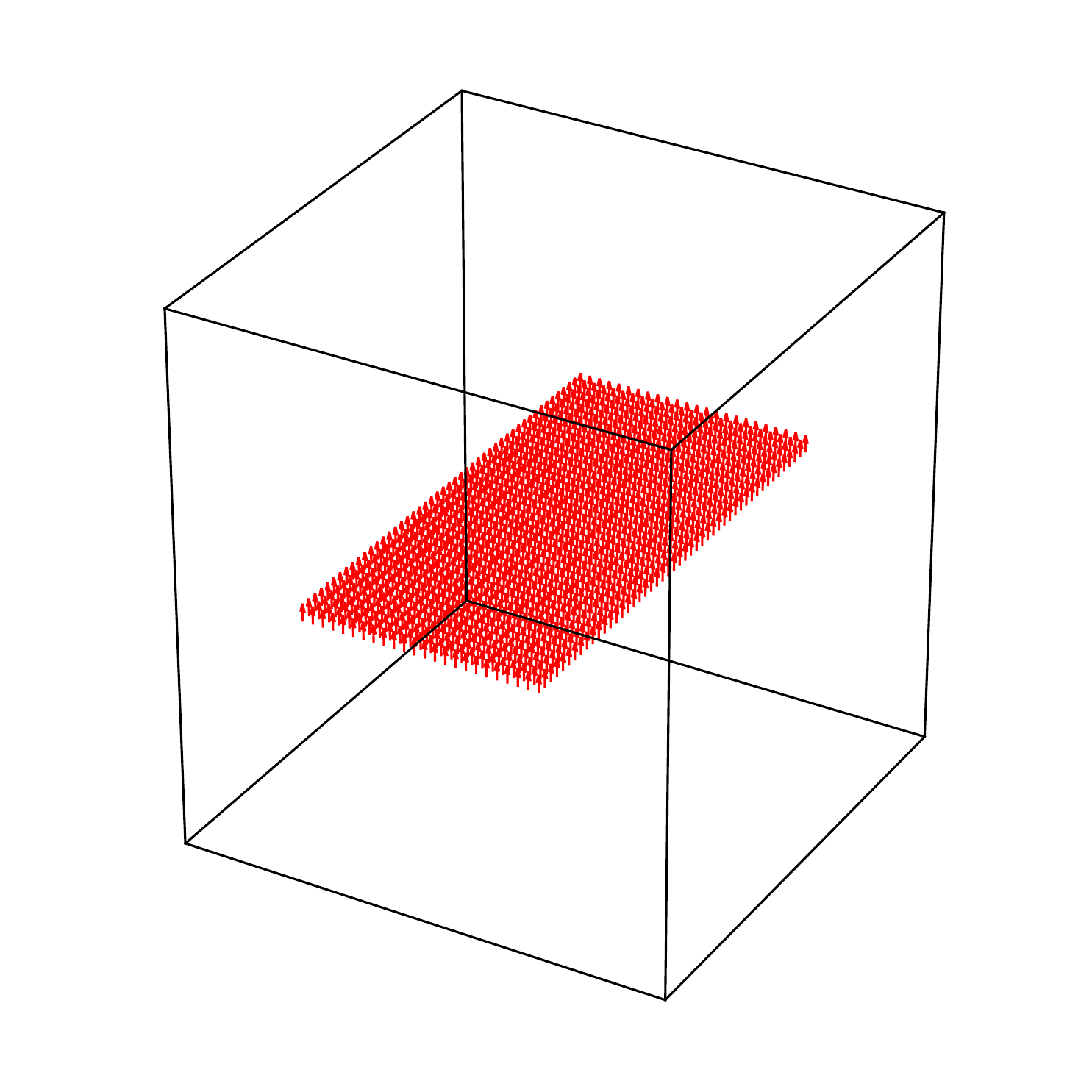}}
        \put(15,45){\normalsize \text{(c)}}
	  \put(60,25){\Large{$\Leftrightarrow$}}
        \put(80,5){\includegraphics[width=1.5cm]{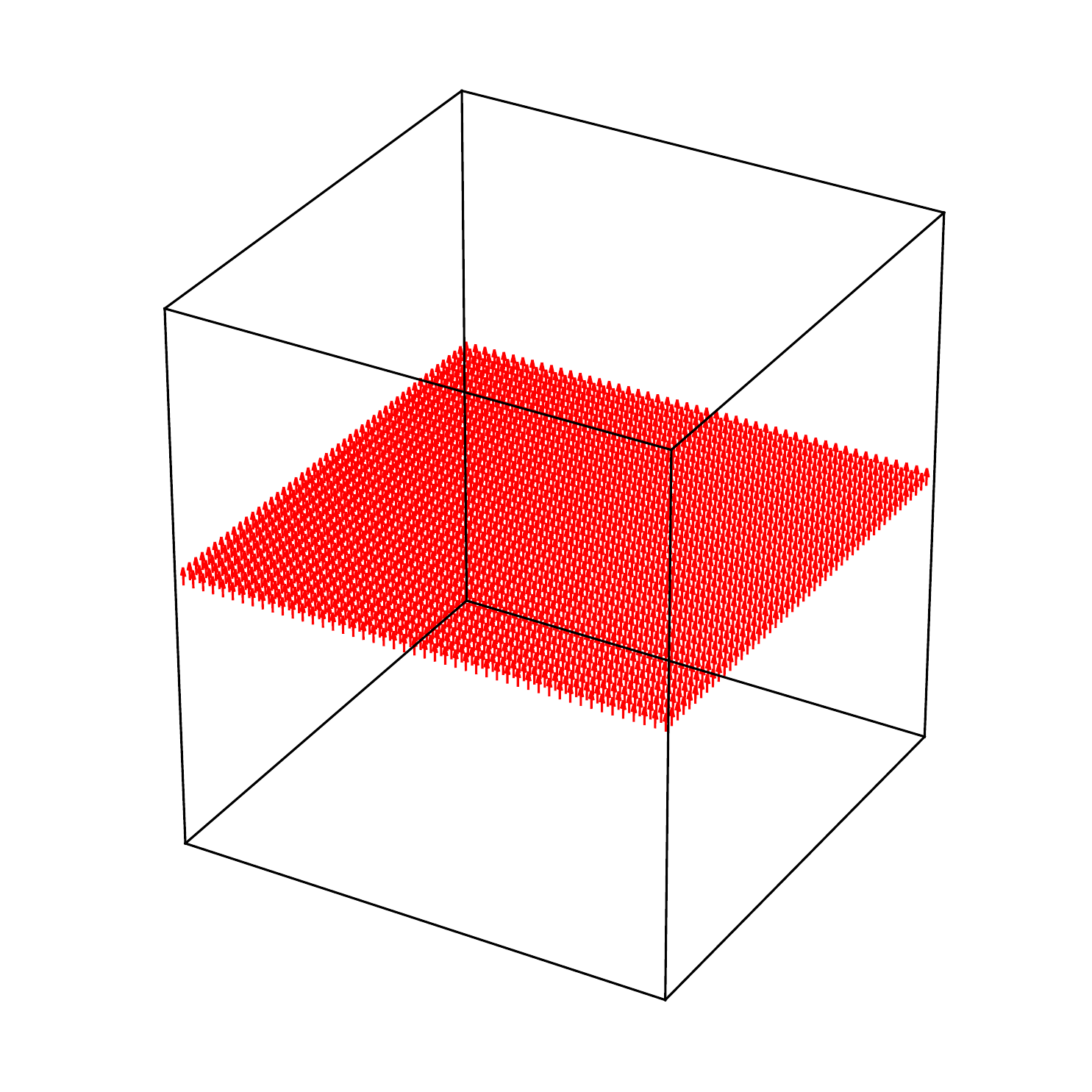}}
        \put(80,45){\normalsize \text{(d)}}
	  \put(125,25){\Large{$\Leftrightarrow$}}
        \put(145,5){\includegraphics[width=1.5cm]{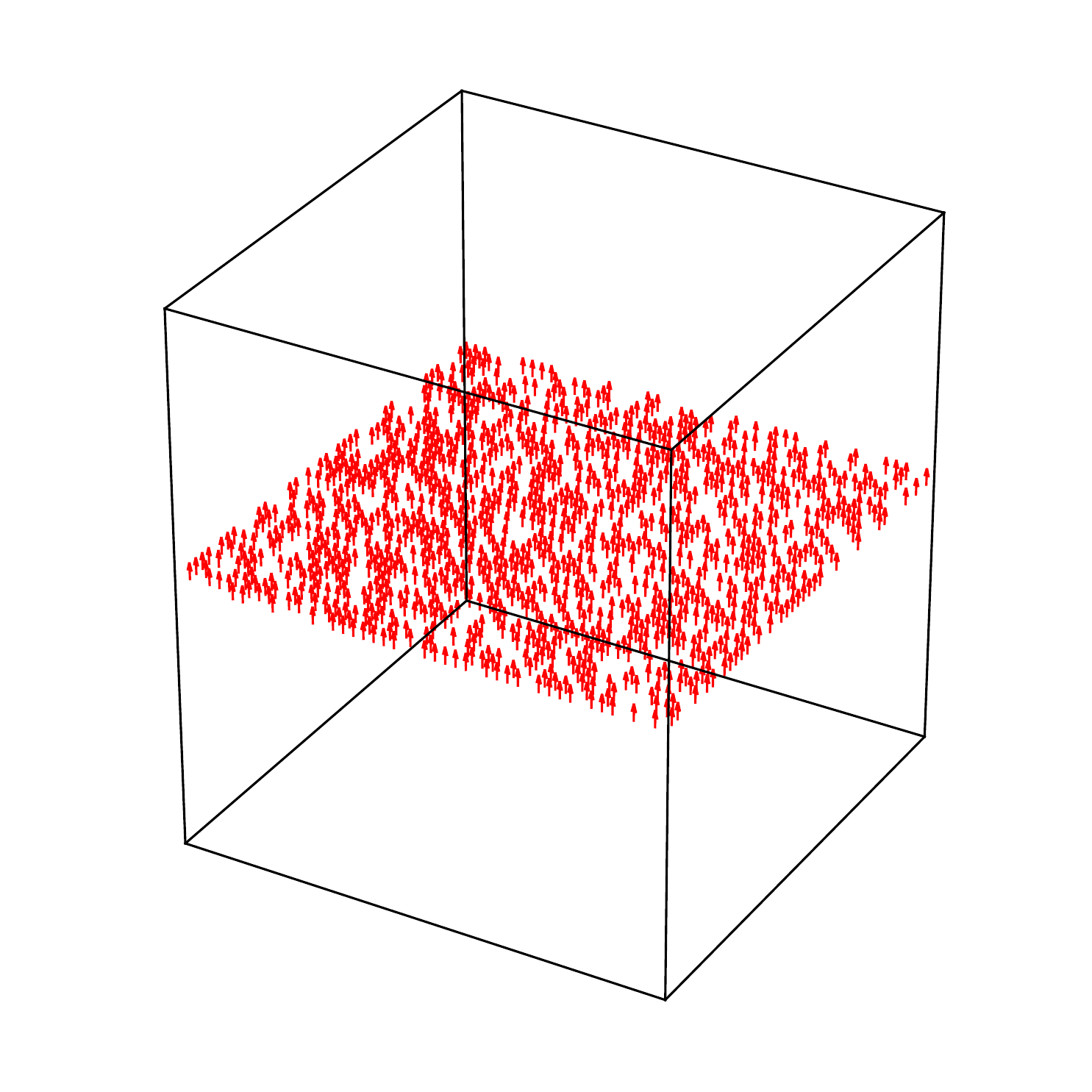}}
	\put(145,45){\normalsize \text{(e)}}
    \end{Overpic}
    }
    \resizebox{.8\linewidth}{!}{
    \begin{tabular}{|l|c|ccc|}
        \hline
        \multicolumn{1}{|c|}{\multirow{3}{*}{\textbf{Distribution}}} & \multirow{3}{*}{\begin{tabular}[c]{@{}c@{}}\textbf{Concentration}
        \\[0em](\%)\end{tabular}} & \multicolumn{3}{c|}{\textbf{Active surface area}} \\
        & & \multicolumn{3}{c|}{(each unit with an area of 15.9~\AA$^2$)} \\ \cline{3-5}
        & & \multicolumn{1}{c|}{$x$} & \multicolumn{1}{c|}{$y$} & $z$ \\
        \hline
        (a)~Cube & \begin{tabular}[c]{@{}c@{}}0.058--1.9\\[0em]($a$=4--13~u.c.)\end{tabular} 
        & \multicolumn{1}{c|}{32--338} & \multicolumn{1}{c|}{32--338} & 32--338 \\
        (b)~Rand.3D & 0.5--2 & \multicolumn{1}{c|}{1106--4424} & \multicolumn{1}{c|}{1106--4424} & 1106--4424\\
        (c)~Stripe & \begin{tabular}[c]{@{}c@{}}0.26--1\\[0em]($w$=6--24~u.c.)\end{tabular} & \multicolumn{1}{c|}{*} & \multicolumn{1}{c|}{0} & 576--2304 \\
        (d)~Plane & \begin{tabular}[c]{@{}c@{}}2\\[0em]($w=\infty$)\end{tabular} & \multicolumn{1}{c|}{0} & \multicolumn{1}{c|}{0} & 4608 \\
        (e)~Rand.2D & 0.26--1 & \multicolumn{1}{c|}{*} & \multicolumn{1}{c|}{*} & 576--2304 \\
        \hline
    \end{tabular}
    }
    \caption{Sampled distributions as well as considered concentrations of defects and their {\it{active surface areas}} (table) for: (a) cubic agglomerates, (b) random distribution in 3D, (c)--(e) agglomeration in one plane for (c) compact stripes, (d) full planes and (e) randomly. Thereby $a$ and $w$, \rrrv{are} the size and width of the cube and the stripe, respectively, and the surface areas along $x$, $y$ and $z$ are given in units of u.c., cf.~Fig.~\ref{fig:depolarization}.
    Each u.c.\ contributes an active area of about (3.9435~\AA)$^2$.
    \rv{$*$: Negligible effective  \textit{active surface area}, see text.}}
    \label{fig:defect_configurations}
\end{figure}

We take two types of point defects into account:  
First, substituents with different radii or electronic structures, e.g., Ce$^{4+}$ or Zr$^{4+}$ on B-site, may prevent the local off-centering in the stoichiometric material and suppress the local dipole moment  \cite{hillWhyAreThere2000,veerapandiyanOriginRelaxorBehavior2022}.
Second, pairs of a substituent with different valency and an oxygen vacancy may form defect dipoles M+V$_{\text{O}}$, which align with the polarization direction in the tetragonal phase during aging \cite{erhartFormationSwitchingDefect2013} and follow different dynamics than the intrinsic polarization. 
\rv{At ambient temperatures the dynamics of these defect dipoles is orders of magnitude slower compared to the switching of the ferroelectric polarization and can be neglected for the analysis of separate field cycles \cite{renLargeElectricfieldinducedStrain2004, genenkoMechanismsAgingFatigue2015,zhangReorientationMnTiVODefect2008,erhartFormationSwitchingDefect2013}.
Therefore, we} approximate these \rv{defects in fully aged samples} by frozen-in dipoles which are all aligned along one direction, cf. Ref.~\citenum{grunebohmInfluenceDefectsFerroelectric2016}. \rv{Defect dipoles may switch their direction with time and affect phase stability and transition temperatures accordingly, but these dipoles can be re-arranged by proper poling steps \cite{genenkoMechanismsAgingFatigue2015}.}
\rrrv{Figure~\ref{fig:batio3_unit_cell} shows possible microscopic realizations for both scenarios.}

\rrv{The model explicitly includes all long-range interactions between all dipoles and defects.  However, possible changes of the short-range interactions and the strain-polarization coupling 
 in the vicinity of defects are neglected and possible strain fields induced by the defects are not considered.}
\rv{Furthermore, the strength of the defect dipoles is used as a parameter. 
These approximations may change our predictions of transition temperatures and field hysteresis only quantitatively. 
}

To screen the impact of the defect distribution, we focus on \rv{the limiting cases of} defects randomly distributed in the material and clustering in planes and cubes.
Figure~\ref{fig:defect_configurations} collects all defect distributions and concentrations \rv{as well as their surface areas normal to $x$, $y$ and $z$}. 


\section{Results}
\label{sec:Results}
\subsection{Phase diagrams}
\begin{figure}[t]
  \centering
  \resizebox{1\linewidth}{!}{
  \begin{Overpic}[abs]{\begin{tabular}{p{.6\textwidth}}\vspace{4.8cm}\\\end{tabular}}
  	\put(0,0){\includegraphics[scale=.38,trim={0 0 0 0},clip]{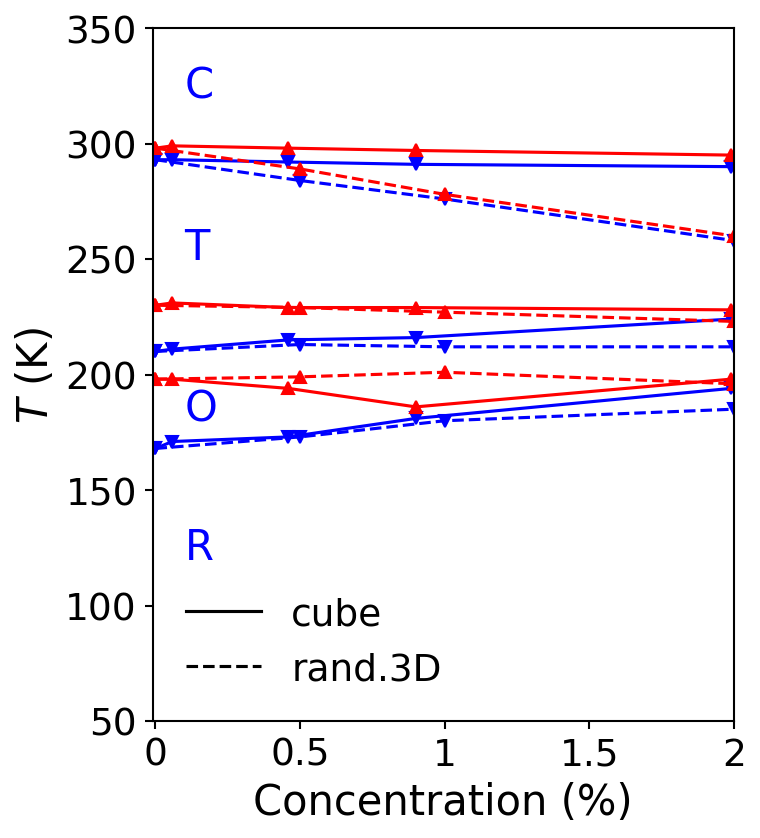}}
  	\put(30,126){\footnotesize 32}
		\put(50,126){\footnotesize 128}
		\put(70,126){\footnotesize 200}
		\put(118,125){\footnotesize 338}
		\put(48,109){\footnotesize \textit{1106}}
		\put(70,105){\footnotesize \textit{2212}}
		\put(113,100){\footnotesize \textit{4424}}
		\put(117,135){\text{(a)}}
		\put(138,0){\includegraphics[scale=.38,trim={2.5cm 0 0 0},clip]{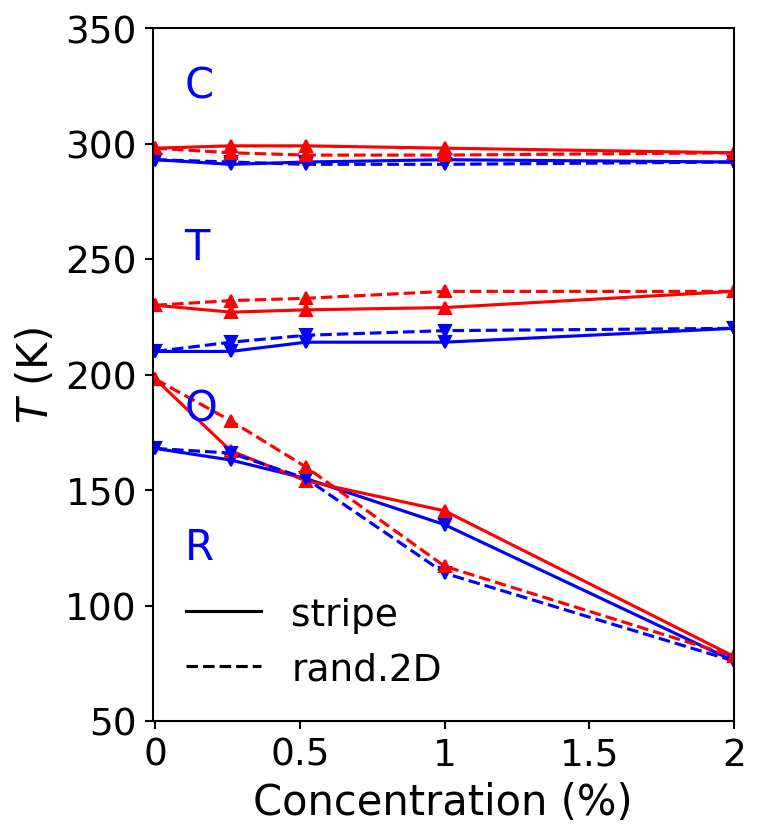}}
		\put(152,99){\footnotesize *\hspace{.3cm}*\hspace{.6cm}*\hspace{1.6cm}*}
		\put(155,76){\footnotesize 576}
		\put(170,68){\footnotesize 1152}
		\put(190,62){\footnotesize 2304}
		\put(224,44){\footnotesize 4608}
		\put(228,135){\text{(b)}}
		\put(125,8){\textcolor{white}{\rule{.4cm}{.4cm}}}
  \end{Overpic}
  \begin{Overpic}[abs]{\begin{tabular}{p{.42\textwidth}}\vspace{4.8cm}\\\end{tabular}}
    \put(0,15){\includegraphics[width=0.45\textwidth]{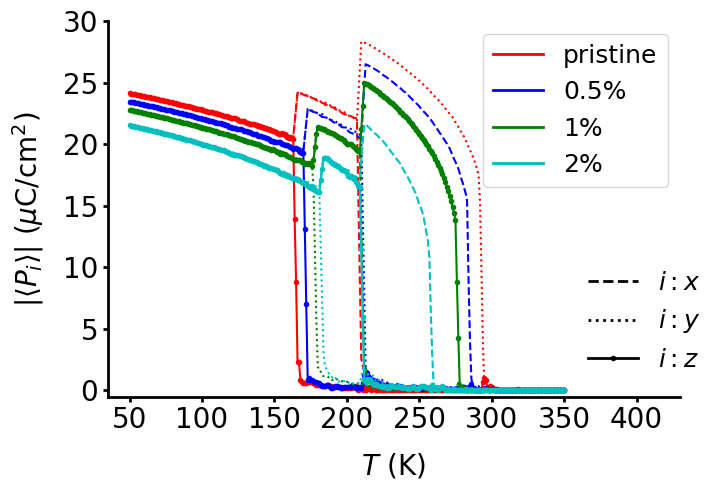}}
    \put(30,123){\text{(c)}}
  \end{Overpic}
  }
  \caption{Temperature - defect concentration phase diagrams in the presence of non-polar defects ($|u_{\text{D}}|=$0.0~\AA) for (a) cubic agglomerates (solid lines) and random distribution in 3D (dashed lines) and (b) stripes of defects (solid lines) and random distribution in 2D (dashed lines). Red and blue lines show transition temperatures for heating and cooling, respectively. Numbers give the approximate number of active surface units \rv{normal to the polarization directions in (a) all phases and (b) for the polarization along $z$ in the R phase}. Note that the defect agglomerates in (b) have no effective {\it{active surface area}} for T and O phases as discussed in the text \rv{(stars)}.
  \rv{(c) Underlying changes of polarization components $|\braket{P_i}|$, with $i: x,y,z$, in cooling simulations  induced by randomly distributed non-polar defects ($|u_{\text{D}}|=$0.0~\AA) of different concentration. }}
  \label{fig:PD_u0}
\end{figure}

We start our analysis with the phase stability without external electrical fields.
For defect-free (Ba$_{0.7}$,Sr$_{0.3}$)TiO$_3$, we find transition temperatures from cubic to tetragonal ($T_C^{\text{C-T}}$) to orthorhombic ($T_C^{\text{T-O}}$), and to rhombohedral ($T_C^{\text{O-R}}$) phases at  293~K, 210~K, and 168~K within the spread of experimental values of about 285--308~K, 217--222~K and 154--172~K, respectively \cite{zhouEffectYttriumDoping2022, vigneshwaranStudyLowTemperaturedependent2020, liouDielectricCharacteristicsDoped1997, lemanovPhaseTransitionsGlasslike1996}. 
These transition temperatures are modified with the concentration of non-polar defects, see Fig.~\ref{fig:PD_u0}. Thereby, changing the distribution of defects from random to 3D agglomerates (dashed and solid lines in subfigure (a)) or to 2D agglomerates (subfigure (b)) has a larger impact on the transition temperatures than changing their concentration. 

First, increasing the concentration of randomly distributed defects induces an approximately linear reduction of $T_C^{\text{C-T}}$, with a reduction of more than 35~K for 2~\% defects.
\rv{This is in qualitative agreement to experimental findings for the substitution of Ti by Zr \cite{acostaBaTiO3basedPiezoelectricsFundamentals2017}.}
\rv{In addition, the macroscopic polarization, see Fig.~\ref{fig:PD_u0}~(c), and the strain, cf.\ Fig.~\ref{app:StrT_rand3d_ag_c0}~(a), are systematically reduced with the number of randomly distributed defects.}
Second, agglomeration in 2D mainly reduces $T_C^{\text{O-R}}$, e.g.,\ by 33~K for 1~\% of defects and by 92~K for a full plane of defects\rv{, and the polarization in the R phase, cf. Fig.~\ref{fig:rand3d_ag_snapshots_c0}~(b)}. 
\rv{The difference between perfect stripes and a random distribution of defects in one plane is minor.}
Third, cubic agglomerates have a minor impact on all phase transition temperatures \rv{even if their overall concentration amounts to 2\%.}
Fourth, large agglomerates in 3D and 2D induce a slight increase of $T_C^{\text{T-O}}$, while $T_C^{\text{R-O}}$ increases slightly in the presence of large 3D agglomerates or a high concentration of randomly distributed defects.

\begin{figure}[t]
    \centering
    \resizebox{1\linewidth}{!}{
    \begin{Overpic}[abs]{\begin{tabular}{p{.24\textwidth}}\vspace{3.5cm}\\\end{tabular}}
        \put(12,0){\includegraphics[width=.24\textwidth]{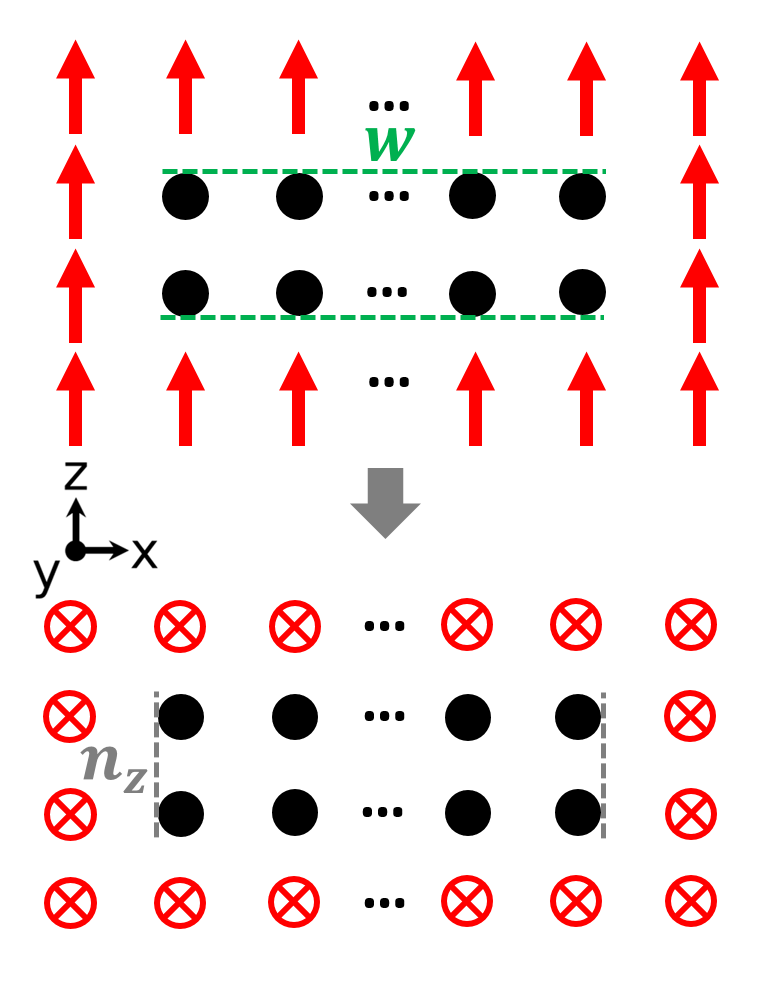}}
        \put(0,105){\text{(a)}}
    \end{Overpic}
    \begin{Overpic}[abs]{\begin{tabular}{p{.33\textwidth}}\vspace{3.5cm}\\\end{tabular}}
        \put(0,0){\includegraphics[height=4cm,clip,trim=0cm 0cm 0cm 0cm]{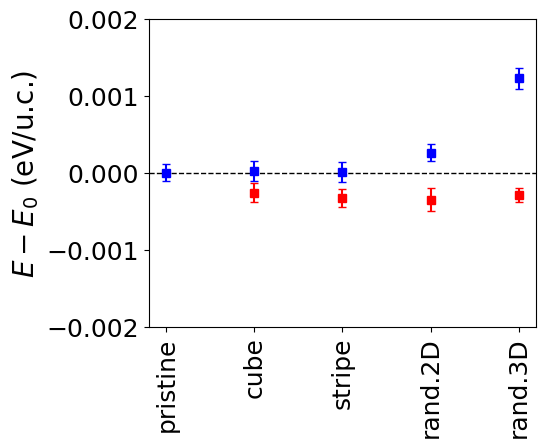}}
        \put(0,105){\text{(b)}}
    \end{Overpic}
    \begin{Overpic}[abs]{\begin{tabular}{p{.24\textwidth}}\vspace{3.5cm}\\\end{tabular}}
        \put(12,0){\includegraphics[width=.24\textwidth]{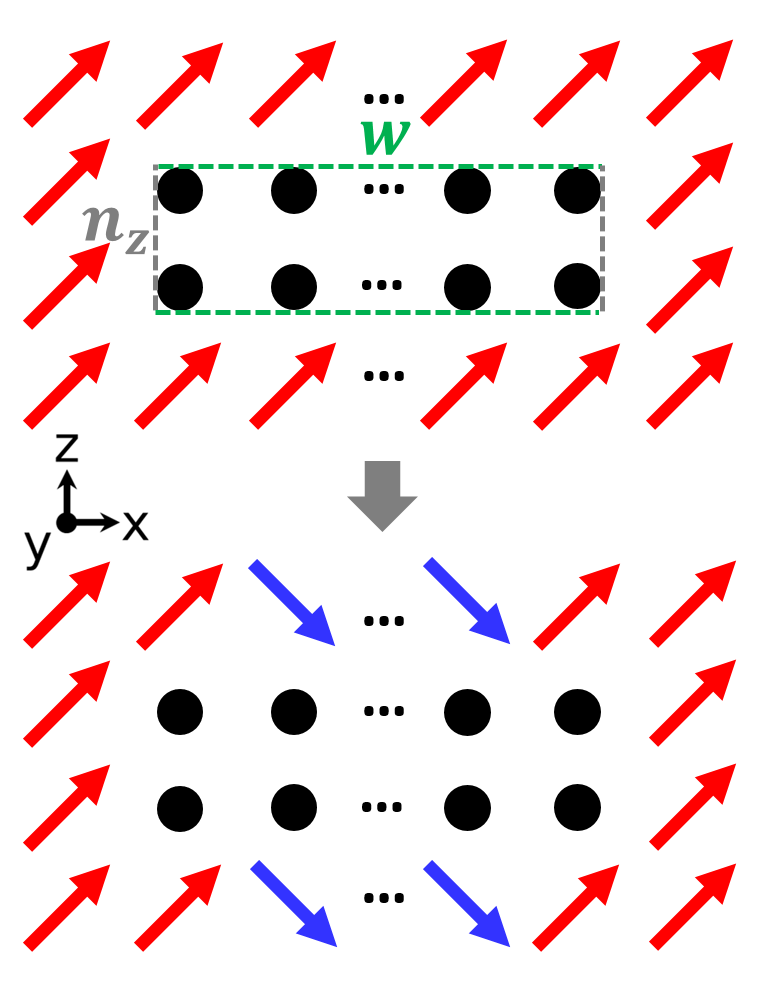}}
        \put(0,105){\text{(c)}}
    \end{Overpic}
    }
    \caption{\rv{Schematic illustration of the {\it{active surface areas}} of a stripe of non-polar defect dipoles (black) of width $w$, height $n_z$ and spanning the whole space along $y$ (not shown). Green and gray lines indicate the {\it{active surface areas}} for the polarization components $P_z$  
    and $P_x$ of the surrounding free dipoles. (a) In the T phase, the polarization aligns along \rrv{$y$} to reduce the {\it{active surface area}} (\rrv{$w>n_z>0$}).
    (c) In the R phase, the polarization component $P_z$ faces a large active area and can lower the depolarization field by the formation of domains. 
    (b) Changes of total energy by 1\% non-polar defects ($|u_{\text{D}}|=0.0$~\AA, blue) and polar defects ($|u_{\text{D}}|=0.1$~\AA~$\hat{z}$, red) in the T phase (250~K). 
    Mean values of 20000 time steps and their standard deviations are given.}}
    \label{fig:depolarization}
\end{figure}

Most of these observations can be understood by the \rv{concept of \rrrv{the} {\it{active surface area}} of the defect configuration: 
At the surface of non-polar defects, a large energy penalty is induced if the surrounding polarization has a normal component which induces a depolarization field ($E_{\text{dep}}\sim \nabla \bm{P}$) \cite{watanabeExaminationPermittivityDepolarization2021}.
On the other hand, this depolarization field is absent if the polarization is parallel to the surface of defects. 
Therefore, it is convenient to define the \textit{active surface \rrrv{area}} of each defect configuration relative to the surrounding polarization as the \rrrv{area} normal to the polarization component.
Figure~\ref{fig:depolarization} illustrates this concept and its consequences for the example of a stripe with a width of $w$, a thickness of $n_z$, and infinite \rrrv{width} along $y$ in the T phase. 
For \rrv{$w>n_z>0$}, the tetragonal state with $P_z$ had a maximum \textit{active surface area}, was high in energy and is not realized. 
Instead, the defect cluster favors polarization along \rrv{$y$} and its {\it{active surface area}} for the T phase \rrv{is zero}.
No changes of $T_C^{\text{C-T}}$, $P_x$\rrrv{,} or the energy of the system, see \rrrv{Fig~\ref{fig:PD_u0}~(b)}, are induced by the defects. 
In case of a random distribution of defects in the plane, their neighbors in the plane have a finite \textit{active surface area} and the energy in the T phase increases slightly. 
However, the induced modifications of $P_x$ are confined to that plane and do not couple to the macroscopic polarization. 
The effective active area marked as ($*$) in Fig.~\ref{fig:PD_u0} is thus negligible as are the differences in phase diagram and macroscopic polarization. 
The same is true for the O phase for all 2D orderings.
As illustrated in Fig~\ref{fig:depolarization}~(c), 2D agglomerates have a maximal \textit{active surface area} in the R phase, as one polarization component has to point along $z$. 
Because of that $T_C^{\text{O-R}}$ is reduced with the defect concentration and if the area of the agglomerate is sufficiently large, the depolarization field is reduced by the formation of a multi-domain structure.

For defects randomly distributed or agglomerated in 3D, the \textit{active surface area} is equal along $x$, $y$ and $z$ and impacts all three ferroelectric phases and phase transitions equally.
For a given defect concentration, \rrrv{the \textit{active surface area}} is maximal for a distribution without direct defect neighbors.
As the \textit{active surface area} along direction $i$ does not depend on $n_i$, 3D clustering of defects reduces the \textit{active surface area}, e.g.\ by a factor of 10 for 1~\% defects and the defect-induced change in energy at 250~K is negligible. 
Even for 2~\% defects in such a cluster, changes of macroscopic polarization and strain are barely visible, cf.\ Fig.~\ref{app:cc_c0}.
}

\rv{As perfect agglomerates are unlikely in experimental samples, it is important to test the robustness of the results, with respect to less perfect ordering. 
For 2D ordering, the one stripe, two stripes with different relative positions, as well as the random distribution of defects in one or two layers indeed result in the same changes of phase diagrams with the {\it{active surface area}} of the defects, cfs. Fig.~\ref{app:pt_2L}~(a) and Fig.~\ref{app:pt_asa}~(a)/(c)). 
}

While the main trends of the \rv{impact of defects on the system} are thus related to the {\it{active surface areas}} of the \rv{individual} defects, \rv{this concept is only exact for well-separated defects.}
\rv{In addition, also the distances and dilution of defects} impact phase stability and dipole configuration.
\rv{Defects randomly distributed in 3D mainly reduce the dipole moments of their nearest neighbors along the polarization direction, e.g.\ about 4.4\% dipoles are reduced to about 7.2~\rrrv{$\mu$C/cm$^2$} in the presence of 2\% defect, cf. Fig.~\ref{fig:rand3d_ag_snapshots_c0}.
As the defects start to cluster in 2D, more and more dipoles are reduced and the coupling between defects and dipoles becomes more long-ranged.  
Already a stripe with $w=6$~u.c. (i.e.\ 0.5\% defects) induces gradual rotations of dipoles in at least 3 layers next to the agglomerate. 
If a second defect cluster is within this distance, no bulk-like macroscopic polarization can form in the gap between them and both clusters act as one cluster with a joint effective \textit{active surface area}, cf. Fig.~\ref{app:pt_asa}.
Above a critical width of about $w=12$~u.c., negative domains with a diameter of $8$~u.c.\ form at the defect agglomerate, and for $w=24$~u.c.\ the dipoles normal to the defect are modified in the whole system by the formation of domains with R71 walls along $[110]$ due to depolarization fields~\cite{grunebohmInterplayDomainStructure2021}, cf.\ Fig.~\ref{app:snapshot_ag_cc_c0}. 
}


While a multi-domain structure forms for more than 25~\% coverage of compact agglomerates, more than 50~\% coverage is needed in a diluted plane. Thereby also the change of $T_C^{\text{O-R}}$ for small defect concentrations \rv{during heating} is larger in the former case.
On the other hand, the cross-section of the diluted defect plane is larger for a given defect concentration. 
Once domain walls form, the whole system splits into 2 equidistant R71 stripes and $T_C^{\text{O-R}}$ decreases faster for intermediate defect strengths.
These changes of the interaction range with the dilution of the defects can be related to the dilution of the depolarization field: 
On the one hand, the depolarization field by separated defects only couples to the local environment. 
On the other hand, the fields of dense defects add up, which increases the interaction length and results in the formation of domains above a critical size and density of the agglomerate.
Also large 3D agglomerates, e.g.\ $a=13$~u.c.\ with an active surface area of 338~u.c.\ (cf.\ Fig.~\ref{app:snapshot_ag_cc_c0}) induce rotations of the dipole in several unit cells and multi-domain structures, in this case for all three polarization directions. 

Non-polar defects also modify the character of the transitions and their thermal hysteresis. In the pristine material, all transitions are of first order and we find thermal hysteresis of 5~K, 20~K and 30~K, correspondingly. 
Generally, the hysteresis is systematically reduced with the concentration of defects, for example to 2~K, 11~K and 11~K for 2~\% of randomly distributed non-polar defects. 
Particularly, the character of the O--R transition can be modified by agglomerated defects, cf. continuous change of polarization for wide agglomerates in Fig.~\ref{fig:PD_u0}.  
For these configurations, hysteresis closes for less than 0.5~\% of defects due to the reduction of $T_C^{\text{O-R}}$ during heating. 
Analogously, thermal hysteresis is below our resolution for about 0.5~\% and 1~\% defects randomly distributed in a plane or clustered in a cube, respectively.
This can be understood by the discussed change of the polarization component normal to the agglomerate which induces monoclinic distortions and by domain wall formation in the R phase, as both these modifications may reduce the energy barrier for the transition \cite{nohedaPolarizationRotationMonoclinic2001,grunebohmInterplayDomainStructure2021}.
Note that a hysteresis of about 15~K remains at the T--O transition even for a full defect plane due to the weak coupling between defects and the polarization parallel to the plane. 

In summary, the impact of non-polar defects on the ferroelectric phase diagram is dominated by the effective {\it{active surface area}} of the defect configuration and, in case of compact defect agglomerates, depolarization fields modify the domain structure.
Similar but less pronounced changes of the phase stability have to be expected for locally reduced, but finite local dipoles switching by polar off-centering. 

\begin{figure}[t]
  \centering
  \resizebox{.7\linewidth}{!}{
	\begin{Overpic}[abs]{\begin{tabular}{p{.6\textwidth}}\vspace{4.8cm}\\\end{tabular}}
  	\put(0,0){\includegraphics[scale=.38,trim={0 0 0 0},clip]{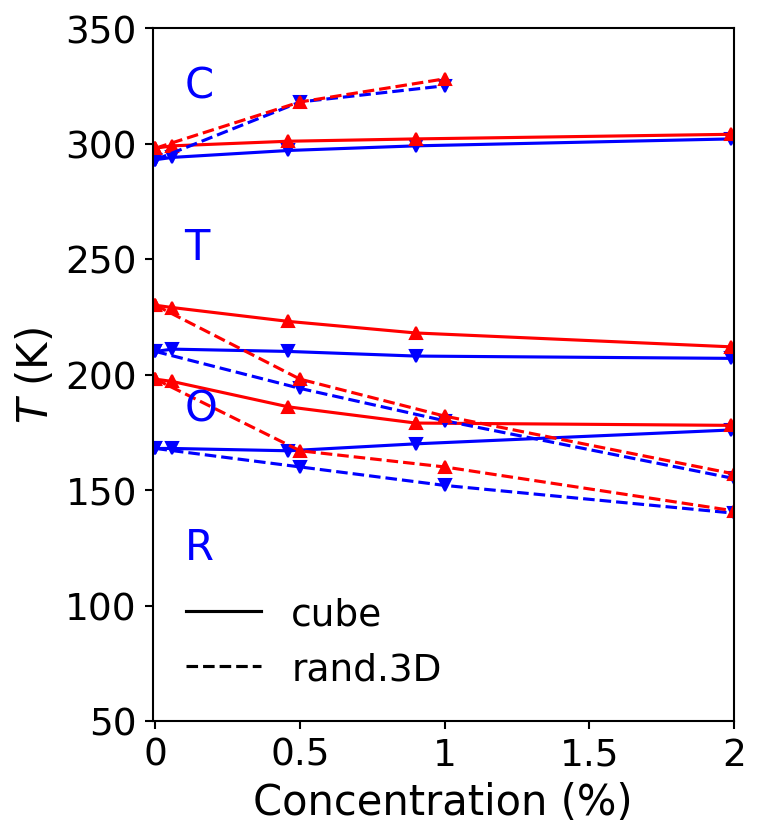}}
  	\put(30,114){\footnotesize 32}
		\put(48,114){\footnotesize 128}
		\put(70,114){\footnotesize 200}
		\put(118,114){\footnotesize 338}
		\put(48,58){\footnotesize \textit{1106}}
		\put(74,54){\footnotesize \textit{2212}}
		\put(113,50){\footnotesize \textit{4424}}
		\put(117,135){\text{(a)}}
    \put(138,0){\includegraphics[scale=.38,trim={2.5cm 0 0 0},clip]{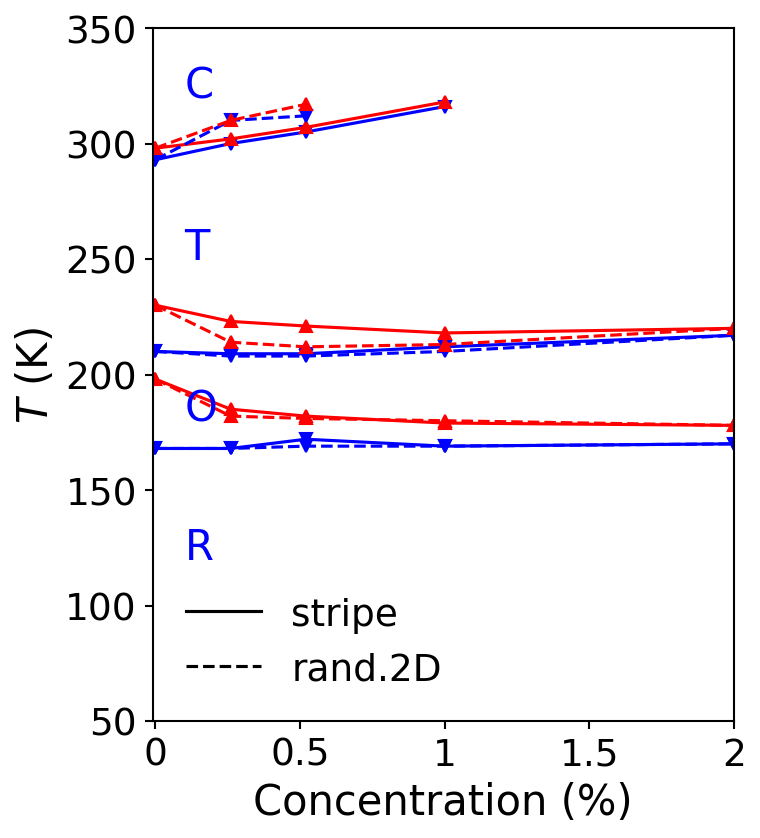}}
		\put(145,116){\footnotesize 576}
		\put(162,118){\footnotesize 1152}
		\put(186,120){\footnotesize 2304}
		\put(225,95){\footnotesize 4608}
  	\put(228,135){\text{(b)}}
		\put(125,8){\textcolor{white}{\rule{.4cm}{.4cm}}}
  \end{Overpic}
  }
  \caption{Temperature - defect concentration phase diagrams for polar defects ($|u_{\text{D}}|=$0.1~\AA $\hat{z}$). Symbols mark transition temperatures for cooling (blue) and heating (red). Note that no C-T transition can be detected for large defect concentrations and magnitudes beyond the critical point. Random ordering (solid lines) and agglomeration (dashed lines) are compared for (a) defects in 3D and (b) defects in a $z$-plane. Numbers give the approximate number of active surface units in all phases for (a) cube and random distribution (italic) and (b) for stripes.}
  \label{fig:PD_u0.1}
\end{figure}


Even larger changes of the phase stability are possible in the presence of defect dipoles related to M-V$_{\text{O}}$ complexes with different switching kinetics, as shown exemplary for defects with $|u_{\text{D}}|=0.1$~(\AA) $\hat{z}$, i.e.\ $u_x=u_y=0$, in Fig.~\ref{fig:PD_u0.1}. 
Again, small cubic agglomerates have a minor impact on all phases due to their small {\it{active surface area}}. 
Besides that, the trends differ from the non-polar case.
\rv{In the T phase, the polarization aligns with the direction of the defect dipoles and all other orderings have similar {\it{active surface areas}}.}
With the active area, $T_C^{\text{C-T}}$ increases until a critical point is induced by about 1\% defects. 
\rv{The polar defects lower the energy of the T phase at a given temperature, see Fig~\ref{fig:depolarization}~(b).}
Additionally, the transition temperatures under heating for $T_C^{\text{T-O}}$ and $T_C^{\text{O-R}}$ decrease for small defect concentrations.
Already 0.5~\% of defects randomly distributed in 2D or 3D thus reduce the thermal hysteresis of the T-O transition to 4~K. 
Analogous trends have been reported for external fields along $\langle 100\rangle$ \cite{maratheFirstprinciplesbasedCalculationElectrocaloric2016,maratheElectrocaloricEffectBaTiO32017,novakImpactCriticalPoint2013}.
In the \rv{T} phase, the defect dipoles thus act as an internal electrical field along the direction of the defect dipoles whose strength can be adjusted with the {\it{active surface area}} of the defects. 

\rv{In case of 2D agglomeration, the effective {\it{active surface area}} along $x$ and $y$ is again zero and the transition temperatures to O or R phases under cooling are not sensitive to the defect concentration. 
On the other hand, randomly distributed defects have an {\it{active surface area}} at their non-polar interfaces and thus induce a reduction of $T_C^{\text{T-O}}$ of more than $55$~K for 2~\% defects and also $T_C^{\text{O-R}}$ goes down with the defect concentration, see Fig.~\ref{fig:PD_u0.1}~(a). 
For 2D agglomerates with, e.g., $|u_{\text{D}}|=0.1$~(\AA) $\hat{x}$, the defects act similar to non-polar defects, cf.\ Fig.~\ref{app:PD_ux0.1}, with polarization along $x$ and $xy$ in the T and O phases and small changes of the transition temperatures, while $T_C^{\text{O-R}}$ decreases by $36$~K ($1$~\% of defects) and by $98$~K for a full defect plane for the R phase with polarization along $z$.}
Comparison between these defects and the non-polar ones shows that the former induces a slight increase of $T_C^{\text{C-T}}$ and a reduction of the thermal hysteresis at the T-O transition, while only in the latter case $T_C^{\text{T-O}}$ increases slightly for large defect concentrations. 
All these observations can be related to the small internal fields along $x$.

\rv{In addition to the main trends, also the defect dilution impacts their coupling to the material. 
For example, the change of $T_C^{\text{C-T}}$ and the critical defect concentration, vary slightly between compact stripes and random defect distributions. 
The main differences between compact stripes and less perfect ordering are slightly weaker defect fields in the T phase for the less perfect ordering and reduced macroscopic polarizations parallel to the defects (cf. Fig.~\ref{app:pt_2L}~(b)), or slightly decreased $T_C^{\text{O-R}}$ (cf. Fig.~\ref{app:pt_asa}~(b)/(d)). 
However, the main trends are identical.}


\begin{figure}[t]
  \centering
  \resizebox{1.\linewidth}{!}{
	\begin{Overpic}[abs]{\begin{tabular}{p{.6\textwidth}}\vspace{4.8cm}\\\end{tabular}}
  	\put(0,0){\includegraphics[scale=.38,trim={0 0 0.6cm 0},clip]{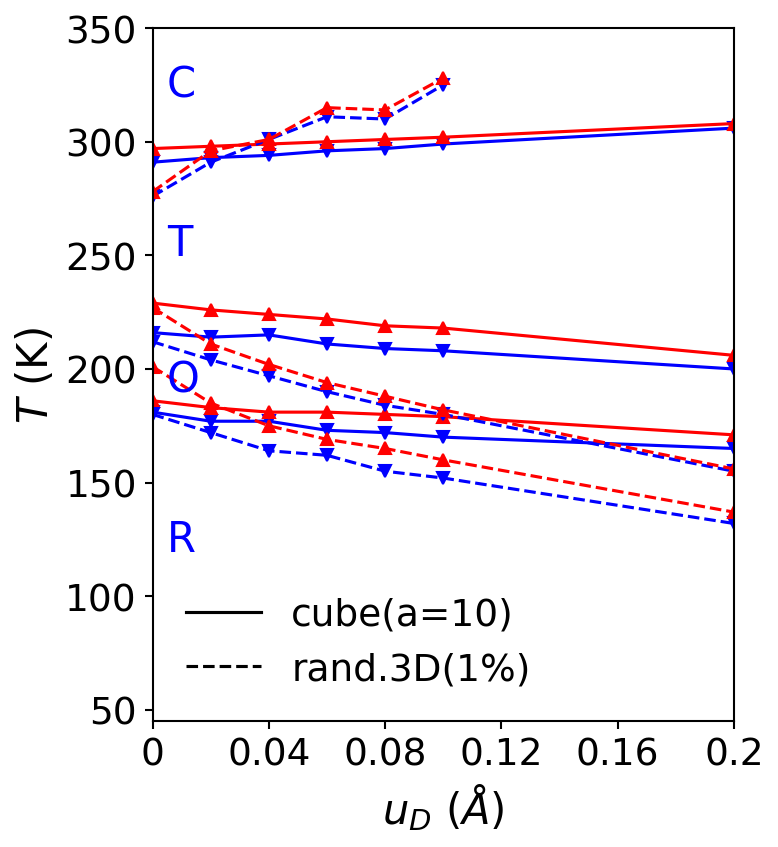}}
  	\put(122,8){\textcolor{white}{\rule{.5cm}{.45cm}}}
  	\put(117,30){\text{(a)}}
  	\put(138,0){\includegraphics[scale=.38,trim={2.4cm 0 0 0},clip]{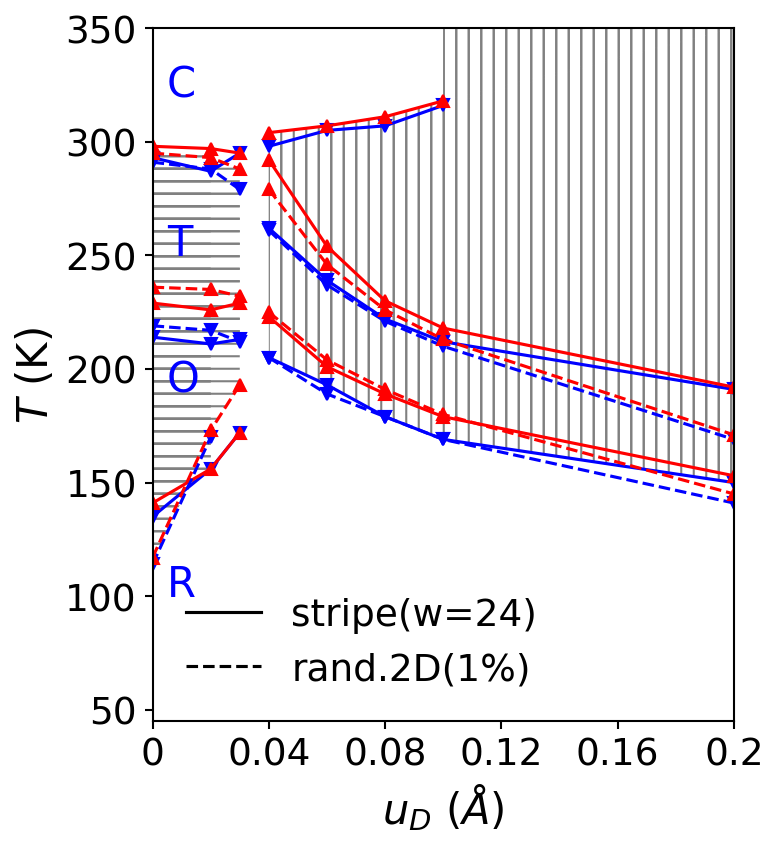}}
  	\put(160,135){\large $\ast$}
  	\put(228,30){\text{(b)}}
  \end{Overpic}
  \begin{Overpic}[abs]{\begin{tabular}{p{.15\textwidth}}\vspace{5cm}\\\end{tabular}}
    \put(4,86){\includegraphics[height=2.2cm,trim={.8cm .8cm 0 .8cm},clip]{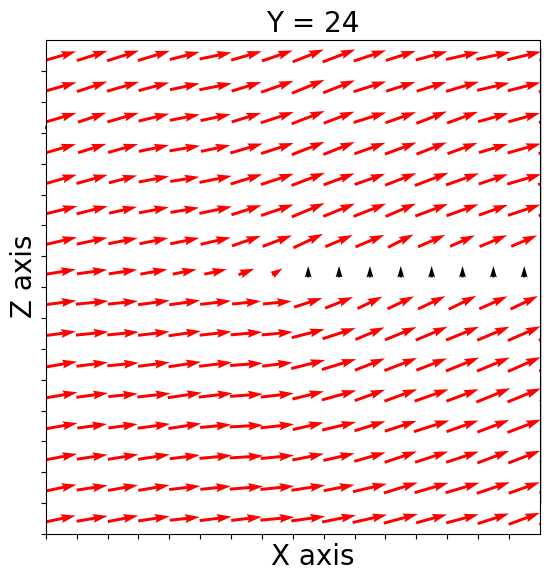}}
    \put(49,96){\text{(c)}}
    \put(4,20){\includegraphics[height=2.2cm,trim={.8cm .8cm 0 .8cm},clip]{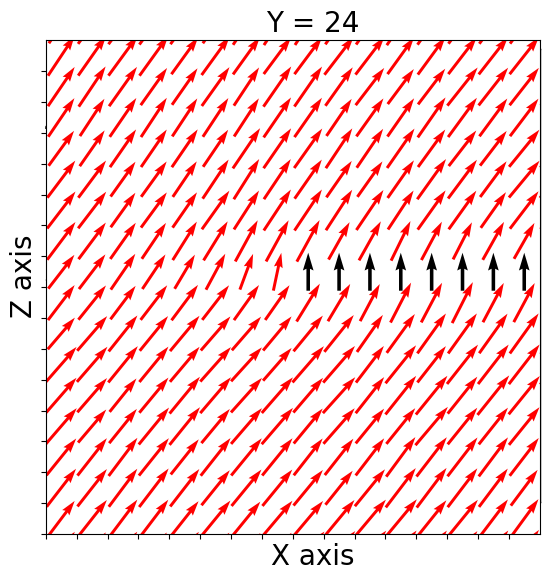}}
    \put(49,30){\text{(d)}}
  \end{Overpic}
  }
  \caption{Impact of the strength of 1~\% defects ($|u_{\text{D}}|$ $\hat{z}$) on the ferroelectric phase diagram. 
  Colors highlight transition temperatures for cooling (blue) and heating (red). Agglomeration (solid lines) and random ordering (dashed lines) are compared for (a) defects in a cube or 3D and (b) defects in one $z$-plane. \rrv{For both the T and O phases in (b),} horizontal ($=$) and vertical ($||$) hatches mark $P_z=0$ and $|P_z|>0$, respectively. (c) and (d) show corresponding time-averaged local dipoles at 200~K for weak defect dipoles ($|u_{\text{D}}|=0.03$~\AA{} $\hat{z}$) and strong defect dipoles ($|u_{\text{D}}|=0.1$~\AA{} $\hat{z}$), respectively. $^{\ast}$Note that beyond $0.04$~\AA{} the sample with 2D random defects is beyond the critical point of the C-T transition.}
  \label{fig:PD_um}
\end{figure}

The second obvious route to modify the internal field induced by the defect dipoles is tuning of the defect strength.
For random defect distribution and cubic agglomerates, the increase of the internal field with defect strength induces the same changes of the phase diagram as the change of the defect concentration, see Fig.~\ref{fig:PD_um}~(a).

For 2D agglomerates \rv{and defects randomly distributed in plane} with polarization along $z$, where is a cross-over from weak to strong defect dipoles, see Fig.~\ref{fig:PD_um}~(b). 
Below $u_z=0.03$~\AA{}, e.g.\ about 27\% of the saturation polarization of the pristine material, the defect dipoles act similar to non-polar defects with polarization in the T and O phases pointing along $P_y$ and $P_y=P_x$, no effective {\it{active surface area}} and no change of $T_C^{\text{C-T}}$ and $T_C^{\text{T-O}}$. 
Only in the R phase\rv{,} one polarization component points parallel to the defects and $T_C^{\text{O-R}}$ increases with their strength. 
Thereby the O--R transition becomes continuous, cf.\ Fig.\ref{app:ag_24_ds_24x48_cp}~\rv{(a)--(b)}.
\rv{These weak defects are smaller than the surrounding free dipoles above $T_C^{\text{C-T}}$, thus inducing depolarization fields.}
For strong defects ($u>0.03$~\AA{}), the polarization in the T phase point along $z$ and $T_C^{\text{C-T}}$ increases \rv{with defect strength} up to the critical point, while $T_C^{\text{O-R}}$ and $T_C^{\text{T-O}}$ decrease with the internal defect field.
\rv{As soon as their magnitude is larger than the typical values of free dipoles slightly above $T_C^{\text{C-T}}$, these defects start to act as an internal electric field, promoting the formation of the ferroelectric phase along that direction, cf.\ discussion in \ref{sec:app_macro}.} 

At the cross-over point, one has to expect that states with $P_y$ and $P_z$ are degenerated which may induce multi-critical points with coexisting C, T\rv{,} and O as well as with coexisting R, O\rv{,} and T phases.
\rv{This is independent of whether defects agglomerate into perfect stripes or are randomly distributed in 2D.}
The exact cross-over points are beyond our scope, for the chosen resolution of  0.01~\AA, the transition temperatures of the C, T and O as well as of T, O and R phases indeed approach each other.

In summary, in case of 2D agglomeration,  non-polar and weak defect dipoles have the same impact on the phase stability while strong defects act as an internal bias field, while defects randomly distributed in 3D induce an internal bias field for all field strengths.

\subsection{P-E Hystereses}
\begin{figure}[t]
    \centering
    \resizebox{\linewidth}{!}{
    \begin{Overpic}[abs]{\begin{tabular}{p{.43\textwidth}}\vspace{4cm}\\\end{tabular}}
        \put(0,0){\includegraphics[scale=.47,clip,trim=0cm 0cm 4cm 0cm]{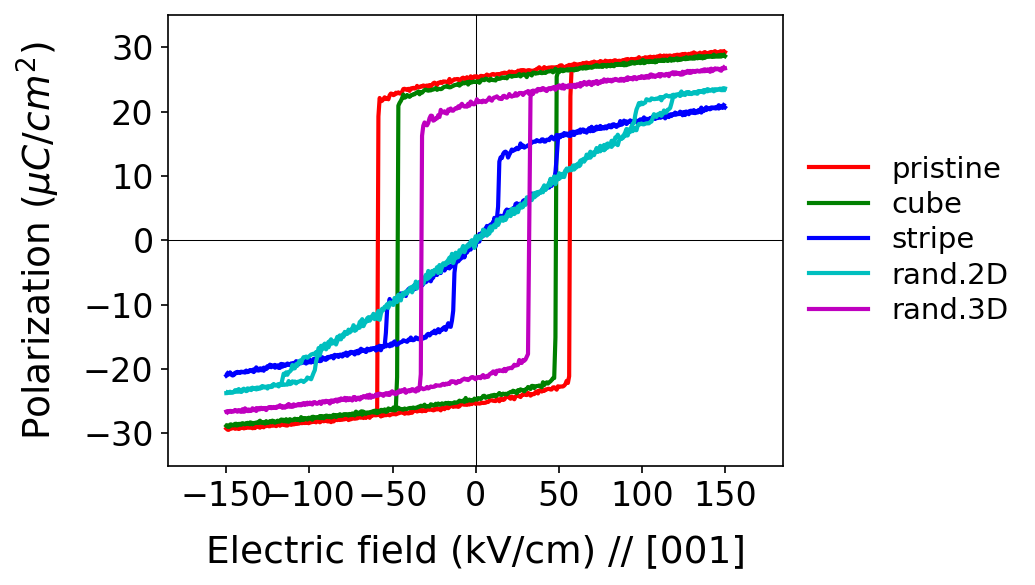}}
        \put(133,5){\includegraphics[scale=.47,trim={14cm 2.5cm 0 2cm},clip]{PEhys_1p_cpz0.0_ez_250K.png}}
        \put(0,10){\textcolor{white}{\rule{.45cm}{4cm}}}
        \put(2,75){\rotatebox[origin=c]{90}{\text{$P_i$ (\rrrv{$\mu$C/cm$^2$})}}}
        \put(28,0){\textcolor{white}{\rule{5cm}{.45cm}}}
        \put(78,4.5){\text{$E_z$ (\rrrv{kV/cm})}}
        \put(40,118){\text{(a)}}
    \end{Overpic}
    \begin{Overpic}[abs]{\begin{tabular}{p{.34\textwidth}}\vspace{4cm}\\\end{tabular}}
        \put(0,0){\includegraphics[scale=.47,clip,trim=2.6cm 0cm 4cm 0cm]{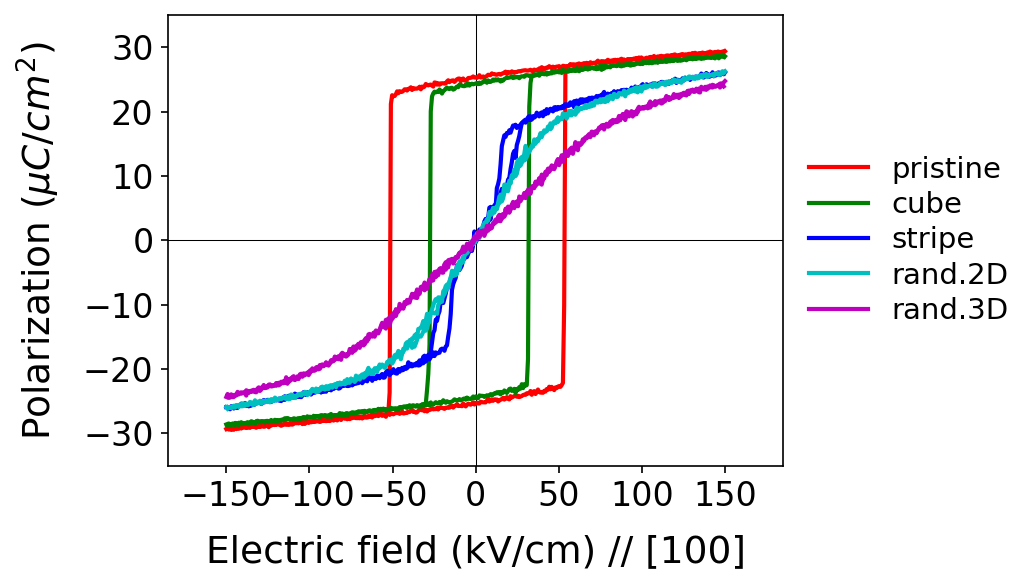}}
        \put(97,5){\includegraphics[scale=.47,trim={14cm 2.5cm 0 2cm},clip]{PEhys_1p_cpz0.1_ex_250K.png}}
        \put(0,0){\textcolor{white}{\rule{5cm}{.45cm}}}
        \put(43,4.5){\text{$E_x$ (\rrrv{kV/cm})}}
        \put(5,118){\text{(b)}}
    \end{Overpic}
    \begin{Overpic}[abs]{\begin{tabular}{p{.4\textwidth}}\vspace{4cm}\\\end{tabular}}
        \put(0,0){\includegraphics[scale=.47,clip,trim=2.6cm 0cm 4cm 0cm]{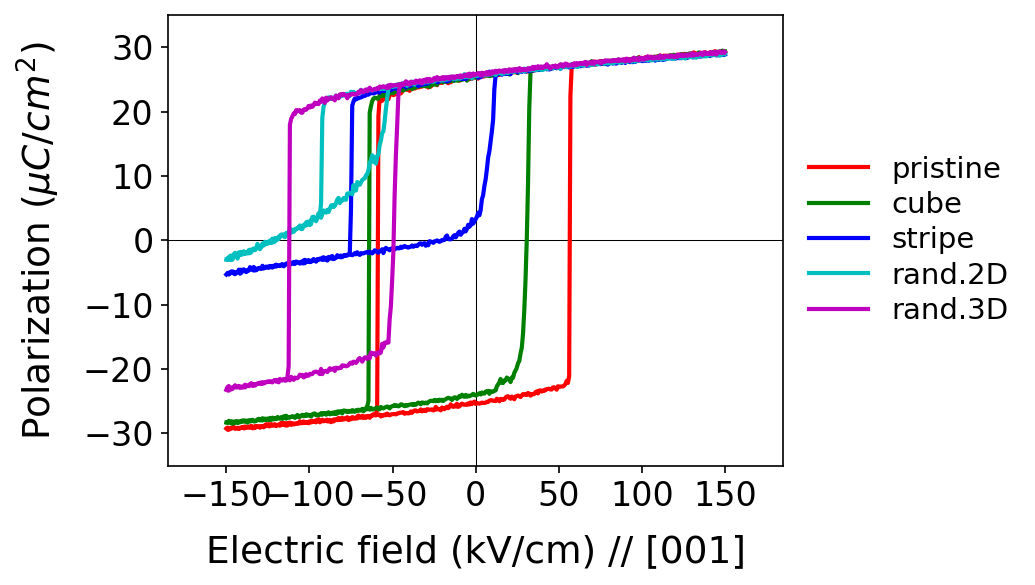}}
        \put(97,5){\includegraphics[scale=.47,trim={14cm 2.5cm 0 2cm},clip]{PEhys_1p_cpz0.1_ez_250K.png}}
        \put(0,0){\textcolor{white}{\rule{5cm}{.45cm}}}
        \put(43,4.5){\text{$E_z$ (\rrrv{kV/cm})}}
        \put(5,118){\text{(c)}}
    \end{Overpic}
    }
  \caption{Changes of the field hysteresis at $T=250$~K if going from the pristine material (red) to systems with different distribution of 1~\% defects  (green: cube; blue and cyan: stripes and randomly in $\hat{z}$-planes; purple: randomly) for (a) non-polar defects and (b)--(c) polar defect dipoles ($|u_{\text{D}}|=$0.1~\AA $\hat{z}$) for the electric field (b)  perpendicular ($E_x$)  and (c) collinear ($E_z$) to the defect dipoles. Only the polarization component $P_i$ along the field direction is shown.}
  \label{fig:pehys_1p}
\end{figure}

How do the different defect configurations affect the field response of the material? 
As reference, red lines in Fig.~\ref{fig:pehys_1p} show the field hysteresis of the pristine material in the T phase at 250~K. 
The $P-E$ curve is symmetric, centered at the origin, and has large coercive fields and remanent polarization. 
\rv{With temperature, hysteresis and polarization are reduced, and finally in the paraelectric phase, the field response is linear, see Fig.~\ref{app:pehys_para} for 350~K.}
\rv{The characteristics in the ferroelectric phase change considerably with the strength and ordering of defects, } see other colors in Fig.~\ref{fig:pehys_1p} and the corresponding strain--field curves in Fig.~\ref{app:sehys_1p}.
Non-polar or weak defect dipoles, reduce remanent polarization ($|P_r^{\pm}|$) and particularly the width of the hysteresis loops ($\Delta E_c$) for all orderings while the hysteresis stays symmetric as shown in subfigure~(a).
Analogous to the phase diagrams, the differences between the configurations are related to the {\it{active surface area}} of the defects. The {\it{active surface area}} and the impact on the field hysteresis is small for cubic agglomerates and 2D agglomerates with the field applied in the defect plane. 
In these cases the changes of $\Delta E_c$ (and $P_r$) are below $18$~\% ($3$~\%) and $20$~\% ($3$~\%), respectively.

Larger modifications are induced by randomly ordered defects that have a larger {\it{active surface area}}. 
The largest changes are induced by 2D agglomerates if the field is applied along \rv{their normal, i.e.\ } along $z$. 
\rv{For this geometry, the depolarization field induced by the defects acts as restoring force on the polarization direction. 
Therefore, the polarization switches reversibly between the tetragonal phase in plane} with $P_x$ and a monoclinic state with $\pm P_z>P_x$, cf.\ Fig.~\ref{app:pehys_ag_ds_pxyz}. 
\rv{The $P_z(E_z)$ hysteresis is pinched to a double-loop.}
Thereby, for the chosen defect concentration of 1\%, the randomly distributed defects in one plane impact the surrounding dipoles more than stripes, cf.\ discussion on Fig.~\ref{fig:rand3d_ag_snapshots_c0}, and stabilize a reversible linear change of $P_z$ with the field for more than $\pm 96$~kV/cm. 

As shown in \rv{Fig.~\ref{fig:pehys_1p}}~(b), strong defect dipoles along $z$ furthermore induce full pinching of the hysteresis for a perpendicular field along $x$. 
In this case, internal and external fields compete with each other for all orderings. 
For both 2D agglomerates and a random distribution of defects, the remanent polarization along the field direction is zero and the strain response is maximized, cf.\ Fig.~\ref{app:sehys_1p} and Fig.~\ref{app:pehys_ag_ds_pxyz}. 
There is no hysteresis for randomly distributed defects and a double hysteresis loop is induced in field ranges of less than 10~kV/cm by compact stripes.
\rv{In qualitative agreement, a reversible strain-field response has been measured in the presence of aged defect dipoles \cite{renLargeElectricfieldinducedStrain2004}.}
Strong defect dipoles collinear to the field, shown in \rv{Fig.~\ref{fig:pehys_1p}}~(c), reduce the width of the field hysteresis, e.g.\ by about 28~\% for the cubic agglomerates and by about 45~\% for random ordering in 3D. 
Furthermore, these defects act as internal bias fields and shift the center of the hysteresis to negative $E_z$. 
Thereby larger changes are induced in $E_c^+$, i.e.\ parallel to the defect direction, and $E_c^+$ is shifted to negative field directions for random defect distributions.  
For the cube, with its small {\it{active surface area}} only this positive coercive field is reduced.
Increasing the {\it{active surface area}}, not only $E_c^+$ shifts further, but also $P_r^{-}$ is reduced. 
These trends can be understood by pinned dipoles.
Already for the cube, its neighbors along the polarization direction do not switch for the chosen field strength and thus act as nucleation centers for the macroscopic switching towards the positive polarization direction lowering the coercive field. 
Going from cube to random in 3D and finally to agglomeration in 2D, the interaction length between defects and surrounding dipoles and thus the pinned volume increase, cf.\ Fig.~\ref{app:snapshots_field}. 
Additionally, the increase of $|E_c^-|$ induced by the internal bias field, increases from stripe to random in plane to random in 3D. 
This again shows that the {\it{active surface area}} governs the main trends of polarization and phase stability, while the subtle details also depend on the compactness or dilution of defects.

\begin{figure*}[t]
  \centering
  \resizebox{.9\linewidth}{!}{
    \begin{Overpic}[abs]{\begin{tabular}{p{.6\textwidth}}\vspace{6.5cm}\\\end{tabular}}
      \put(7,98){\includegraphics[scale=.27]{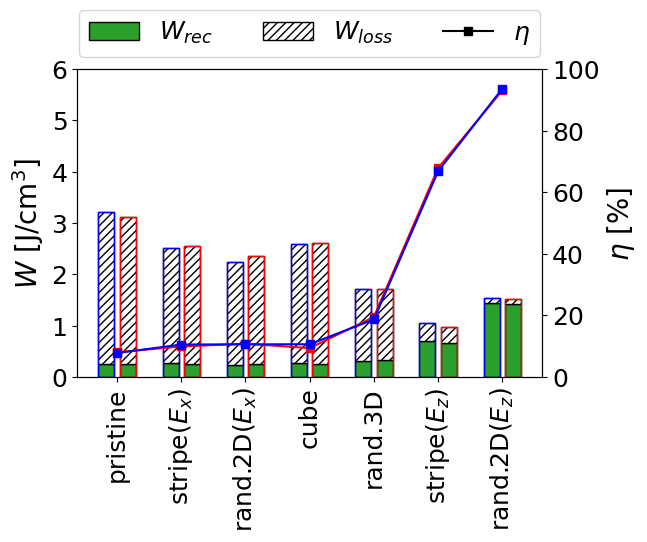}}
      \put(22,98){\textcolor{white}{\rule{3.1cm}{1.1cm}}}
      \put(24,180){\scriptsize\text{(a)}}
      \put(-2,0){\includegraphics[scale=.27]{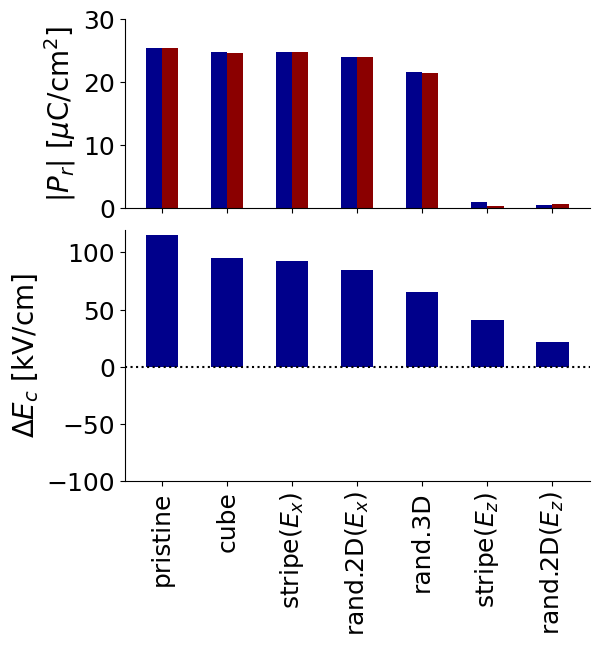}}
      \put(99.5,118){\scriptsize\text{(c)}}
      \put(99.5,38){\scriptsize\text{(e)}}
      \put(135.5,98){\includegraphics[scale=.27]{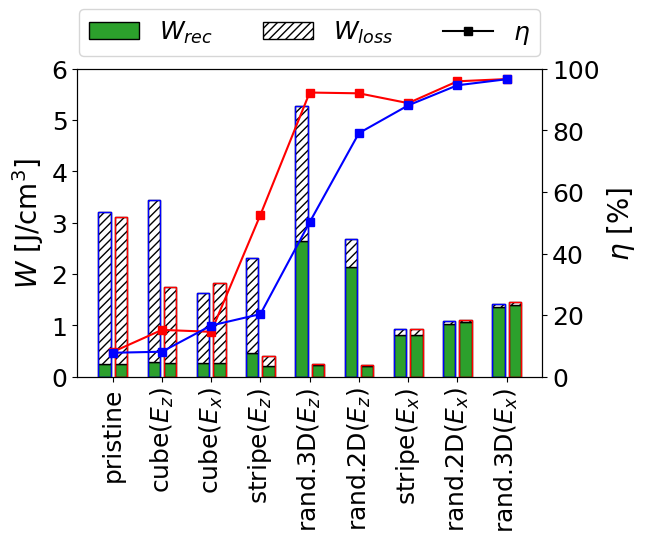}}
      \put(150.5,98){\textcolor{white}{\rule{3.1cm}{1.1cm}}}
      \put(152.5,180){\scriptsize\text{(b)}}
      \put(126,0){\includegraphics[scale=.27]{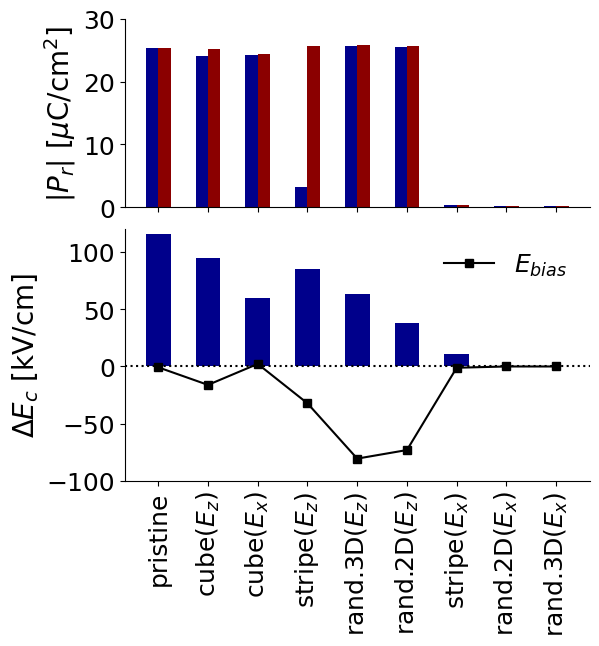}}
      \put(227,118){\scriptsize\text{(d)}}
      \put(227,38){\scriptsize\text{(f)}}
  \end{Overpic}
  }
  \caption{Characteristics of the field hysteresis for 1~\% (a,c,e) non-polar and (b,d,f) polar defects ($|u_{\text{D}}|=$0.1~\AA $\hat{z}$) at $T=250$~K for $E=150$~kV/cm sorted by efficiency. 
  In (a)--(b) $W_{\text{rec}}$ (green) and $W_{\text{loss}}$ (gray hatched) and the efficiency $\eta$ (black dots, right axes), in (c)--(d) the remanent polarization $|P_r|$ and in (e)--(f) the width of the hysteresis ($\Delta E_c)$ and the bias field $E_{\text{bias}}$ for different defect distributions are compared to the pristine material. 
  Red and blue correspond to positive and negative field directions which differ for the collinear alignment of defects and the external field ($E_z$).}
  \label{fig:w_eta_ec_pr}
\end{figure*}
Figure~\ref{fig:w_eta_ec_pr} compares the characteristics of the hysteresis loops for all defect configurations \rv{at 250~K. 
At this temperature, the pristine material is in the T phase and} a work loss of about 3~J/cm\rrrv{$^{3}$} is dissipated if the polarization switches from $P_r^-$ to the saturation polarization $P_{\text{max}}^x$ and vice versa for negative field directions, while the recoverable work as well as reversible changes of polarization and strain without switching are negligible, in qualitative agreement with previous reports \cite{luoFirstprinciplesEffectiveHamiltonian2016}.
On the one hand, non-polar or weak defects \rv{agglomerated in cubes or 2D with parallel applied fields} induce a small reduction of the loss \rv{in the ferroelectric phases} only. 
The efficiency $\eta$ is thus approximately constant \rv{between these configurations} and increases slightly for random defects \rv{and 2D defects in perpendicular applied fields}. 
On the other hand, large efficiencies are possible for the double loop hysteresis found for 2D ordering. 
Thereby, the smallest hysteresis in case of defects randomly distributed in one plane allows for an efficiency of about 94~\%.
Analogously, pinched hysteresis with reduced losses and maximal $\Delta P$ and $\eta$ are stabilized by strong defects along $z$ in a field along $x$ as long as the {\it{active surface area}} is not reduced by clustering in 3D. 
Maximal recoverable energy and efficiency of 1.07~J/cm\rrrv{$^{3}$} and 96~\% are possible for defects randomly distributed in the sample.
Finally, for strong defects and fields along $z$, one has to distinguish between positive and negative field directions. In the former case shown in red, no polarization switching can be induced during unipolar field loops for randomly distributed defects and thus the used definition of loss and efficiency would predict optimal configurations. 
However, this is not relevant as the possible recoverable energy as well as changes of $P$ and strain are negligible. 
In case of the negative field shown in blue, maximal values of $W_{\text{rec}}$ and a moderate efficiency are possible due to the shifted and reduced hysteresis for random defect distributions.

In summary, the field characteristics in the ferroelectric phase depend strongly on the defect arrangement and the relative direction of the field.
Weak defects induce the largest modification of the hysteresis if they agglomerate in 2D, while the pinching of the hysteresis for strong defect dipoles perpendicular to the field is maximal for randomly oriented dipoles. 
Finally, strong defects parallel to the field induce the largest shift of hysteresis for random distribution, while 2D agglomerates induce the largest asymmetry.

\rv{It is important that these large changes of the field hysteresis by defects are expected in the full temperature range of the ferroelectric phase. 
Well above $T_C^{T-C}$, in the paraelectric phase, losses are also absent, but at the same time the saturation polarization is small.
Therefore, $W_{\text{rec}}$ (cf. Eqn~\eqref{eq:rec}) is only 0.24~J/cm$^3$ (0.75~J/cm$^3$) for $E_{\text{max}}=50$~kV/cm ($E_{\text{max}}=100$~kV/cm) at 350~K.
These values of $W_{\text{rec}}$ can be outperformed in the presence of 1\% polar defects ($|u_{\text{D}}|=0.1$~\AA~$\hat{z}$) randomly distributed in 2D with perpendicular applied fields $E_x$ (0.42 and 0.77~J/cm$^3$ for $E_{max}=50$ and $100$~kV/cm, respectively), see Fig.~\ref{app:pehys_para}.}


\section{Conclusion}

While it is commonly accepted that point defects and defect complexes in ferroelectric perovskites modify ferroelectric phase diagrams and field hysteresis, the impact of defect distribution and agglomeration is so far underrepresented.
In this study, we have used $ab\ initio$ based simulation to close this knowledge gap.

\rv{We have analyzed a broad range of defect concentrations and strengths of defect dipoles. 
Thereby we found a cross-over between non-polar or weak defect dipoles, which mainly induce local depolarization fields, and strong defect dipoles (larger than about 30\% of the saturation polarization), which induce internal electric fields. 
Already 1\% of both types of defects reduce thermal hysteresis and coercive fields and affect the phase diagram.
Furthermore, our results show that the control of defect dipole strengths (e.g., by using different dopants) is one promising route to tap the full potential of defect engineering in ferroelectric materials.  
It was particularly interesting to explore if there are atomistic realizations of defect dipoles close to the cross-over strength, which could potentially introduce multi-critical points in the phase diagram.}
\rv{This requires the exact treatment of defects and their interaction in density functional theory simulations. Furthermore, future work can include defect dynamics by combining the effective Hamiltonian with kinetic Monte Carlo simulations.}

\rv{Importantly, the spatial distribution of defects and their agglomeration have an even larger impact on phase stability and field hystereses than their exact concentration or strength. 
Specifically, we show how depolarization and internal fields are associated with the {\it{active surface area}} of the defect configuration. 
For example, randomly distributed defects have the largest impact on the paraelectric to ferroelectric phase transition.}
\rv{The change of this {\it{active surface area}} with time by defect agglomeration, or if defect clusters decompose, can explain many aspects of functional fatigue in experimental samples.} 

Most interesting are defect agglomerates in two dimensions, which reduce the symmetry of the system \rv{and have a maximal active area along their normal. 
As all types of defects favor particular polarization directions,} the field hysteresis in the presence of 2D agglomerates depends on the relative direction between defect agglomerate and external field.
Pinched hysteresis loops and large recoverable stored energy are possible for one \rv{field} direction, while the defects are \emph{invisible} for fields \rv{along the other ones}.  
\rv{Most importantly, we have shown that the pinched double-loop hysteresis induced by only 1\% defects allows for large recoverable energy storage, outperforming the response of the paraelectric phase of (Ba,Sr)TiO$_3$.}
\rv{Thus, one promising route to tailor material properties by defects may be the design of 2D defect-rich areas, e.g., by ion bombardment \cite{saremiLocalControlDefects2018}, or by ion migration and segregation, e.g.\ by heat treatment and interfaces \cite{kleinFermiEnergyCommon2023, genenkoMechanismsAgingFatigue2015, suzukiDislocationLoopFormation2001, batukTrappingOxygenVacancies2015}.}


\section*{CRediT authorship contribution statement}
\textbf{Sheng-Han Teng}: Writing – review \& editing, Writing – original draft, Visualization, Data curation, Conceptualization. \textbf{Anna Gr\"{u}nebohm}: Writing – review \& editing, Writing – original draft, Supervision, Resources, Project administration, Funding acquisition, Conceptualization.

\section*{Acknowledgments}
Funding: This work was supported by the German research foundation (DFG) 412303109. We thank Prof. Dr. Markus Stricker for his helpful comments and fruitful discussions.

\section*{Declaration of generative AI in scientific writing}
During the preparation of this work, the authors used DeepL and Grammarly for wording suggestions and checking of language. After using these tools, we reviewed and edited the content as needed and take full responsibility for the content of the publication.


\bibliographystyle{elsarticle-num} 
\bibliography{references}

\begin{figure}[htbp]
  \centering
  \resizebox{.7\linewidth}{!}{
  \begin{Overpic}[abs]{\begin{tabular}{p{.6\textwidth}}\vspace{4.8cm}\\\end{tabular}}
  	\put(0,0){\includegraphics[scale=.38,trim={0 0 0 0},clip]{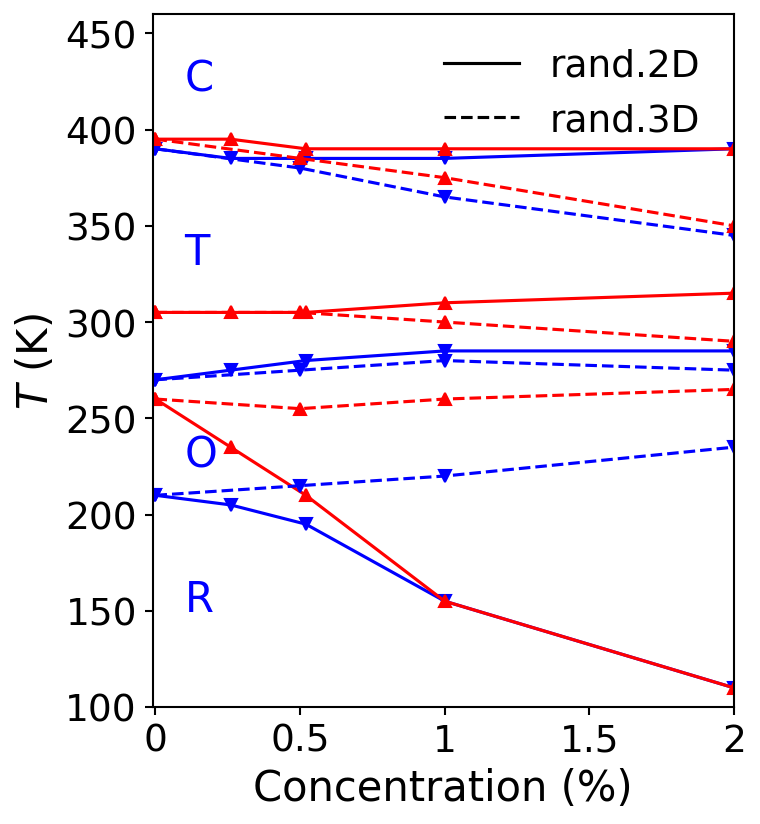}}
		\put(117,35){\text{(a)}}
		\put(138,0){\includegraphics[scale=.38,trim={2.5cm 0 0 0},clip]{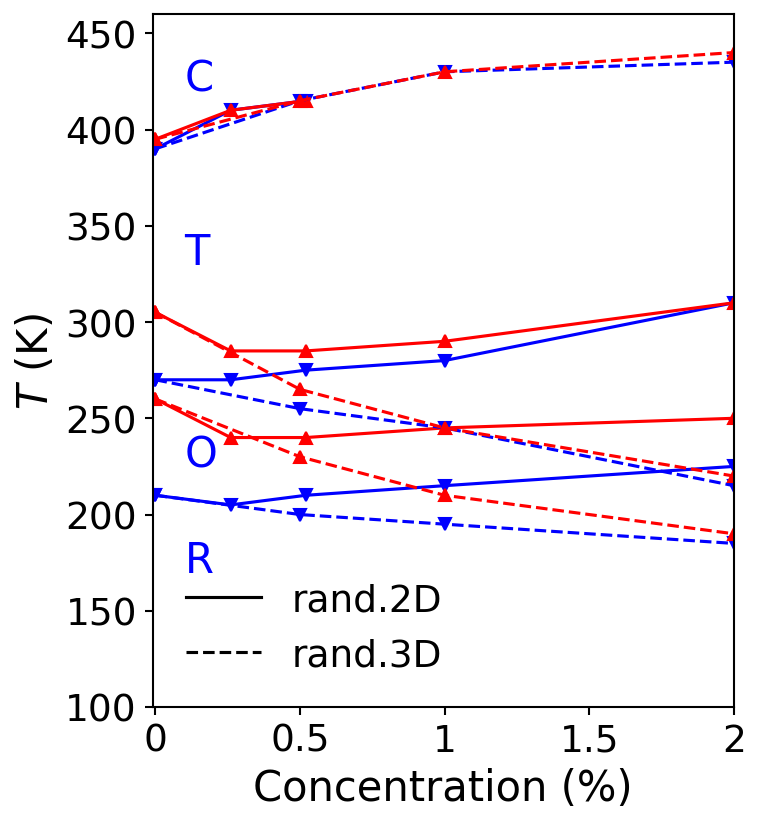}}
		\put(228,35){\text{(b)}}
		\put(125,8){\textcolor{white}{\rule{.4cm}{.4cm}}}
  \end{Overpic}
  }
  \caption{\rv{Temperature - defect concentration phase diagrams of BaTiO$_3$ in the presence of (a) non-polar defect dipoles ($|u_{\text{D}}|=$0.0~\AA) and (b) polar defects ($|u_{\text{D}}|=$0.1~\AA $\hat{x}$) for randomly distributed defects in 2D (solid lines) and in 3D (dashed lines). Red and blue lines show transition temperatures for heating and cooling, respectively.}}
  \label{app:PD_bto}
\end{figure}

\begin{figure}[htb]
    \centering
    \resizebox{\linewidth}{!}{
    \begin{Overpic}[abs]{\begin{tabular}{p{.43\textwidth}}\vspace{4cm}\\\end{tabular}}
        \put(0,0){\includegraphics[scale=.47,clip,trim=0cm 0cm 4cm 0cm]{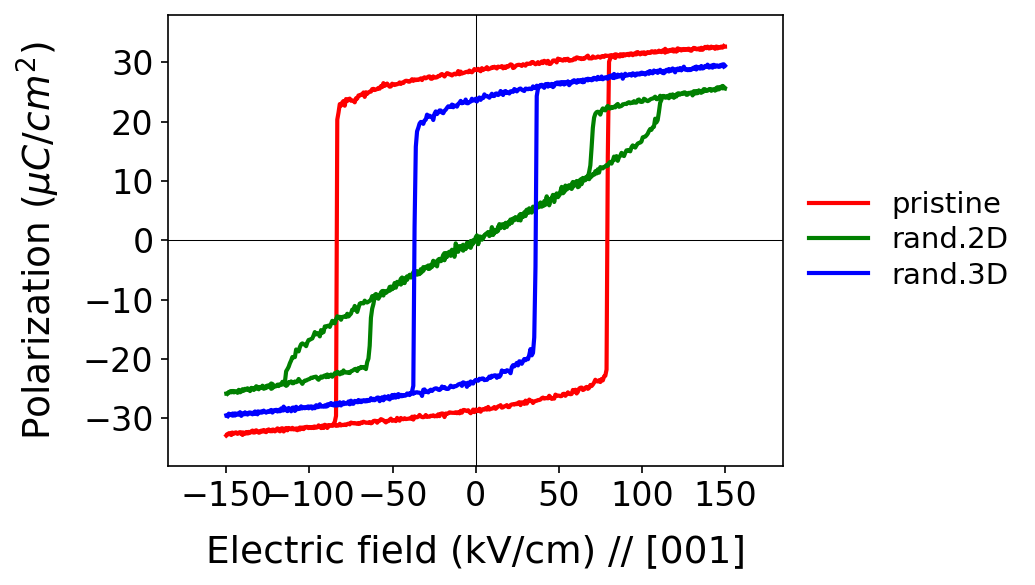}}
        \put(139,28){\includegraphics[scale=.47,trim={14.3cm 5cm 0.4cm 3cm},clip]{PEhys_BTO_1p_cpz0.0_ez_350K.png}}
        \put(0,10){\textcolor{white}{\rule{.45cm}{4cm}}}
        \put(2,75){\rotatebox[origin=c]{90}{\text{$P_i$ (\rrrv{$\mu$C/cm$^2$})}}}
        \put(28,0){\textcolor{white}{\rule{5cm}{.45cm}}}
        \put(78,4.5){\text{$E_z$ (\rrrv{kV/cm})}}
        \put(40,118){\text{(a)}}
    \end{Overpic}
    \begin{Overpic}[abs]{\begin{tabular}{p{.34\textwidth}}\vspace{4cm}\\\end{tabular}}
        \put(0,0){\includegraphics[scale=.47,clip,trim=2.6cm 0cm 4cm 0cm]{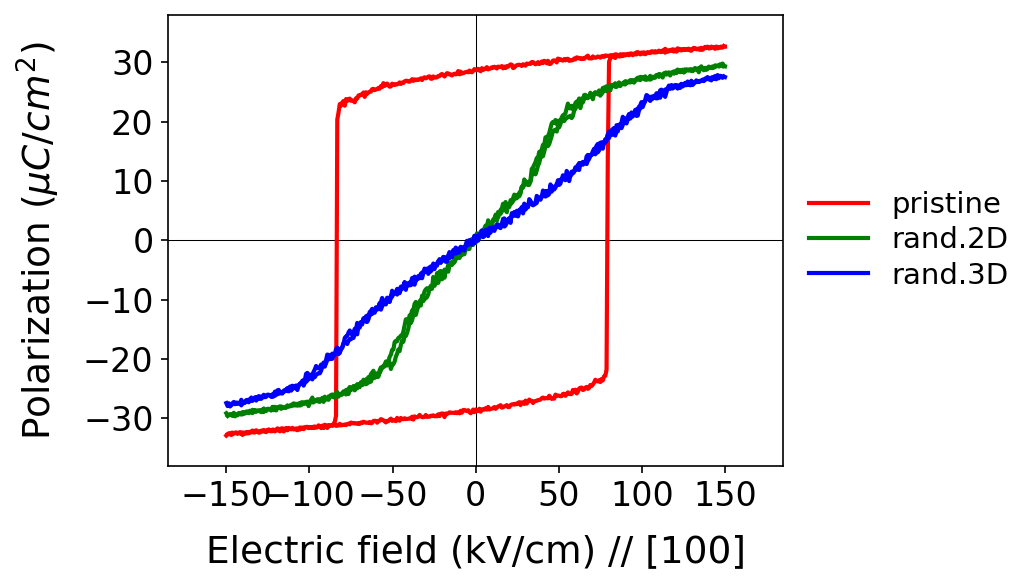}}
        \put(104,28){\includegraphics[scale=.47,trim={14.3cm 5cm 0.4cm 3cm},clip]{PEhys_BTO_1p_cpz0.1_ex_350K.png}}
        \put(0,0){\textcolor{white}{\rule{5cm}{.45cm}}}
        \put(43,4.5){\text{$E_x$ (\rrrv{kV/cm})}}
        \put(5,118){\text{(b)}}
    \end{Overpic}
    \begin{Overpic}[abs]{\begin{tabular}{p{.4\textwidth}}\vspace{4cm}\\\end{tabular}}
        \put(0,0){\includegraphics[scale=.47,clip,trim=2.6cm 0cm 4cm 0cm]{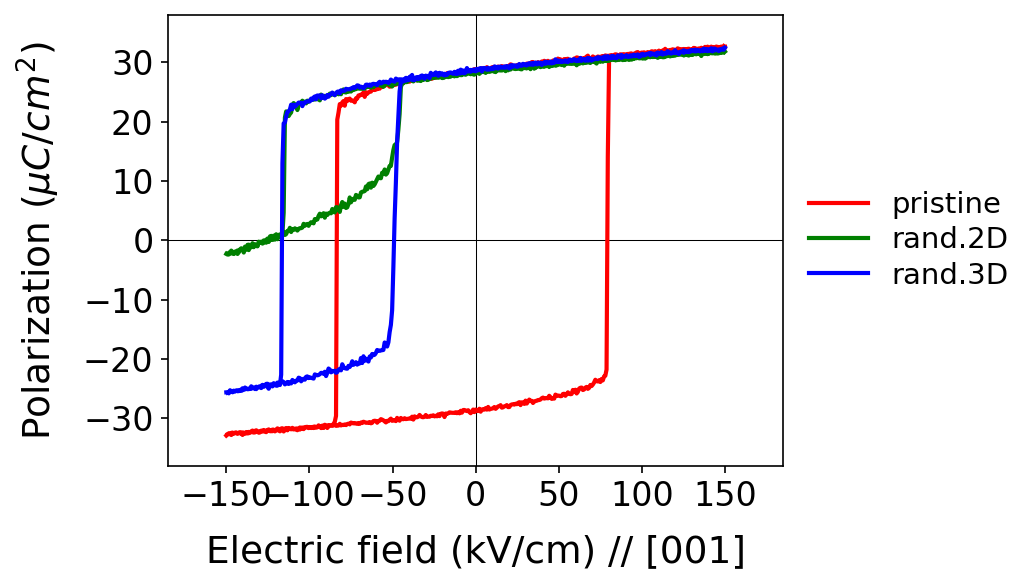}}
        \put(104,28){\includegraphics[scale=.47,trim={14.3cm 5cm 0.4cm 3cm},clip]{PEhys_BTO_1p_cpz0.1_ez_350K.png}}
        \put(0,0){\textcolor{white}{\rule{5cm}{.45cm}}}
        \put(43,4.5){\text{$E_z$ (\rrrv{kV/cm})}}
        \put(5,118){\text{(c)}}
    \end{Overpic}
    }
  \caption{\rv{Changes of the field hysteresis for BaTiO$_3$ at $T=350$~K if going from the pristine material (red) to systems with different distribution of 1~\% defects (green: randomly in $\hat{z}$-planes; blue: randomly) for (a) non-polar defects and (b)--(c) polar defect dipoles ($|u_{\text{D}}|=$0.1~\AA $\hat{z}$) for the electric field (b) perpendicular ($E_x$) and (c) collinear ($E_z$) to the defect dipoles. Only the polarization component $P_i$ along the field direction is shown.}}
  \label{app:pehys_bto_1p}
\end{figure}

\newpage
\appendix
\rv{
\section{Impact of Sr-substitution}
\label{sec:app_pristine}

In the following, we compare the impact of defects on BaTiO$_3$ with and without Sr substitution.
Analogous to the results for (Ba,Sr)TiO$_3$ presented in Fig.~\ref{fig:PD_u0} and Fig.~\ref{fig:PD_u0.1}, Figure~\ref{app:PD_bto} shows for BaTiO$_3$ that non-polar defects randomly distributed in 3D mainly reduce $T_C^{\text{C-T}}$ (45~K, with 5~K resolution, for 2\%), while their distribution in planes reduces $T_C^{\text{O-R}}$ (100~K for 2\%) and the thermal hysteresis of the transition.
Polar defects ($u_{\text{D}}=0.1$~\AA~$\hat{z}$) randomly distributed in 3D increase $T_C^{\text{C-T}}$ (by 45~K for 2\%) and reduce $T_C^{\text{T-O}}$ and $T_C^{\text{O-R}}$ by 55~K and 25~K in cooling simulations, and by 85~K or 70~K in heating simulation, respectively.
Polar defects randomly distributed in 2D, increase $T_C^{\text{T-O}}$ by 5~K and 40~K and decreases $T_C^{\text{O-R}}$ by 10~K and 15~K, during heating and cooling, respectively.  

Fig.~\ref{app:pehys_bto_1p} shows the changes of the field hysteresis of the T phase of BaTiO$_3$ at 350~K by defects.
Analogous to (Ba,Sr)TiO$_3$ (cf. Fig.~\ref{fig:pehys_1p}), non-polar defects or polar defects in a perpendicular electric field reduce and pinch the hysteresis, while polar defects parallel to the field act as bias fields.
}

\newpage

\begin{figure}[htb]
    \centering
    \resizebox{\linewidth}{!}{
    \begin{Overpic}[abs]{\begin{tabular}{p{.38\textwidth}}\vspace{3.8cm}\\\end{tabular}}
        \put(0,0){\includegraphics[height=4cm]{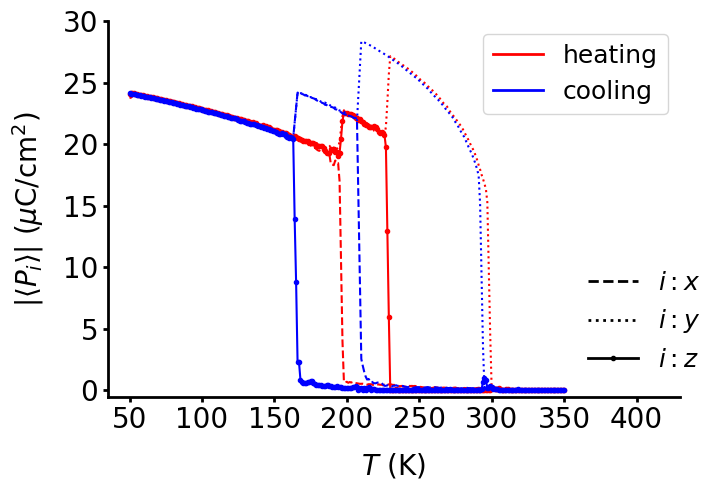}}
        \put(29,101){\text{(a)}}
    \end{Overpic}
    \begin{Overpic}[abs]{\begin{tabular}{p{.38\textwidth}}\vspace{3.8cm}\\\end{tabular}}
        \put(0,0){\includegraphics[height=4cm]{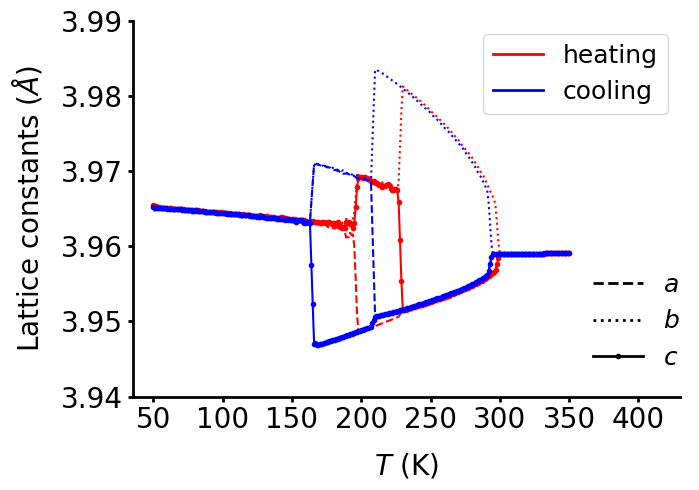}}
        \put(35,101){\text{(b)}}
    \end{Overpic}
    }
        \resizebox{\linewidth}{!}{
    \begin{Overpic}[abs]{\begin{tabular}{p{.38\textwidth}}\vspace{3.8cm}\\\end{tabular}}
        \put(0,0){\includegraphics[height=4cm]{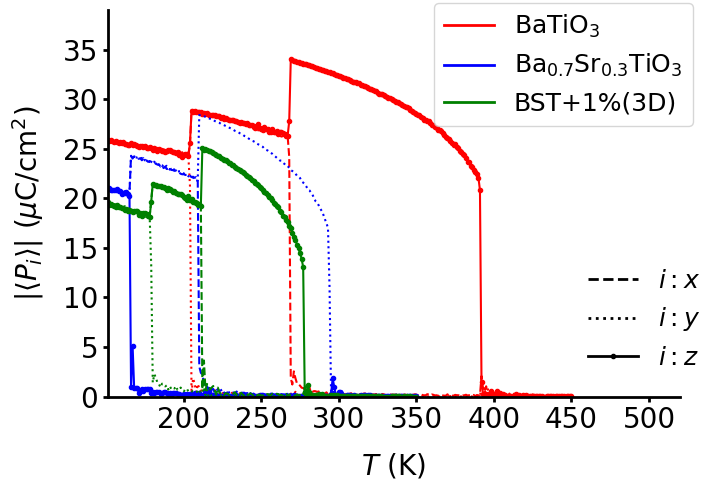}}
        \put(27,101){\text{(c)}}
    \end{Overpic}
    \begin{Overpic}[abs]{\begin{tabular}{p{.38\textwidth}}\vspace{3.8cm}\\\end{tabular}}
        \put(0,0){\includegraphics[height=4cm]{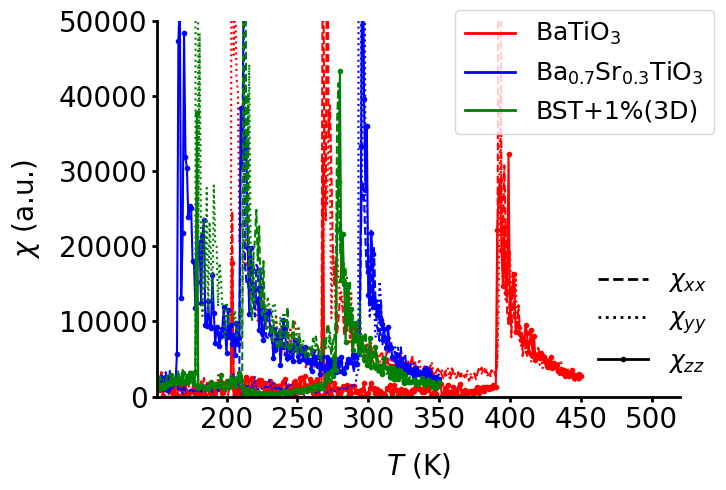}}
        \put(0,101){\text{(d)}}
        \put(0,60){\textcolor{white}{\rule{.45cm}{1cm}}}
    \end{Overpic}
    }
    \caption{\rv{Phase transitions of defect-free  Ba$_{0.7}$Sr$_{0.3}$TiO$_3$ found in cooling simulations compared to (a)--(b) heating simulations and  to (c)-(d) BaTiO$_3$ and Ba$_{0.7}$Sr$_{0.3}$TiO$_3$ with 1\% non-polar defects randomly distributed in 3D.
    (a)/(c) Polarization components $|\braket{P_i}|$, (b) lattice constants $a,b,c$ and (d) susceptibility $\chi_{ii}$, with $i: x,y,z$.
    Note that the maxima of $\chi_{ii}$ at $T_C$ are not fully shown due to their large errorbars.}}
   \label{app:pt_st_pristine}
\end{figure}

\rv{Figure~\ref{app:pt_st_pristine}--(a)--(c) compares the changes of polarization components and lattice parameters with temperature between BaTiO$_3$, Ba$_{0.7}$Sr$_{0.3}$TiO$_3$, and the system with Sr and 1\% randomly distributed non-polar defects.
For the defect-free materials, our model can well reproduce the three first-order phase transitions with abrupt changes of polarization and lattice parameters, their thermal hysteresis between cooling and heating simulations, as well as the decrease of transition temperatures and polarization with Sr addition. 
As discussed in the main paper, the addition of 1\% non-polar defects further reduces the polarization in all phases while $T_C^{\text{C-T}}$ and $T^{\text{O-R}}$ decrease and increase, respectively.

Figure~\ref{app:pt_st_pristine}~(d) compares the susceptibilities determined by
\begin{equation}
\chi_{ij}=\frac{1}{\epsilon_0 V k_B T}(\langle P_i P_j \rangle-\langle P_i \rangle \langle P_j \rangle), 
\label{eq:chi}
\end{equation}
where $\epsilon_0$ is the dielectric constant in vacuum, $V$ is the volume of the system, $k_B$ is the Boltzmann constant, $T$ is the temperature, $\langle P_i\rangle$, and $\langle P_i P_j \rangle$ are the average of the polarization components and their product for $i,j: x,y,z$, respectively. 
For all materials $\chi_{ij}$ is maximal in a temperature range of about 50~K above the phase transition temperatures. 
Therefore, the reduction of the ferroelectric transition temperature by Sr substitution enhances the absolute value of this response around room temperature.
Note that 1\% non-polar defects reduce the temperature with maximal $\chi_{ij}$ further and a maximal value at room temperature in the presence of defects demands a smaller Sr concentration. 
}

\begin{figure}[tbp]
    \centering
    \begin{Overpic}[abs]{\begin{tabular}{p{.4\textwidth}}\vspace{4cm}\\\end{tabular}}
        \put(0,0){\includegraphics[scale=.47, clip, trim=0 0 5cm 0]{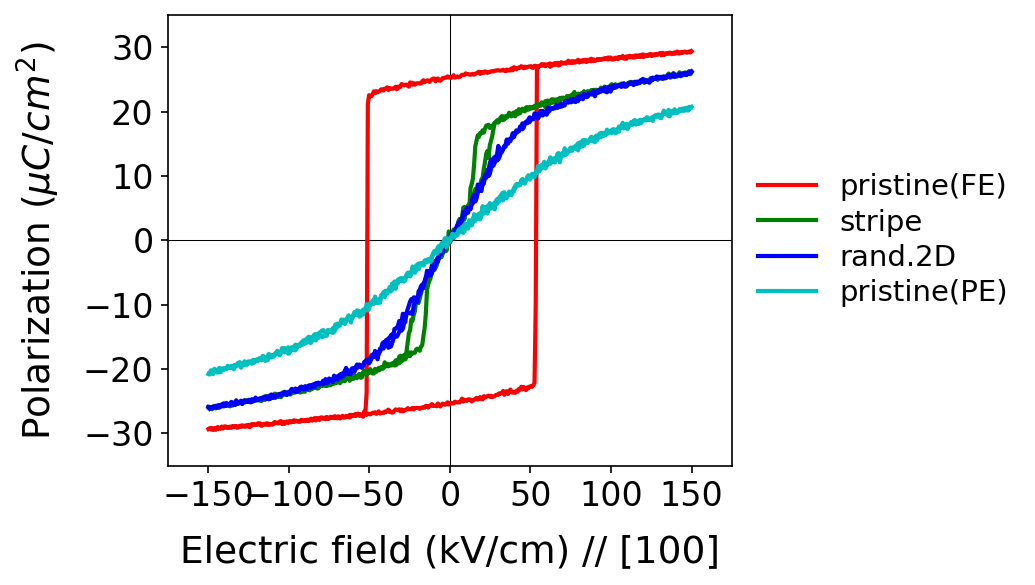}}
        \put(0,10){\textcolor{white}{\rule{.45cm}{4cm}}}
        \put(2,75){\rotatebox[origin=c]{90}{\text{$P_i$ (\rrrv{$\mu$C/cm$^2$})}}}
        \put(28,0){\textcolor{white}{\rule{5cm}{.45cm}}}
        \put(78,4.5){\text{$E_x$ (\rrrv{kV/cm})}}
        \put(0,120){\text{(a)}}
    \end{Overpic}
    \begin{Overpic}[abs]{\begin{tabular}{p{.4\textwidth}}\vspace{4cm}\\\end{tabular}}
        \put(0,0){\includegraphics[height=4.68cm]{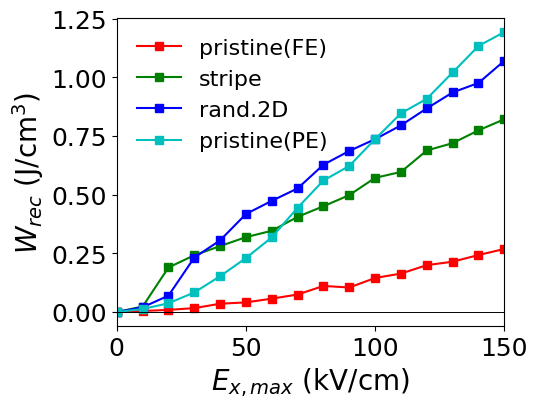}}
        \put(28,0){\textcolor{white}{\rule{5cm}{.46cm}}}
        \put(70,4.5){\text{$E_{x,max}$ (\rrrv{kV/cm})}}
        \put(0,120){\text{(b)}}
    \end{Overpic}
    \caption{\rv{Comparisons of (a) field hysteresis and (b) the energy density which can be reversibly stored ($W_{\text{rec}}$)  between pristine Ba$_{0.7}$Sr$_{0.3}$O$_3$ in its ferroelectric ($T=250$~K, red) and paraelectric ($T=350$~K, cyan) phases to Ba$_{0.7}$Sr$_{0.3}$O$_3$ with 1~\% defects ($|u_{\text{D}}|=$0.1~\AA $\hat{z}$) at 250~K for defect stripes (green) and random distribution in 2D (blue) in $\hat{z}$-planes. 
    Only the polarization component $P_x$ along the field direction is shown in (a).}}
   \label{app:pehys_para}
\end{figure}

\rv{The energy which can be reversibly stored in the material depends on the full field-hysteresis rather than on this small field response, see Eqns.~\eqref{eq:rec}--\eqref{eq:loss}. 
Figure~\ref{app:pehys_para} compares the field hystereses and $W_{\text{rec}}$ of Ba$_{0.7}$Sr$_{0.3}$TiO$_3$ between the ferroelectric  ($T=250$~K, red) and paraelectric ($T=350$~K, cyan) phases of the pristine material with those in the presence of 1\% polar defects ($|u_{\text{D}}|=$0.1~\AA $\hat{z}$ in $z$-planes) in the ferroelectric phase (green: stripe, blue: randomly in 2D).

For the pristine material, $W_{\text{rec}}$ is considerably larger in the paraelectric than in the ferroelectric phase. 
As the Sr concentration allows to adjust $T_C$, thus a large increase of $W_{\text{rec}}$ at any temperature of interest can potentially be realized by substitution. 
However, as the saturation polarization decreases in the paraelectric phase and with Sr concentration, defect dipoles may outperform this design strategy. 
The defects stabilize a double-loop hysteresis without remanent polarization but with large saturation polarization and thus enhance $W_{\text{rec}}$ well in the ferroelectric phase. 
For the tested Sr and defect concentrations and temperatures, $W_{\text{rec}}$ in the presence of defects is larger than the response of the parelectric phase of (Ba,Sr)TiO$_3$ for field strengths below 100~kV/cm.
}

\newpage
\rv{\section{Relative direction of 2D ordering and polarization}}
\label{sec:app_pd_x}

\begin{figure}[htbp]
  \centering
  \resizebox{.7\linewidth}{!}{
  \begin{Overpic}[abs]{\begin{tabular}{p{.6\textwidth}}\vspace{4.8cm}\\\end{tabular}}
  	\put(0,0){\includegraphics[scale=.38,trim={0 0 0 0},clip]{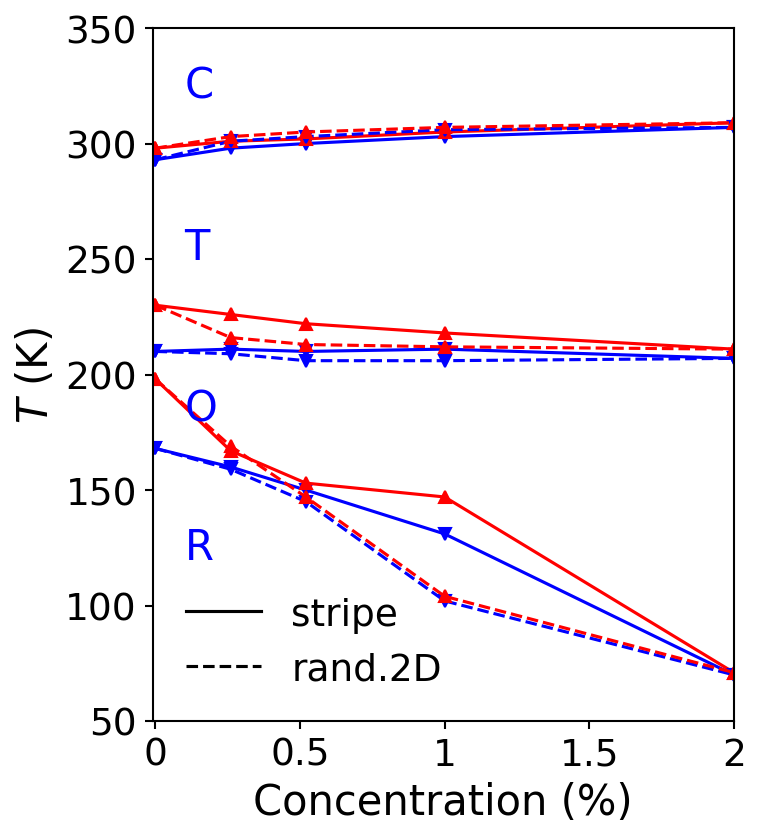}}
		\put(40,95){\footnotesize	*\hspace{.2cm}*\hspace{.5cm}*\hspace{1.4cm}*}
		\put(43,73){\footnotesize	576}
		\put(58,68){\footnotesize	1152}
		\put(78,64){\footnotesize 2304}
		\put(116,42){\footnotesize	4608}
		\put(117,135){\text{(a)}}
		\put(138,0){\includegraphics[scale=.38,trim={2.5cm 0 0 0},clip]{PD_ag_ds_u0.png}}
		\put(152,99){\footnotesize *\hspace{.3cm}*\hspace{.6cm}*\hspace{1.6cm}*}
		\put(155,76){\footnotesize 576}
		\put(170,68){\footnotesize 1152}
		\put(190,62){\footnotesize 2304}
		\put(224,44){\footnotesize 4608}
		\put(228,135){\text{(b)}}
		\put(125,8){\textcolor{white}{\rule{.4cm}{.4cm}}}
  \end{Overpic}
  }
  \caption{Temperature - defect concentration phase diagrams in the presence of (a) polar defect dipoles ($|u_{\text{D}}|=$0.1~\AA $\hat{x}$) and (b) non-polar defects ($|u_{\text{D}}|=$0.0~\AA) for stripes of defects (solid lines) and random distribution in 2D (dashed lines). Red and blue lines show transition temperatures for heating and cooling, respectively. Numbers give the approximate number of active surface units in R phases. Note that the defect agglomerates have no effective {\it{active surface area}} for T and O phases as discussed in the text. 
  \rv{Subfigure~(b) is the same as shown in Fig.~\ref{fig:PD_u0} in the main paper.}}
  \label{app:PD_ux0.1}
\end{figure}

Figure~\ref{app:PD_ux0.1} compares phase diagrams in the presence of (a) polar defect dipoles in the plane ($|u_{\text{D}}|=0.1$~\AA{}$\hat{x}$) and (b) non-polar defects ($|u_{\text{D}}|=0$~\AA) for 2D agglomerates (solid lines) and random distribution in 2D (dashed lines).
\rv{While both defects have similar \textit{active surface areas} and thus induce similar trend in $T_C^{\text{O-R}}$, the in-plane polar defects only increase $T_C^{\text{C-T}}$ by 11~K, decrease $T_C^{\text{T-O}}$ by 19~K (for 2\% defects during heating), and reduce thermal hysteresis for T-O transition due to stronger internal bias fields.}
\newpage

\section{Defect configurations}
\label{sec:app_deng}
\rv{
\begin{figure}[tpbh]
    \centering
    \resizebox{\linewidth}{!}{
    \begin{Overpic}[abs]{\begin{tabular}{p{.315\textwidth}}\vspace{3.7cm}\\\end{tabular}}
        \put(0,0){\includegraphics[height=4cm]{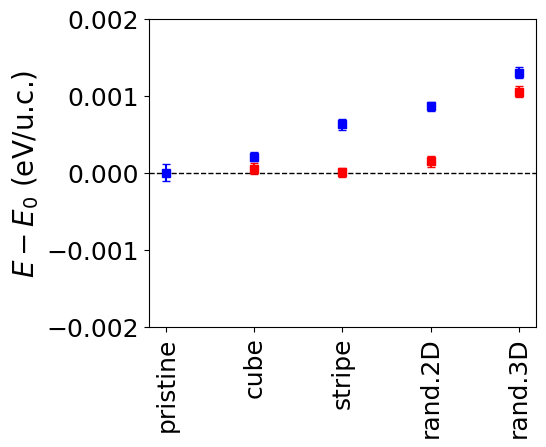}}
        \put(40,96){\normalsize\text{(a)}}
    \end{Overpic}
    \begin{Overpic}[abs]{\begin{tabular}{p{.22\textwidth}}\vspace{3.7cm}\\\end{tabular}}
        \put(0,0){\includegraphics[height=4cm,clip,trim=3.6cm 0cm 0cm 0cm]{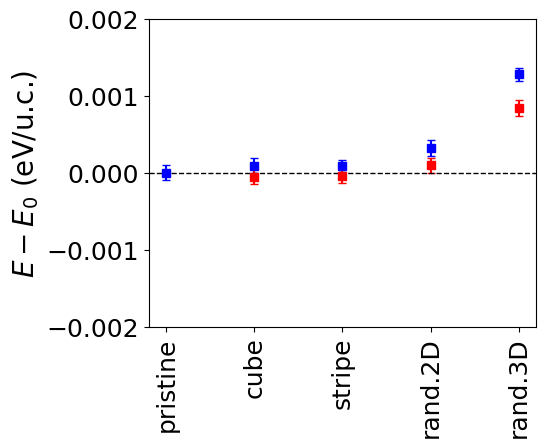}}
        \put(3,96){\text{(b)}}
    \end{Overpic}
    \begin{Overpic}[abs]{\begin{tabular}{p{.22\textwidth}}\vspace{3.7cm}\\\end{tabular}}
        \put(0,0){\includegraphics[height=4cm,clip,trim=3.6cm 0cm 0cm 0cm]{etot_250K.png}}
        \put(3,96){\text{(c)}}
    \end{Overpic}
    }
    \caption{\rv{Changes of total energy induced by different configurations of 1\% for non-polar ($|u_{\text{D}}|=$0.0~\AA, blue) and polar ($|u_{\text{D}}|=$0.1~\AA$\hat{z}$, red) defects in the (a) R phase ($T=150$~K), (b) O phase ($200$~K), and (c) T phase ($250$~K). Mean values of 20000~time steps and their standard deviations are given.}}
   \label{app:deng}
\end{figure}

Figure~\ref{app:deng} compares the changes of the total energy induced by 1\% non-polar (blue) and polar (red) defects of different ordering in R, O, and T phases.
\rrv{Their energy dependency} on agglomeration and phase is in line with the concept of their \textit{active surface area}: 
First, the \textit{active surface area} of all cubic agglomerates, of all agglomerates in O phase and of all non-polar agglomerates in T phase are too small to induce sizeable changes in energy. 
Second, the randomly distributed non-polar defects induce the largest energy penalties. 
Third, the non-polar 2D agglomerates increase the energy of the R phase. 
Fourth, the polar 2D agglomerates act as internal fields and thus lower the energy of the T phase. 
Note that the difference of defect dipoles and free dipoles in the R phase is minor and thus no sizeable energy differences are induced. 
Note that the local energy of pairs of defects may be larger. 
A closer inspection of the defect-defect interaction is out of the scope of the present work. 

\begin{figure}[tpbh]
    \centering
    \resizebox{\linewidth}{!}{
    \begin{Overpic}[abs]{\begin{tabular}{p{.38\textwidth}}\vspace{3.8cm}\\\end{tabular}}
        \put(0,0){\includegraphics[height=4cm]{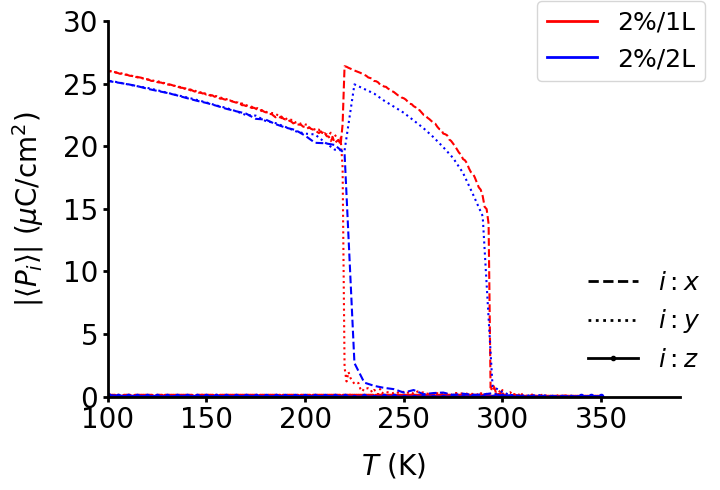}}
        \put(29,101){\text{(a)}}
    \end{Overpic}
    \begin{Overpic}[abs]{\begin{tabular}{p{.38\textwidth}}\vspace{3.8cm}\\\end{tabular}}
        \put(0,0){\includegraphics[height=4cm]{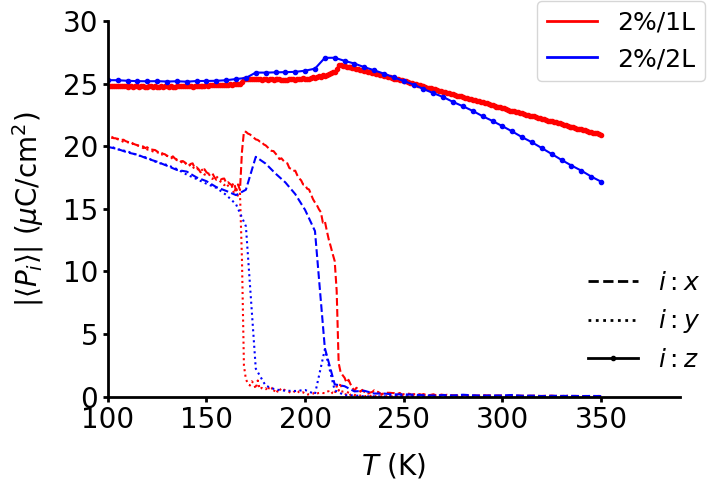}}
        \put(29,101){\text{(b)}}
    \end{Overpic}
    }
    \caption{\rv{Changes of polarization components ($|\braket{P_i}|$, where $i: x,y,z$) in cooling simulations induced by (a) non-polar defects ($|u_{\text{D}}|=$0.0~\AA) and (b) polar defects ($|u_{\text{D}}|=$0.1~\AA$\hat{z}$) concentrated in one plane (red) or distributed over 2 layers (blue).}}
   \label{app:pt_2L}
\end{figure}

Figure~\ref{app:pt_2L} shows the changes of the polarization-temperature phase diagrams for 2D agglomerates with 2\% defects if going from $n_z=1$ (red) to $n_z=2$. 
The trends discussed in the main paper for $n_z=1$ are robust against this distribution of the defects in a larger volume: 
Non-polar defects stabilize T and O phases parallel to the defect plane and a multi-domain R phase without macroscopic polarization and polar defects act as internal fields along $z$. 
The main difference is a slight reduction of the polarization components not aligned with polar defects for the less compact defect distribution.
}

\begin{figure}[tpb]
    \centering
    \resizebox{\linewidth}{!}{
    \begin{Overpic}[abs]{\begin{tabular}{p{.38\textwidth}}\vspace{3.8cm}\\\end{tabular}}
        \put(0,0){\includegraphics[height=4cm]{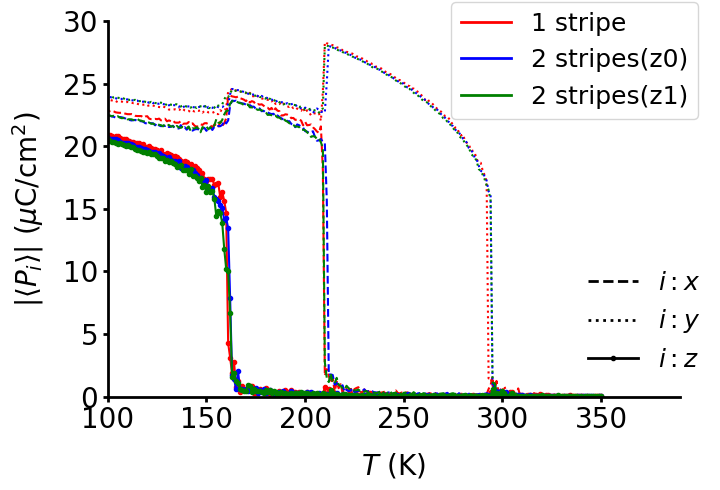}}
        \put(29,101){\text{(a)}}
    \end{Overpic}
    \begin{Overpic}[abs]{\begin{tabular}{p{.38\textwidth}}\vspace{3.8cm}\\\end{tabular}}
        \put(0,0){\includegraphics[height=4cm]{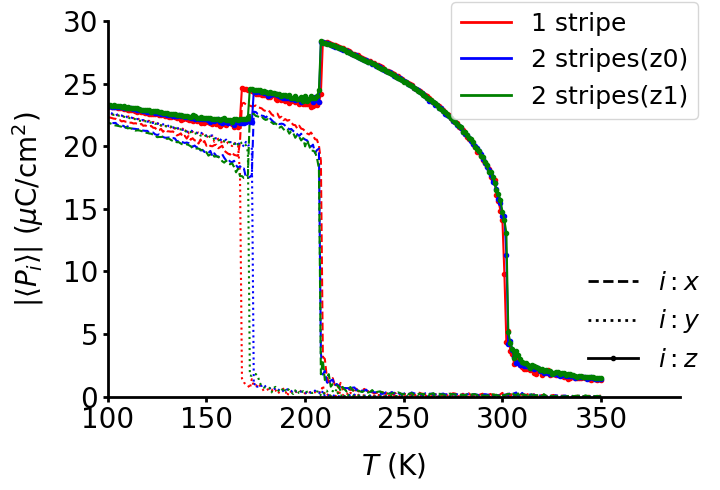}}
        \put(29,101){\text{(b)}}
    \end{Overpic}
    }
    \resizebox{\linewidth}{!}{
    \begin{Overpic}[abs]{\begin{tabular}{p{.38\textwidth}}\vspace{3.8cm}\\\end{tabular}}
        \put(0,0){\includegraphics[height=4cm]{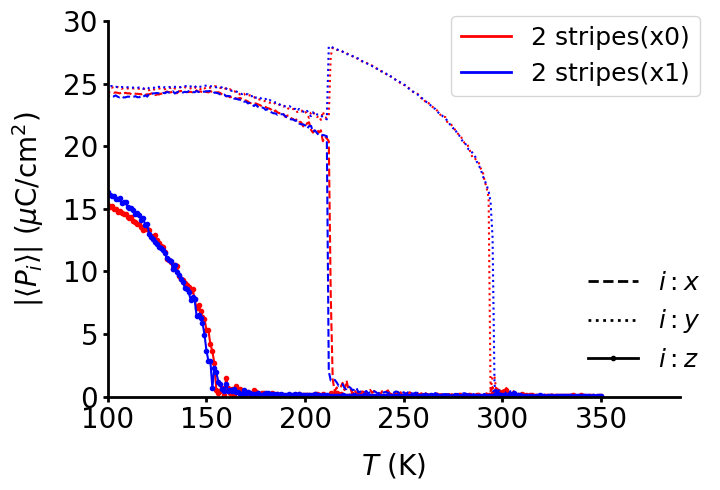}}
        \put(29,101){\text{(c)}}
    \end{Overpic}
    \begin{Overpic}[abs]{\begin{tabular}{p{.38\textwidth}}\vspace{3.8cm}\\\end{tabular}}
        \put(0,0){\includegraphics[height=4cm]{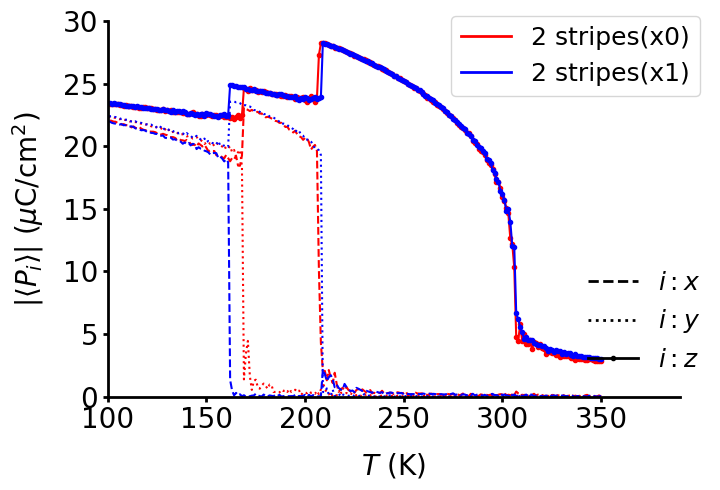}}
        \put(29,101){\text{(d)}}
    \end{Overpic}
    }
    \caption{\rv{Changes of polarization components ($|\braket{P_i}|$, where $i: x,y,z$) in cooling simulations induced by (a)/(c) non-polar defects ($|u_{\text{D}}|=$0.0~\AA) and (b)/(d) polar defects ($|u_{\text{D}}|=$0.1~\AA$\hat{z}$) in 2D stripes for (a)--(b) 1 stripe of $w=6$~u.c. (red), 2 stripes of $w=6$~u.c.\ not separated ($\Delta z=0$, blue) or with a gap ($\Delta z=1$, green), and (c)--(d) 2 stripes of $w=6$~u.c. with $\Delta x=0$ (red) or $\Delta x=1$ (blue).}}
   \label{app:pt_asa}
\end{figure}

\rv{Figure~\ref{app:pt_asa}~(a)--(b) compares the macroscopic polarization in the presence of one and two defect stripes with height $n_z=1$ and width $w=6$ on top of each other. 
Although the addition of the second stripe doubles the number of defects in the system, both stripes together correspond to one stripe with $n_z=2$. Thus the {\it{active surface areas}} for polarization along $y$ and $z$ are not modified and the change along $x$ (from 1 to 2) is negligible. 
In turn, changes of the phase diagram by non-polar defects are barely visible. 
Also, the increasing number of polar defects in this geometry does not modify the T phase. 
The doubled number of polar defects only slightly modifies the polarization in the O and R phases and increases $T_C^{O-R}$ by 6~K. 
Even if the defect stripes are separated by a plane of free dipoles no large changes of the phase diagram are induced as no macroscopic bulk-like polarization can form in the gap. 

Figure~\ref{app:pt_asa}~(c)--(d) shows the phase diagrams if two stripes with $w=6$ and $n_z=1$ are placed next to each other, either without gap (red) or with a gap of one x-plane (blue). 
For non-polar defects, the polarization points parallel to the stripes in T and O phases and their {\it{active surface areas}} do not change with $w$ nor do polarization or phase transition temperatures. 
However, the larger \textit{active surface area} along z changes the character of the O-R transition and reduces the macroscopic polarization in the R phase. 
These results are not modified by a gap of one $x$-plane between the stripes. 
Increasing the number of polar defects in one defect plane increases their active surface area, $T_C$, and polarization in T phase and induced $P_z$ at high temperature increase.
For O/R phases, only $T_C^{\text{O-R}}$ increases slightly by 4~K and $P_z$ increases slightly due to the increased \textit{active surface area} of the defects.
Again, there is no impact by one gap in plane.
}

\newpage

\section{Defect-induced changes of polarization and strain}
\label{sec:app_macro}

\begin{figure}[htbp]
    \centering
    \resizebox{.45\linewidth}{!}{
    \begin{Overpic}[abs]{\begin{tabular}{p{.64\textwidth}}\vspace{10cm}\\\end{tabular}}
        \put(2,130){\includegraphics[width=9cm]{PT_rand3d_c0.png}}
        \put(200,170){\textcolor{white}{\rule{1.8cm}{1.5cm}}}
        \put(43,290){\large\text{(a)}}
        \put(94,237.5){\LARGE $\ast$}
        \put(0,0){\includegraphics[height=4.5cm,trim={0 0 3.2cm 0},clip]{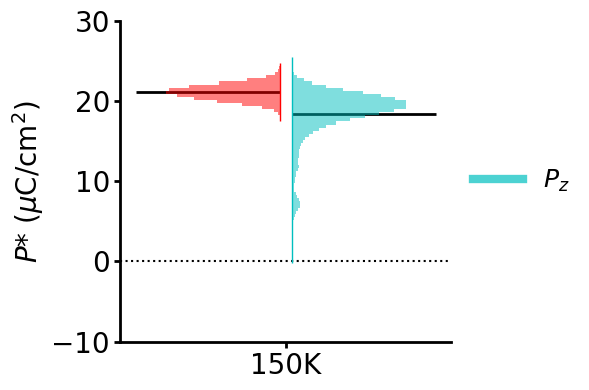}}
        \put(1,28){\textcolor{white}{\rule{0.4cm}{0.9cm}}}
        \put(4,40){\rotatebox[origin=c]{90}{$P_i$}}
        \put(72,0){\textcolor{white}{\rule{1.5cm}{0.45cm}}}
        \put(38,2){\text{Normalized histogram}}
        \put(95,18){\text{$T=150K$}}
        \put(43,118){\large\text{(c)}}
        \put(153,14){\includegraphics[height=3.5cm,trim={1cm 1cm 0 1cm},clip]{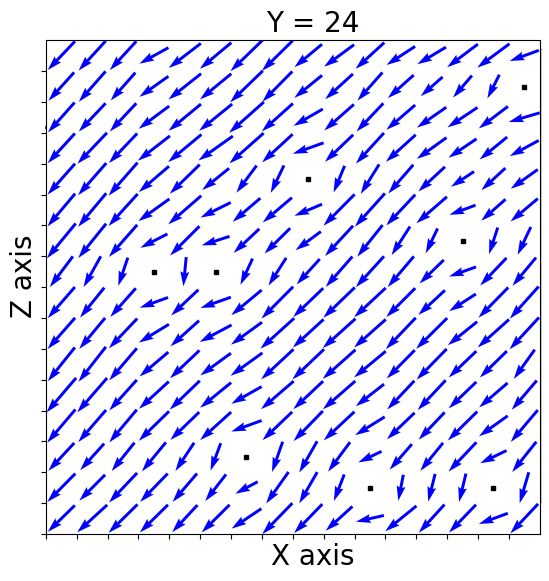}}
        \put(154,118){\large\text{(d)}}
    \end{Overpic}
    }
    \resizebox{.45\linewidth}{!}{
    \begin{Overpic}[abs]{\begin{tabular}{p{.64\textwidth}}\vspace{10cm}\\\end{tabular}}
        \put(2,130){\includegraphics[width=9cm]{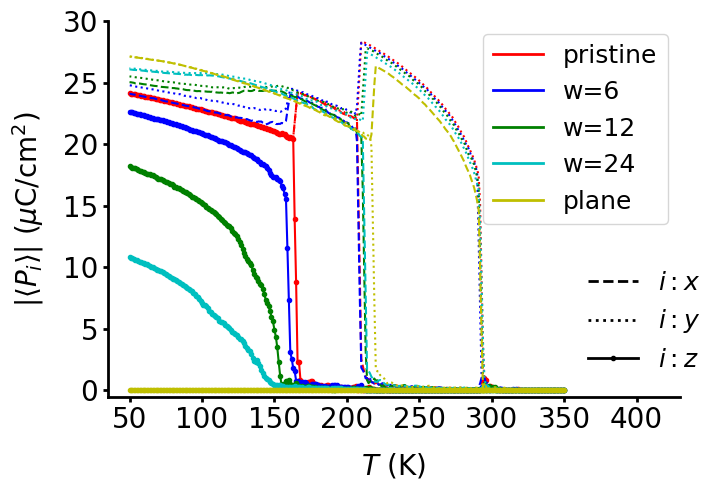}}
        \put(200,170){\textcolor{white}{\rule{1.8cm}{1.5cm}}}
        \put(43,290){\large\text{(b)}}
        \put(68,193){\LARGE $\ast$}
        \put(0,0){\includegraphics[height=4.5cm,trim={0 0 3.2cm 0},clip]{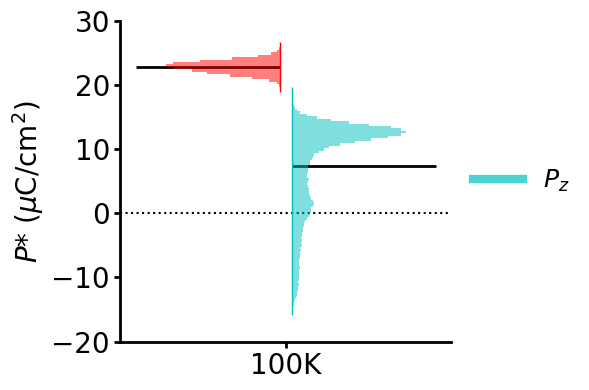}}
        \put(1,28){\textcolor{white}{\rule{0.4cm}{0.9cm}}}
        \put(4,40){\rotatebox[origin=c]{90}{$P_i$}}
        \put(72,0){\textcolor{white}{\rule{1.5cm}{0.45cm}}}
        \put(38,2){\text{Normalized histogram}}
        \put(95,18){\text{$T=100K$}}
        \put(43,118){\large\text{(e)}}
        \put(153,14){\includegraphics[height=3.5cm,trim={.8cm .8cm 0 .8cm},clip]{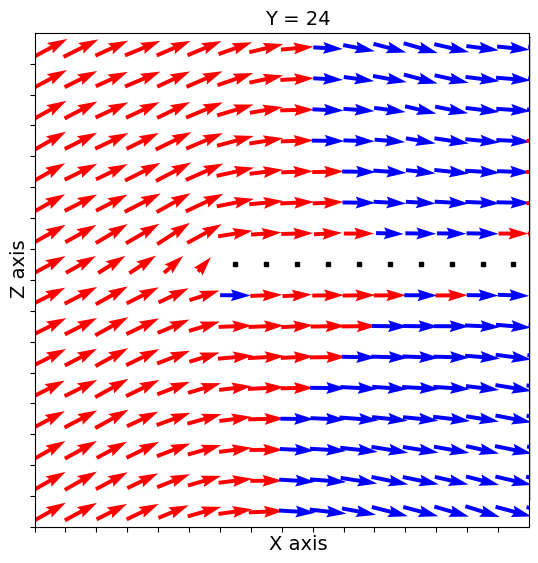}}
        \put(154,118){\large\text{(f)}}
    \end{Overpic}
    }
    \caption{Changes of (a)-(b) polarization components $|\braket{P_i}|$, with $i: x,y,z$, in cooling simulations and (c),(e) of dipole histograms, induced by non-polar defects ($|u_{\text{D}}|=$0.0~\AA) for (a),(c),(d) random distribution and (b),(e),(f)  stripes in $z$-planes with varying width ($w$). 
    For (c) 2~\% defects at 150~K and (e) 1~\% defects ($w=24$~u.c.)  at 100~K, histograms of local $P_z$ (right) are compared to the pristine reference system (left).
    (d) and (f) show corresponding time-averages of local dipoles close to the defects color encoded by the sign of $P_z$.}
    \label{fig:rand3d_ag_snapshots_c0}
\end{figure}

\rv{This appendix collects additional information of defect-induced changes of polarization and strain.}

\rv{Figure~\ref{fig:rand3d_ag_snapshots_c0} compares the change of macroscopic polarization, underlying histograms of one polarization component, and time-averages of local dipoles near the defects in the presence of 2\% and 1\% of non-polar defects being randomly distributed in 3D and agglomerated in 2D stripes, respectively.
The discussion of their association with \textit{active surface areas} can be found in the main paper.}

\begin{figure}[tpb]
    \centering
    \resizebox{.9\linewidth}{!}{
    \begin{Overpic}[abs]{\begin{tabular}{p{.38\textwidth}}\vspace{3.8cm}\\\end{tabular}}
        \put(0,0){\includegraphics[height=4cm]{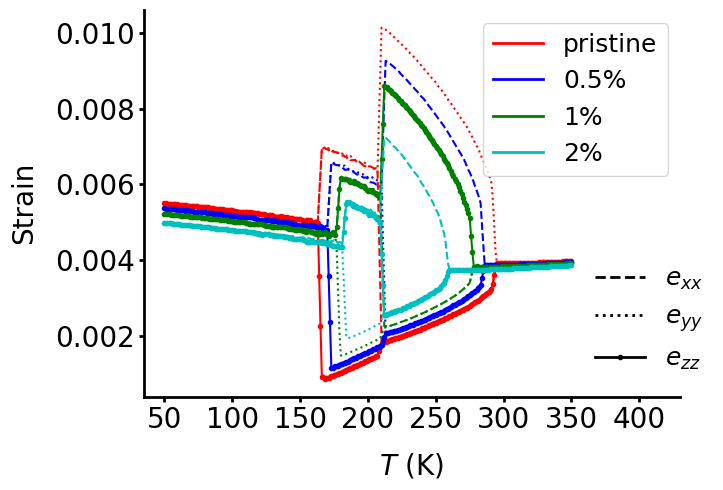}}
        \put(36,101){\text{(a)}}
    \end{Overpic}
    \begin{Overpic}[abs]{\begin{tabular}{p{.38\textwidth}}\vspace{3.8cm}\\\end{tabular}}
        \put(0,0){\includegraphics[height=4cm]{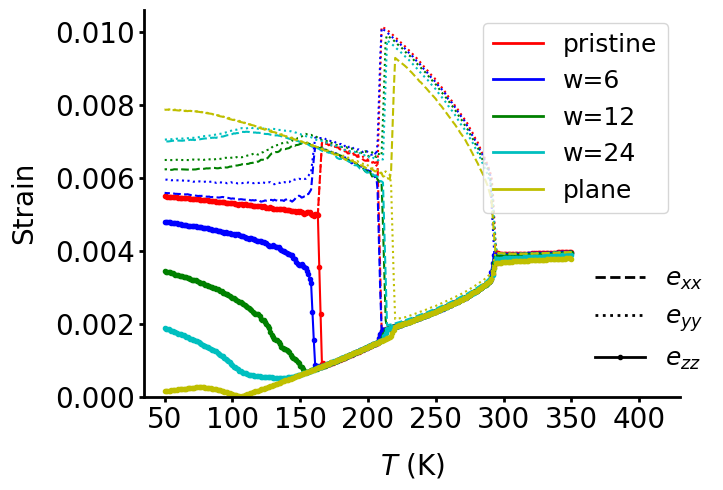}}
        \put(36,101){\text{(b)}}
    \end{Overpic}
    }
    \caption{Changes of strain components $\eta_{ii}$, with $i: x,y,z$, in cooling simulations induced by non-polar defects ($|u_{\text{D}}|=$0.0~\AA) for (a) random distribution and (b)  stripes in $z$-planes with varying width ($w$).}
   \label{app:StrT_rand3d_ag_c0}
\end{figure}

Figure~\ref{app:StrT_rand3d_ag_c0} shows the changes of macroscopic strain components ($\eta_{ii}$ with $i=x,y,z$) induced by \rv{non-polar defects (a) randomly distributed in 3D or agglomerated in 2D stripes of varying width. 
The corresponding changes of macroscopic polarization components ($|\langle P_i\rangle|$, where $i=x,y,z$) are shown in Fig.~\ref{fig:PD_u0}~(c) and Fig.~\ref{fig:rand3d_ag_snapshots_c0}~(b).} 

\newpage
\begin{figure}[tpb]
    \centering
    \resizebox{.9\linewidth}{!}{
    \begin{Overpic}[abs]{\begin{tabular}{p{.38\textwidth}}\vspace{3.8cm}\\\end{tabular}}
        \put(0,0){\includegraphics[height=4cm]{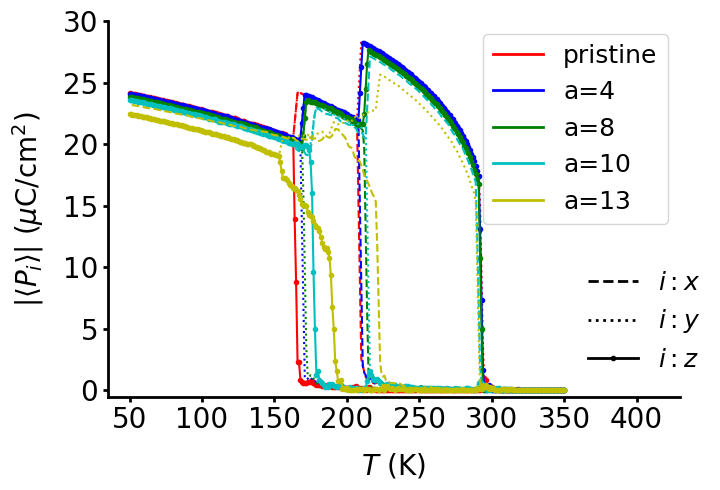}}
        \put(29,101){\text{(a)}}
    \end{Overpic}
    \begin{Overpic}[abs]{\begin{tabular}{p{.38\textwidth}}\vspace{3.8cm}\\\end{tabular}}
        \put(0,0){\includegraphics[height=4cm]{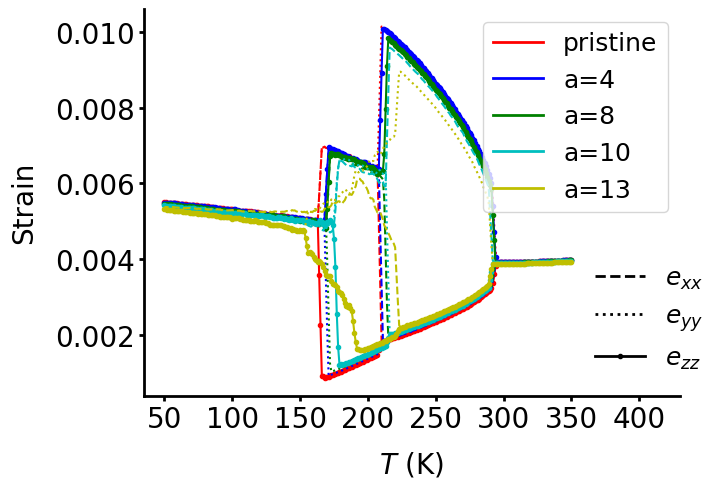}}
        \put(36,101){\text{(b)}}
    \end{Overpic}
    }
    \caption{Changes of (a) polarization components $|\braket{P_i}|$ and (b) strain components $\eta_{ii}$, with $i: x,y,z$, in cooling simulations induced by non-polar defects ($|u_{\text{D}}|=$0.0~\AA) with varying length ($a$) of cube. }
   \label{app:cc_c0}
\end{figure}

\begin{figure}[htb]
  \centering
  \begin{overpic}[abs,width=.45\linewidth]{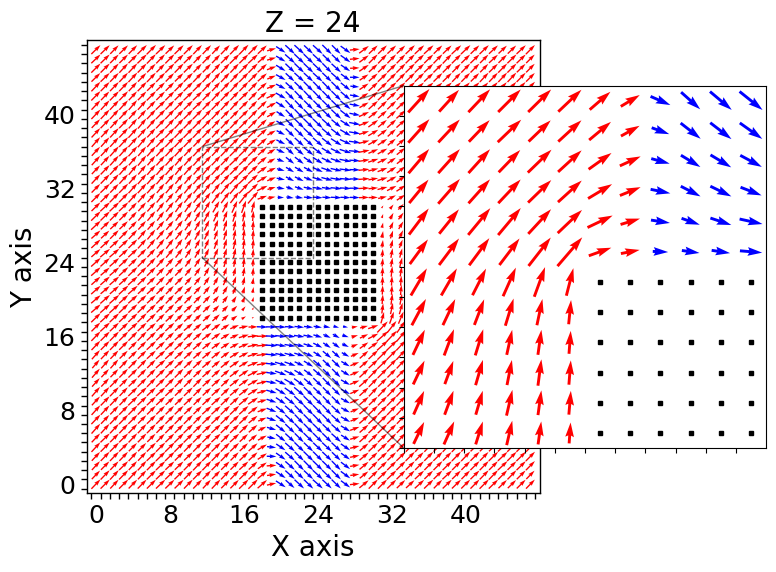}
    \put(0,123){\text{(a)}}
  \end{overpic}
  \begin{overpic}[abs,width=.45\linewidth]{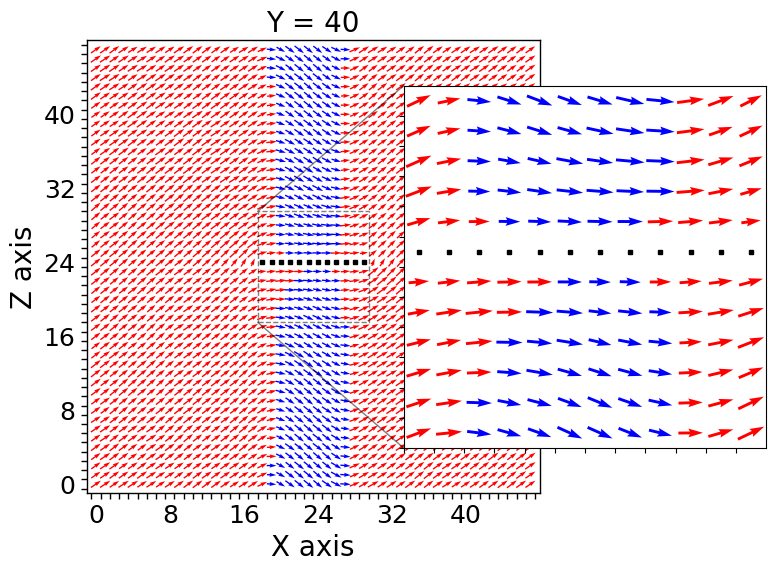}
    \put(0,123){\text{(b)}}
  \end{overpic}
	\caption{\rv{Cross section of time-averaged local dipoles in the presence of (a) a cube with 2~\% ($a=13$~u.c) or (b) a 2D agglomerate with 0.5~\% ($w=12$~u.c.) of non-polar defects at 100~K (R phase). Colors indicate the sign of $P_z$s and highlight the formation of R71 domain walls. The insets show the polarization rotation at the interface of the agglomerate.}}
   \label{app:snapshot_ag_cc_c0}
\end{figure}
\rv{Figure~\ref{app:cc_c0} (a) and (b) illustrate how macroscopic polarization and strain change with the size of cubic agglomerates of non-polar defects. 
As shown in Fig.~\ref{app:snapshot_ag_cc_c0}~(a), the cube with $a=13$~u.c. induces a multi-domain structure with R71 domain walls above/below the defect agglomerate. 
At the same time, the O--R phase transition becomes continuous.} 
Figure~\ref{app:snapshot_ag_cc_c0}~(b) shows the formation of R71 walls at a 2D agglomerates with $w=12$~u.c.
\newpage

\begin{figure}[htb]
    \centering
    \resizebox{.9\linewidth}{!}{
    \begin{Overpic}[abs]{\begin{tabular}{p{.38\textwidth}}\vspace{3.8cm}\\\end{tabular}}
        \put(0,0){\includegraphics[height=4cm]{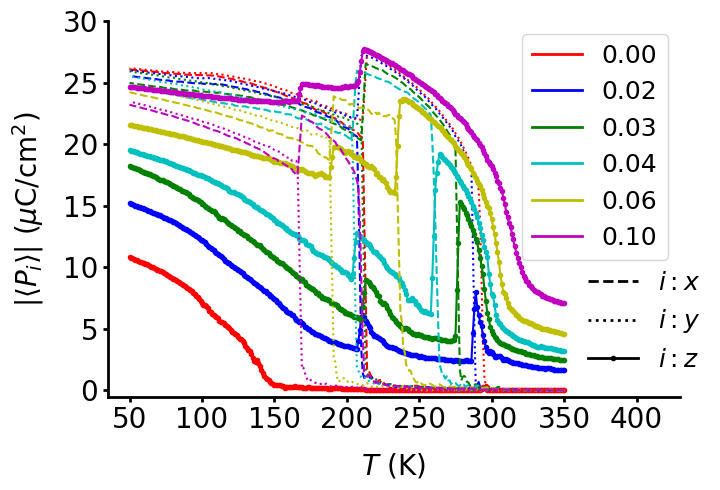}}
        \put(29,101){\text{(a)}}
    \end{Overpic}
    \begin{Overpic}[abs]{\begin{tabular}{p{.38\textwidth}}\vspace{3.8cm}\\\end{tabular}}
        \put(0,0){\includegraphics[height=4cm]{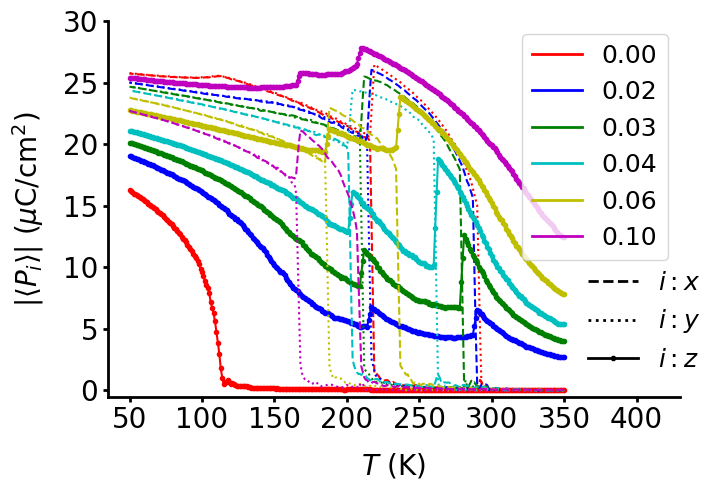}}
        \put(29,101){\text{(b)}}
    \end{Overpic}
    }
    \resizebox{.9\linewidth}{!}{
    \begin{Overpic}[abs]{\begin{tabular}{p{.38\textwidth}}\vspace{3.5cm}\\\end{tabular}}
        \put(0,0){\includegraphics[height=3.5cm]{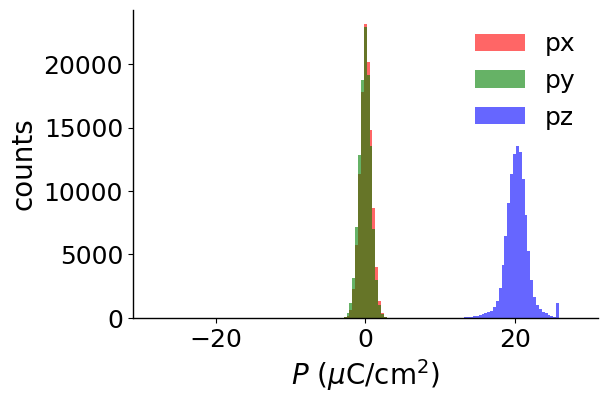}}
        \put(35,85){\text{(c)}}
    \end{Overpic}
    \begin{Overpic}[abs]{\begin{tabular}{p{.38\textwidth}}\vspace{3.5cm}\\\end{tabular}}
        \put(0,0){\includegraphics[height=3.5cm]{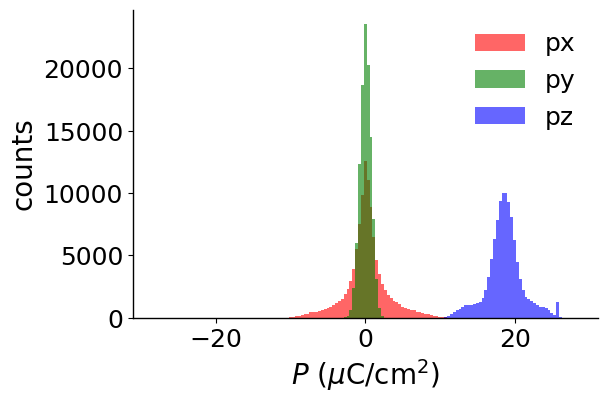}}
        \put(35,85){\text{(d)}}
    \end{Overpic}
    }
    \caption{Impact of the strength of 1~\% defects ($|u_{\text{D}}|$ $\hat{z}$) on the changes of polarization components $|\braket{P_i}|$, with $i: x,y,z$, in cooling simulations for (a) random distribution \rv{in 2D} and (b) stripes. Colors encode the strength of the defect dipoles from 0.00 (red), 0.02 (blue), 0.03 (green), 0.04 (light blue), 0.06 (light green) to 0.10\rv{~\AA{}} (purple).
    \rv{(c)--(d) Histograms of local polarization at 300~K for $|u_{\text{D}}|=0.1$~\AA~$\hat{z}$ for (c) random distribution in 2D and (d) stripes.}}
   \label{app:ag_24_ds_24x48_cp}
\end{figure}
The impact of the dipole strength on the macroscopic polarization is summarized in \rv{Fig.~\ref{app:ag_24_ds_24x48_cp} for 1~\% defects (a) randomly distributed in 2D and (b) in stripes of $w=24$~u.c.
Subfigures (c) and (d) show the corresponding histograms of the local polarization for the example of $|u_{\text{D}}|=$0.1~\AA{} at 300~K.
In case of random distribution, the system is in a single domain T phase, with one peak of $P_z$ centered around 20~\rrrv{$\mu$C/cm$^2$}. 
In case of defect stripes, this peak shows broadened distribution due to polarization rotation around the stripes.
Both cases show a second small peak at 25.6~\rrrv{$\mu$C/cm$^2$}, which belongs to the defect dipoles.
The cross-over as shown in Fig.~\ref{fig:PD_um}~(b) is therefore determined by the relative strength of the free and defect dipoles.
For weak defects ($|u_{\text{D}}|\leq 0.03$~\AA{}$\hat{z}$), the defect dipole strength is smaller than the free dipoles, thus depolarization fields dominate.
Once the strength of the defect dipole is larger than the typical values of free dipoles at slightly above $T_C^\text{T-C}$, these defects start to act as internal bias fields.
}
\newpage

\section{Field hysteresis}
\label{sec:app_hys}
\begin{figure}[tpb]
    \centering
    \resizebox{\linewidth}{!}{
    \begin{Overpic}[abs]{\begin{tabular}{p{.43\textwidth}}\vspace{4cm}\\\end{tabular}}
        \put(0,0){\includegraphics[scale=.47,clip,trim=0cm 0cm 4cm 0cm]{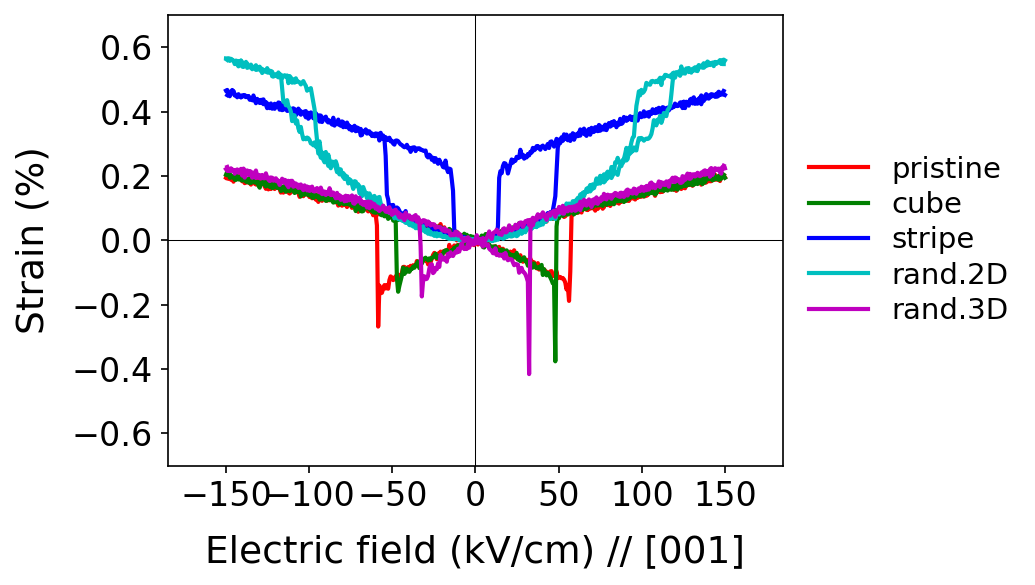}}
        \put(133,5){\includegraphics[scale=.47,trim={14cm 2.5cm 0 2cm},clip]{StrEhys_1p_cpz0.0_ez_250K.png}}
        \put(0,10){\textcolor{white}{\rule{.45cm}{4cm}}}
        \put(2,75){\rotatebox[origin=c]{90}{\text{$\eta_{ii}$ (\%)}}}
        \put(28,0){\textcolor{white}{\rule{5cm}{.45cm}}}
        \put(78,4.5){\text{$E_z$ (\rrrv{kV/cm})}}
        \put(90,118){\text{(a)}}
    \end{Overpic}
    \begin{Overpic}[abs]{\begin{tabular}{p{.34\textwidth}}\vspace{4cm}\\\end{tabular}}
        \put(0,0){\includegraphics[scale=.47,clip,trim=2.6cm 0cm 4cm 0cm]{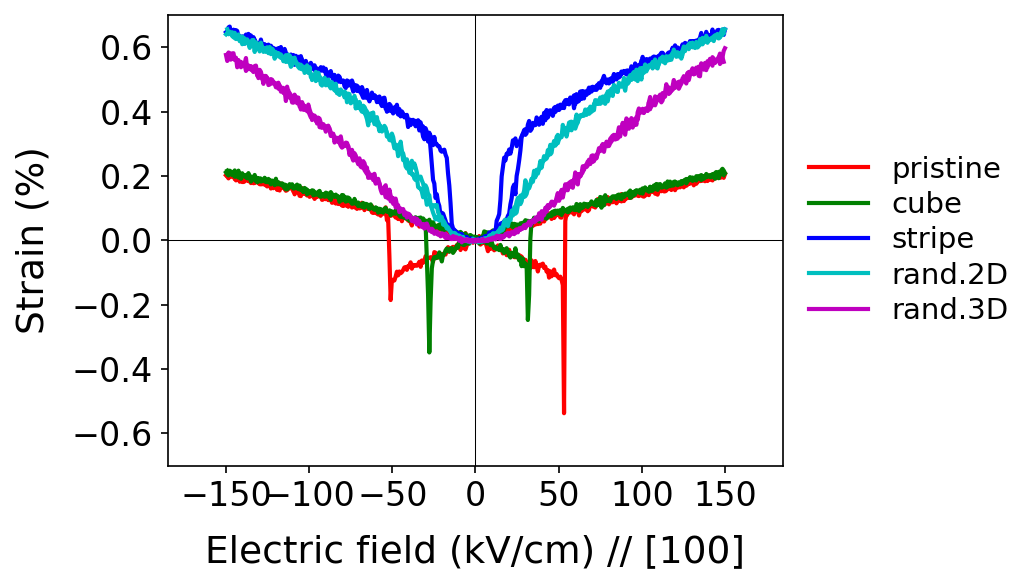}}
        \put(97,5){\includegraphics[scale=.47,trim={14cm 2.5cm 0 2cm},clip]{StrEhys_1p_cpz0.1_ex_250K.png}}
        \put(0,0){\textcolor{white}{\rule{5cm}{.45cm}}}
        \put(43,4.5){\text{$E_x$ (\rrrv{kV/cm})}}
        \put(55,118){\text{(b)}}
    \end{Overpic}
    \begin{Overpic}[abs]{\begin{tabular}{p{.4\textwidth}}\vspace{4cm}\\\end{tabular}}
        \put(0,0){\includegraphics[scale=.47,clip,trim=2.6cm 0cm 4cm 0cm]{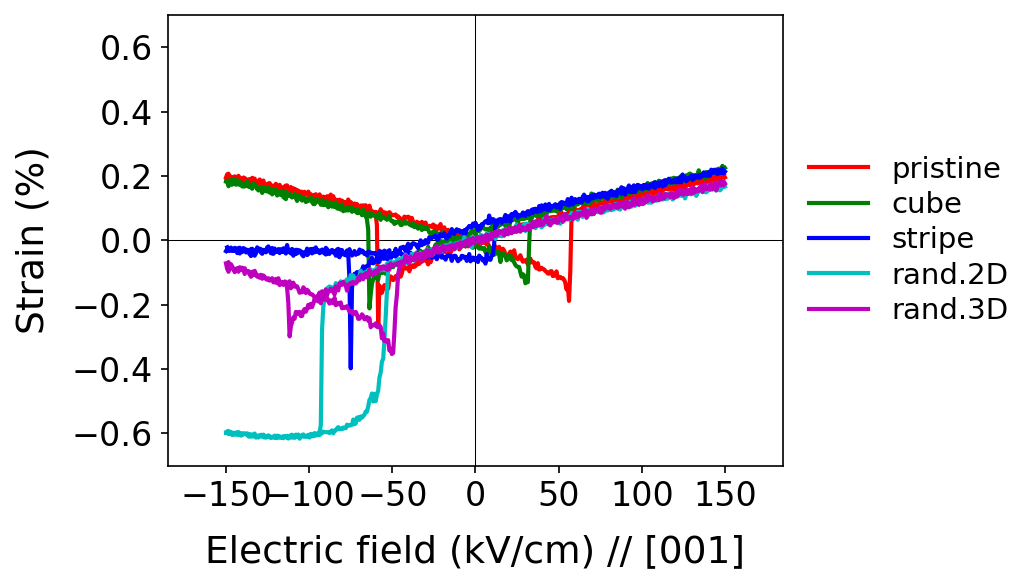}}
        \put(97,5){\includegraphics[scale=.47,trim={14cm 2.5cm 0 2cm},clip]{StrEhys_1p_cpz0.1_ez_250K.png}}
        \put(0,0){\textcolor{white}{\rule{5cm}{.45cm}}}
        \put(43,4.5){\text{$E_z$ (\rrrv{kV/cm})}}
        \put(55,118){\text{(c)}}
    \end{Overpic}
    }
  \caption{Changes of the field hysteresis at $T=250$~K if going from the pristine material (red) to systems with different distribution of 1~\% defects  (green: cube; blue and cyan: stripes and random in $\hat{z}$-planes; purple: random in 3D) for (a) non-polar defects and (b)--(c) polar defect dipoles ($|u_{\text{D}}|=0.1$~\AA $\hat{z}$) for the electric field (b)  perpendicular ($E_x$)  and (c) collinear ($E_z$) to the defect dipoles. Only the diagonal strain component $\eta_{ii}$ along the field direction is shown.}
  \label{app:sehys_1p}
\end{figure}

\begin{figure}[hb]
    \centering
    \resizebox{.9\linewidth}{!}{
    \begin{Overpic}[abs]{\begin{tabular}{p{.48\textwidth}}\vspace{4.5cm}\\\end{tabular}}
        \put(0,0){\includegraphics[scale=.47,clip,trim=0cm 0cm 3.2cm 0cm]{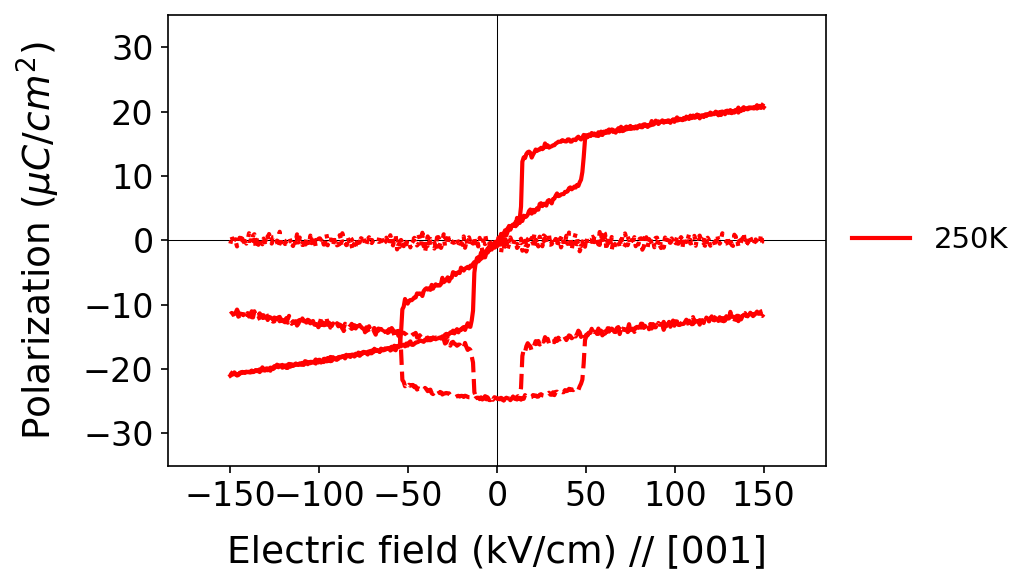}}
        \put(0,10){\textcolor{white}{\rule{.45cm}{4cm}}}
        \put(2,75){\rotatebox[origin=c]{90}{\text{$P_i$ (\rrrv{$\mu$C/cm$^2$})}}}
        \put(30,0){\textcolor{white}{\rule{6cm}{.45cm}}}
        \put(78,4.5){\text{$E_z$ (\rrrv{kV/cm})}}.
        \put(170,50){\small\text{$P_x$}}
        \put(170,83){\small\text{$P_y$}}
        \put(170,112){\small\text{$P_z$}}
        \put(37,118){\text{(a)}}
    \end{Overpic}
    \begin{Overpic}[abs]{\begin{tabular}{p{.48\textwidth}}\vspace{4.5cm}\\\end{tabular}}
        \put(0,0){\includegraphics[scale=.47,clip,trim=0cm 0cm 3.2cm 0cm]{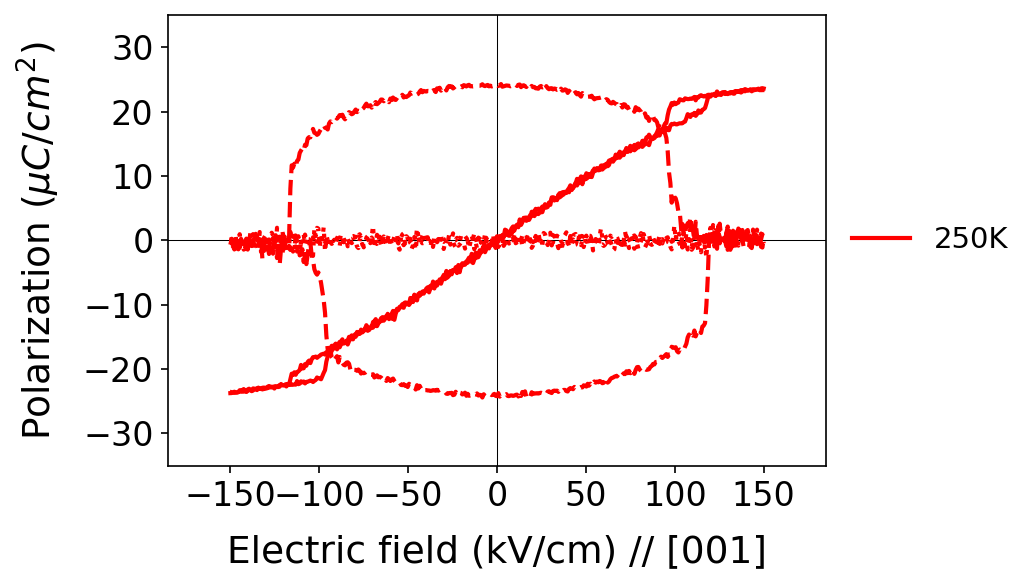}}
        \put(0,10){\textcolor{white}{\rule{.45cm}{4cm}}}
        \put(2,75){\rotatebox[origin=c]{90}{\text{$P_i$ (\rrrv{$\mu$C/cm$^2$})}}}
        \put(30,0){\textcolor{white}{\rule{6cm}{.45cm}}}
        \put(78,4.5){\text{$E_z$ (\rrrv{kV/cm})}}.
        \put(158,50){\small\text{$P_x$}}
        \put(170,83){\small\text{$P_y$}}
        \put(170,115){\small\text{$P_z$}}
        \put(37,118){\text{(b)}}
    \end{Overpic}
    }
    \resizebox{.9\linewidth}{!}{
    \begin{Overpic}[abs]{\begin{tabular}{p{.48\textwidth}}\vspace{4.5cm}\\\end{tabular}}
        \put(0,0){\includegraphics[scale=.47,clip,trim=0cm 0cm 3.2cm 0cm]{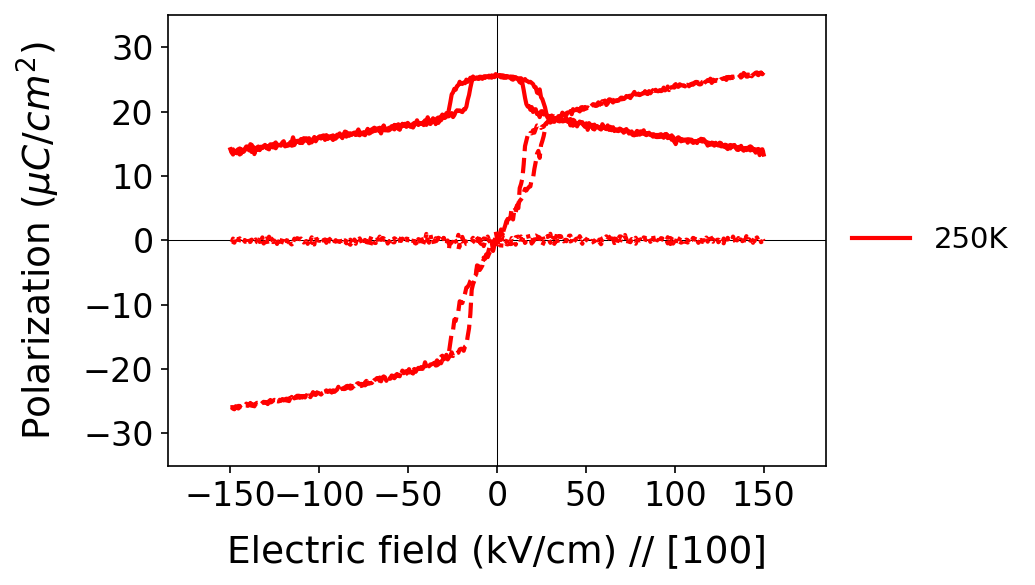}}
        \put(0,10){\textcolor{white}{\rule{.45cm}{4cm}}}
        \put(2,75){\rotatebox[origin=c]{90}{\text{$P_i$ (\rrrv{$\mu$C/cm$^2$})}}}
        \put(30,0){\textcolor{white}{\rule{6cm}{.45cm}}}
        \put(78,4.5){\text{$E_x$ (\rrrv{kV/cm})}}.
        \put(170,90){\small\text{$P_z$}}
        \put(170,67){\small\text{$P_y$}}
        \put(170,115){\small\text{$P_x$}}
        \put(37,118){\text{(c)}}
    \end{Overpic}
    \begin{Overpic}[abs]{\begin{tabular}{p{.48\textwidth}}\vspace{4.5cm}\\\end{tabular}}
        \put(0,0){\includegraphics[scale=.47,clip,trim=0cm 0cm 3.2cm 0cm]{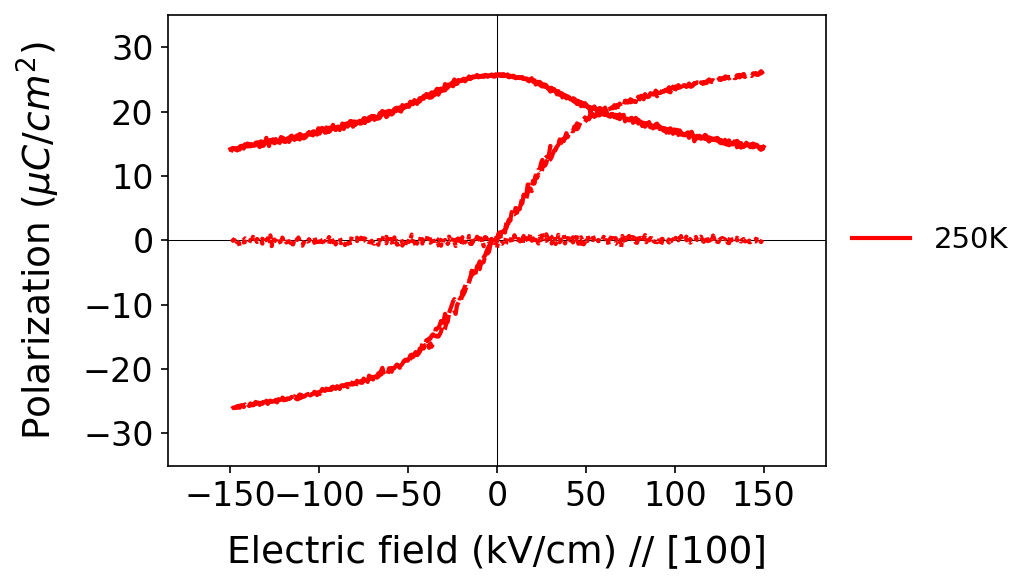}}
        \put(0,10){\textcolor{white}{\rule{.45cm}{4cm}}}
        \put(2,75){\rotatebox[origin=c]{90}{\text{$P_i$ (\rrrv{$\mu$C/cm$^2$})}}}
        \put(30,0){\textcolor{white}{\rule{6cm}{.45cm}}}
        \put(78,4.5){\text{$E_x$ (\rrrv{kV/cm})}}.
        \put(170,90){\small\text{$P_z$}}
        \put(170,67){\small\text{$P_y$}}
        \put(170,115){\small\text{$P_x$}}
        \put(37,118){\text{(d)}}
    \end{Overpic}
    }
    \caption{Field dependence of polarization component $P_i$ with $i=x,y,z$ at $T=250$~K with 1~\% defects for (a,c) stripes and (b,d) randomly in one plane for (a,b) non-polar defects and (c,d) polar defects with ($|u_{\text{D}}|=$0.1~\AA $\hat{z}$) for the electric field (a,b) along the $z$ or (c,d) perpendicular ($E_x$) to the defect dipoles.}
   \label{app:pehys_ag_ds_pxyz}
\end{figure}

\begin{figure}[htb]
    \centering
    \resizebox{.85\linewidth}{!}{
    \begin{Overpic}[abs]{\begin{tabular}{p{.35\textwidth}}\vspace{4.5cm}\\\end{tabular}}
        \put(0,0){\includegraphics[scale=.35]{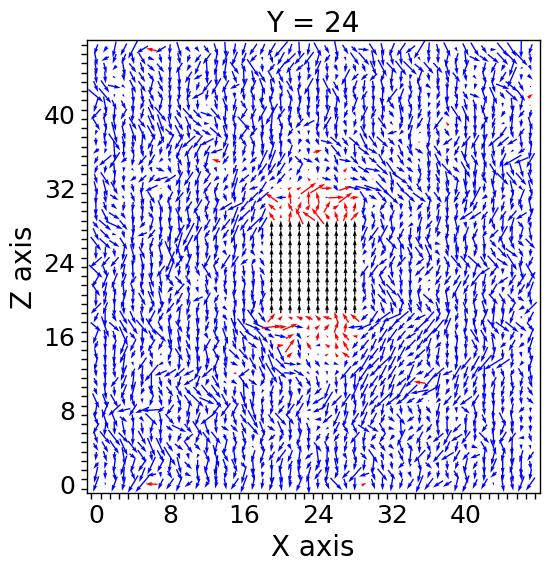}}
        \put(0,125){\text{(a)}}
    \end{Overpic}
    \begin{Overpic}[abs]{\begin{tabular}{p{.35\textwidth}}\vspace{4.5cm}\\\end{tabular}}
        \put(0,0){\includegraphics[scale=.35]{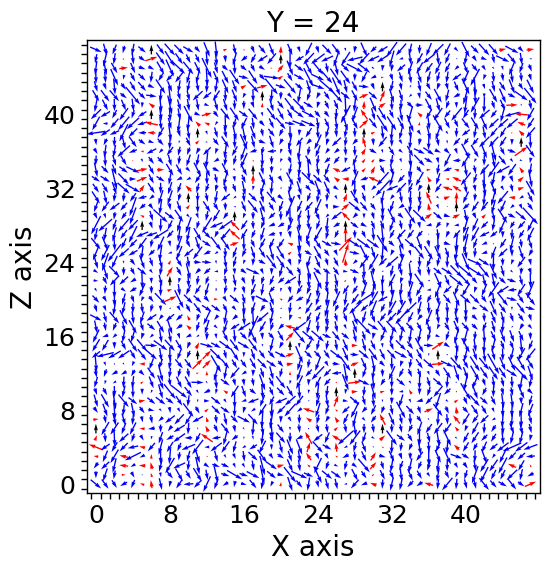}}
        \put(0,125){\text{(b)}}
    \end{Overpic}
    }
    \resizebox{.85\linewidth}{!}{
    \begin{Overpic}[abs]{\begin{tabular}{p{.35\textwidth}}\vspace{4.5cm}\\\end{tabular}}
        \put(0,0){\includegraphics[scale=.35]{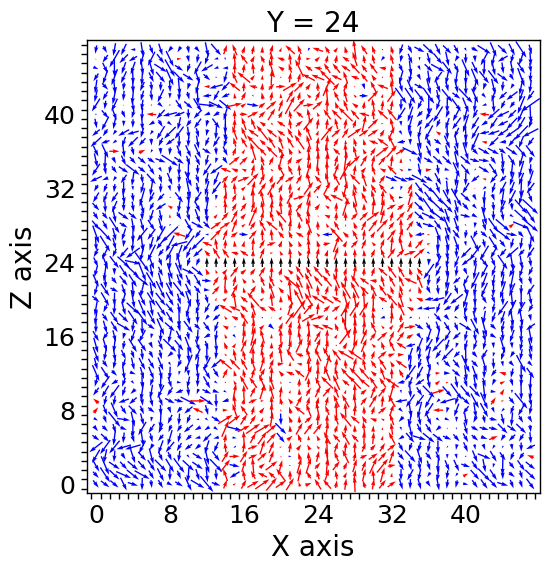}}
        \put(0,125){\text{(c)}}
    \end{Overpic}
    \begin{Overpic}[abs]{\begin{tabular}{p{.35\textwidth}}\vspace{4.5cm}\\\end{tabular}}
        \put(0,0){\includegraphics[scale=.35]{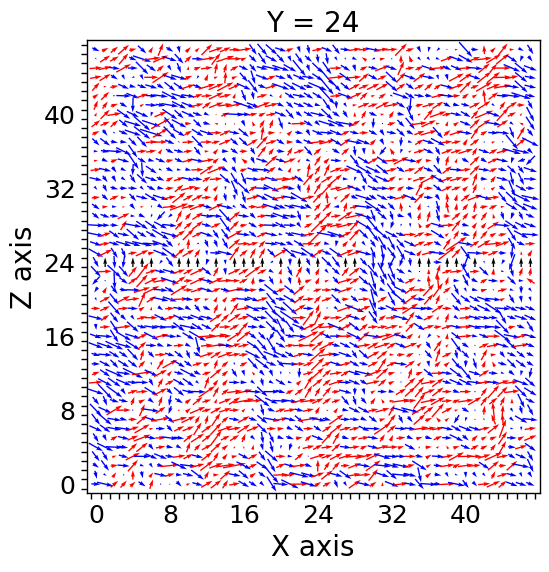}}
        \put(0,125){\text{(d)}}
    \end{Overpic}
    }
    \caption{Snapshots of dipoles at 250~K for 1~\% defects ($|u_{\text{D}}|=$0.1~\AA $\hat{z}$) in (a) cubic agglomerate of $a=10$~u.c., (b) 3D random, (c) 2D agglomerate of $w=24$~u.c. and (d) 2D random in a plane with the applied field $-150$~kV/cm $\hat{z}$. Colors indicate the sign of $P_z$s. Note that all these snapshots are not time-averaged and are thus noisy.}
   \label{app:snapshots_field}
\end{figure}

\rv{
Figure~\ref{app:sehys_1p} (a), (b) and (c) show the strain--field curves for non-polar defects with the electric field $E_z$ and polar defects with the electric field $E_x$ or $E_z$ applied.
Figure~\ref{app:pehys_ag_ds_pxyz} (a) and (b) show all 3 polarization components for non-polar defects with bipolar field $E_z$ applied, while Fig.~\ref{app:pehys_ag_ds_pxyz} (a) and (b) are for polar defects ($|u_{\text{D}}|=0.1$~\AA{}$\hat{z}$) with the bipolar electric field $E_x$  perpendicular to the defect dipoles.
Figure~\ref{app:snapshots_field} shows snapshots of systems with 1~\% defects of cubic agglomerate, 3D random distribution, 2D agglomerate, and 2D random distribution at $E_z=-150$~kV/cm, showing pinned domains near defects that could act as nucleation centers for the macroscopic switching towards the positive polarization direction.}

\end{document}